\documentclass[a4paper,11pt,titlepage,twoside,openright]{report}
\pdfoutput=1
\usepackage[english]{babel}
\usepackage{graphicx}
\usepackage{geometry}

\usepackage{sectsty}
\usepackage{pdfpages}
\usepackage{epstopdf}
\chapterfont{\centering}
\usepackage{chngcntr}
\counterwithout{footnote}{chapter}
\usepackage{verbatim}
\usepackage{makeidx}
\usepackage{color}
\usepackage{amsfonts}
\usepackage{slashed}
\usepackage{amssymb,amsmath} 
\usepackage{dsfont}
\usepackage{multirow} 
\usepackage{cite}
\usepackage{tabularx}
\usepackage{tikz}

\usepackage{colortbl}
\usepackage{xcolor}

\usepackage{fancybox}

\usepackage{feynarts}
\usepackage{bbm}
\allowdisplaybreaks

\global\oddsidemargin=6mm
\global\evensidemargin=-0.5mm

\input epsf
\input pix.sty

\newcommand{\bit}{\begin{itemize}}
\newcommand{\eit}{\end{itemize}}
\newcommand{\bc}{\begin{center}}
\newcommand{\ec}{\end{center}}

\newcommand{\ie}{{i.e.}}

\newcommand{\sumint}[1]{\mbox{$\sum$}\!\!\!\!\!\!\!\int_{#1}}
\renewcommand{\eq}{eq.~}
\renewcommand{\nr}[1]{(\ref{#1})}
\renewcommand{\eqs}{eqs.~}

\renewcommand{\fig}{fig.~}

\newcommand{\dd}{\mathrm{d}}
\newcommand{\tinymsbar}{{\overline{\mbox{\tiny\rm{MS}}}}}
\newcommand{\Lambdamsbar}{{\Lambda_\tinymsbar}}

\newcommand{\alphas}{\alpha_{\rm s}}

\newcommand{\Nc}{N_{\rm c}}

\newcommand{\Tc}{T_{\rm c}}
\newcommand{\gB}{g_\rmii{B}}
\newcommand{\mE}{m_\rmii{E}}

\newcommand{\gammaE}{\gamma_\rmii{E}}

\newcommand{\rmO}{{\mathcal{O}}}
\newcommand{\bmu}{\bar\Lambda} 
\newcommand{\CA}{\Nc}

\def\lsi{\raise0.3ex\hbox{$<$\kern-0.75em\raise-1.1ex\hbox{$\sim$}}}
\def\gsi{\raise0.3ex\hbox{$>$\kern-0.75em\raise-1.1ex\hbox{$\sim$}}}

\newcommand{\nB}{n_\rmii{B}}
\newcommand{\rmii}[1]{{\mbox{\tiny\rm{#1}}}}

\newcommand{\im}{\mathop{\mbox{Im}}}

\newcommand{\Tint}[1]{{\hbox{$\sum$}\!\!\!\!\!\!\!\int\,}_{\!\!\!\!\raise-0.9ex\hbox{$\scriptstyle{#1}$}}}
\newcommand{\Tinti}[1]{{{\Sigma}\!\!\!\!\raise0.3ex\hbox{$\int$}_\rmii{${#1}$}}}
\renewcommand{\Tint}[1]{\sumint{#1}}

\newcommand{\bi}{\begin{itemize}}
\newcommand{\ei}{\end{itemize}}

\newcommand{\hide}[1]{ }

\newcommand{\It}[2]{\mathcal{I}_\rmi{#1}^\rmi{#2}}

\renewcommand{\i}{i}

\renewcommand{\v}{{ii}} 
\newcommand{\iv}{iii} 
\newcommand{\ii}{iv} 

\newcommand{\mychapter}[1]
	{
	\chapter{#1}	
	\markboth{\thechapter.~~{#1}}{\thechapter.~~{#1}}
	}
\newcommand{\mysection}[1]
	{
	\section{#1}	
	\markright{\thesection.~~{#1}}
	}

\newcommand{\myappendix}
	{
	\renewcommand\chaptername{Appendix}
	\renewcommand{\thechapter}{\Alph{chapter}}
	}

\author{Yan Zhu\\
Universit\"at Bielefeld\\}

\title{ \mbox{On the Perturbative Evaluation of Thermal Green's Functions}\\
       \mbox{ in the Bulk and Shear Channels of Yang-Mills Theory}}
\date{\today}

\begin{document}
\pagestyle{empty}


\newgeometry{left=3cm}

\vglue0.2cm


\begin{center}
{ \LARGE{On the Perturbative Evaluation of Thermal Green's Functions}\\
\vglue0.15cm
       \LARGE{ in the Bulk and Shear Channels of Yang-Mills Theory}}
\end{center}

\vglue1cm
{
\begin{center}
 \large Yan Zhu

\end{center}
\vglue1cm
\vglue2cm

{\large
{\noindent  Vollst\"andiger Abdruck der von der Fakult\"at f\"ur Physik der Universit\"at Bielefeld zur Erlangung}
des akademischen Grades eines
\begin{center}
{\textit{Doktors der Naturwissenschaften (Dr. rer. nat.)}}
\end{center}
genehmigten Dissertation.}

\vglue4cm

{\large
{\noindent Pr\"ufer der Dissertation (Gutachter):}

\hspace{1cm}Privatdozent Dr. Aleksi Vuorinen

\hspace{1cm}Prof. Dr. York Schr\"oder

\noindent Weitere Mitglieder des Pr\"ufungsausschusses:

\hspace{1cm}Prof. Dr. Dietrich B\"odeker

\hspace{1cm}Prof. Dr. Walter Pfeiffer

}

\vglue1cm

\noindent{\large Tag der m\"undlichen Pr\"ufung (Disputation): 23 Juli, 2013}

\restoregeometry


\cleardoublepage
\pagestyle{plain}
\pagenumbering{roman}
\setcounter{page}{1}
\tableofcontents

\cleardoublepage
\pagenumbering{arabic}
\setcounter{page}{1}

%
%
%
%
%
%

\chapter*{\centering Preface}
\addcontentsline{toc}{chapter}{Preface}
 
This thesis is based on work carried out at the Faculty of Physics of Bielefeld University between the fall of 2010 and the spring of 2013. Its main purpose has been to develop and apply new methods for the perturbative evaluation of energy momentum tensor correlators --- in particular spectral functions --- within SU($N_c$) Yang-Mills theory. The motivation for these computations originates from heavy ion physics, and more specifically the hydrodynamic description of the evolution of thermalizing quark gluon plasma (QGP), which needs the values of the bulk and shear viscosities of the matter as input parameters. A nonperturbative first principles determination of these quantities has turned out to be a daunting task, and at the moment the most promising approach for it is based on the analytic continuation of lattice data from Euclidean correlation functions to Minkowskian signature. This process is, however, numerically very challenging and necessitates gathering as much analytic information of the spectral densities as possible. This is where perturbative calculations, such as the work presented in this thesis, come into play.

The bulk of the scientific results contained in this thesis were published in three articles \cite{Laine:2011xm,Schroder:2011ht,Zhu:2012be}, on which our presentation largely relies. Before introducing the methods and results of these papers, we, however, want to provide a concise introduction to the general topic of heavy ion physics and thermal Quantum Chromodynamics (QCD). To this end, the thesis is structured as follows:

\begin{itemize}

\item In chapter 1, we begin with a generic introduction to QCD at nonzero temperatures, covering how equilibrium thermodynamic quantities, such as thermodynamic potentials and thermal Green's functions are defined and calculated.

\item In chapter 2, we will specialize to the theoretical description of the QGP, and review in particular the role of hydrodynamical simulations in phenomenological heavy ion physics. Different ways of determining the transport coefficients of the plasma are also discussed.

\item In chapter 3, we will move on to introduce the main quantities studied in the thesis: The bulk and shear channel thermal Green's functions in pure SU($N_c$) Yang-Mills theory. The quantities are first defined and their most central properties discussed, after which it is explained, how one can determine their ultraviolet (UV) behavior in a relatively straightforward way. This chapter is largely based on ref.~\cite{Schroder:2011ht}.

\item In chapter 4, we will introduce the most important technical tools developed in this thesis, \ie~a novel method to evaluate next-to-leading order (NLO) thermal spectral functions. Specific examples are considered both in the bulk \cite{Laine:2011xm} and shear channels \cite{Zhu:2012be}.

\item In chapter 5, we will present and discuss the numerical results obtained for the bulk and shear spectral densities in refs. \cite{Laine:2011xm,Zhu:2012be}. Special attention is paid on the convergence and applicability of the results, as well as their use in the verification of sum rules and the determination of Euclidean imaginary time correlation functions. In the bulk channel case, the latter results are furthermore compared to lattice data.

\item In chapter 6, we finally summarize our findings and discuss their physical implications.

\end{itemize}

In addition to the chapters listed above, appendices \ref{app:master} and \ref{fs} contain several details of the calculations, and are frequently referred to in particular in chapters \ref{CHAP 4} and \ref{CHAP 5}.

%


\setcounter{chapter}{0}
\pagestyle{myheadings}

\mychapter{Introduction to thermal QCD}
\la{CHAP 1}

\mysection{Strong interactions at nonzero temperature}

As one of the four fundamental interactions in nature, the strong interaction of the color charge has attracted a lot of attention in modern particle physics. In the context of the Standard Model, the elementary particles involved in the strong interaction are the quarks and gluons. There are altogether six species, or ``flavors'', of quarks, u, d, s, c, b and t, which stand for the up, down, strange, charm, bottom, and top quarks, respectively. The quarks are Dirac fermions with spin $s=\fr12$, and have either $\fr23$ (u, c, t) or $-\fr{1}{3}$ (d, s, b) units of electric charge. They also carry an extra ``color'' index that takes three values (often called red, green, and blue), and each quark in addition comes with a corresponding antiquark with opposite quantum numbers. The gluons on the other hand are eight electrically neutral gauge bosons with $s=1$ that mediate the strong force between quarks. Unlike in electromagnetism, the gluons feel each others' presence, \ie~also carry color charge.

The fact that particles with fractional charge have not been observed is explained by the ``confining'' nature of the strong interaction. This has to do with the fact that the strong force does not decrease with increasing distance, but if one e.g.~tries to break a meson --- a bound state composed of a quark and an antiquark --- one will eventually create an additional quark-antiquark pair from the vacuum and end up with two mesons. At zero temperature, quarks and gluons indeed always form colorless bound states, hadrons, with integer electric charge; they are either called mesons or baryons, the latter being composed of three quarks.

\begin{figure}
\bc
\includegraphics[width=10cm]{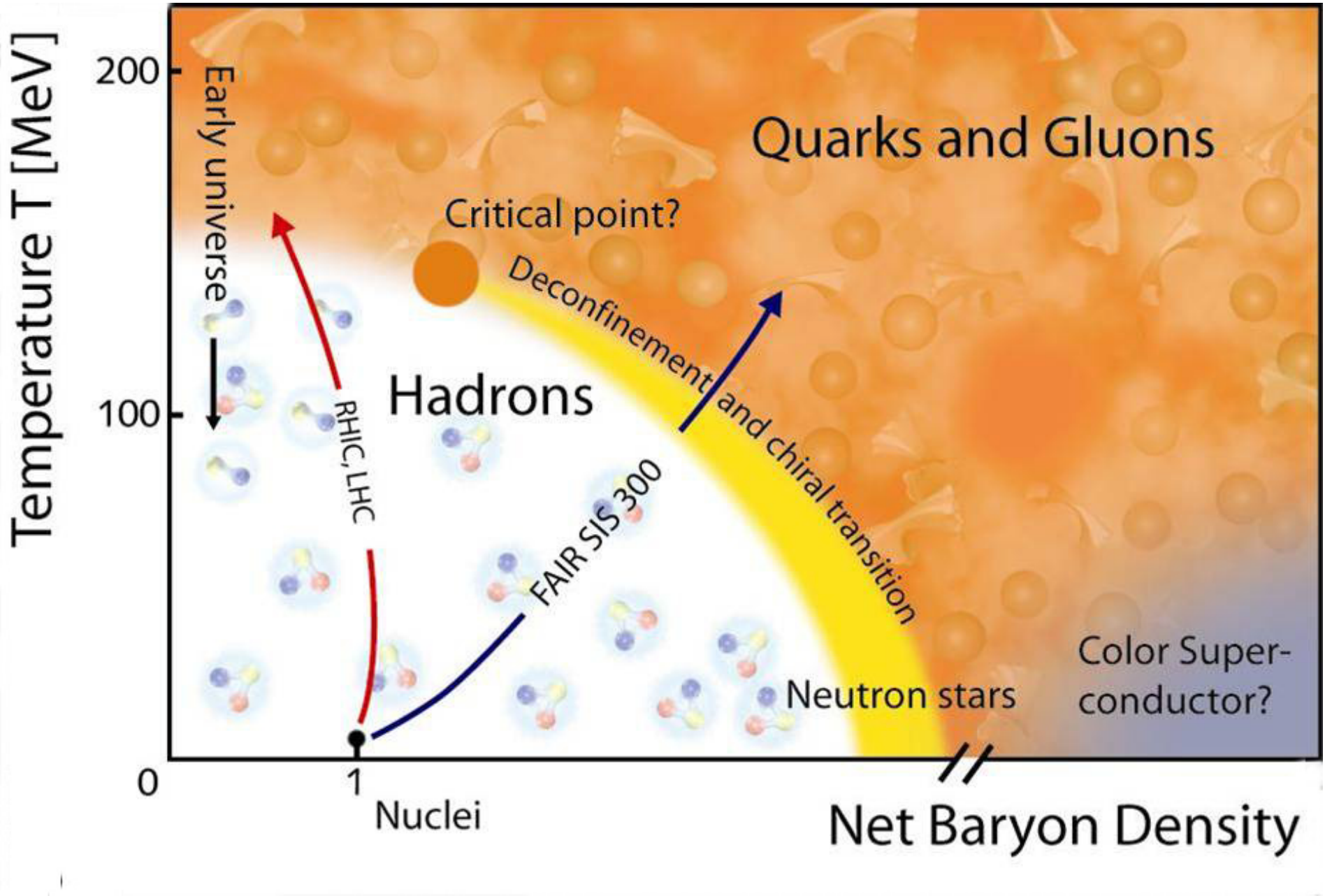}
 \caption{\small The phase diagram of QCD. The figure is taken from a presentation by P. Senger.
 \label{fig1}}
 \ec
\end{figure}

The gauge field theory describing the color interaction is called Quantum Chromodynamics (QCD)\cite{QCD}. It is an SU(3) Yang-Mills theory coupled to fundamental representation Dirac fermions, and successfully reproduces both the observed ``confinement'' and ``asymptotic freedom'' of the strong interactions.  Of these, asymptotic freedom indicates that in the limit of large energies or short distances the strong interaction in fact becomes weak, \ie~quarks and gluons become ``free''. In the context of nonzero temperature and density, this in particular suggests the possibility of a transition to a phase described by ``liberated'' quarks and gluons, once the temperature or density is high enough. 

It is indeed believed that at some critical temperature hadronic matter undergoes both chiral symmetry restoration and a deconfinement phase transition to a new state of matter called the quark gluon plasma (QGP) \cite{Collins:1974ky,Baym:1976yu,Freedman:1977gz,Shuryak:1980tp,Gross:1980br,Satz:1985vb,McLerran:1986zb}. The QCD phase diagram, containing both a confined hadronic phase and the QGP is depicted in \fig \ref{fig1}. At low baryon density (relevant for the early universe as well as collider experiments at Relativistic Heavy Ion Collider (RHIC) and the
Large Hadron Collider (LHC)), no sharp phase transition is encountered, but rather a smooth crossover \cite{Bernard:2004je,Cheng:2006qk,Aoki:2006we}. Increasing the baryon density (the heavy ion synchrotrons SIS300 at the Facility for Antiproton and Ion Research (FAIR)), it is however widely believed that the system exhibits a critical point, from which a first order phase transition line starts, ending only at the $T=0$ axis. In the limit of small temperatures and very high density, deconfined QCD matter --- in this context usually called quark matter --- may be found in the cores of neutron stars. This region of the phase diagram is theoretically rather poorly known, and only in the asymptotic limit of infinite density it is known that the physical ground state of three-flavor QCD is that of a Color-Flavor-Locked (CFL) color superconductor \cite{Alford:1997zt,Rapp:1997zu,Alford:1998mk}. 

In light of the fundamental role of QCD as part of the Standard Model and the role it plays both in the realm of nuclear physics and early universe thermodynamics, it is no doubt of importance to try to improve its theoretical description at nonzero temperature and density. This is indeed the main purpose of the thesis, and to this end we will now move on to review the structure of the theory as well as the main tools available for its study. Here, we mostly follow the references \cite{Laine:basis2013,peskin,qcdbooks}. Due to the fact that we will in this thesis be mostly concerned with the \textit{equilibrium} thermodynamics of the theory, our treatment below will be carried out in Euclidean spacetime, unless otherwise stated.

\mysection{QCD Lagrangian}\la{se:lag}
The gauge field sector of QCD corresponds to that of an SU(3) Yang-Mills theory \cite{Yang:1954ek}, to which the fundamental quarks are minimally coupled. The gauge invariant Euclidean Lagrangian of full QCD \footnote{Scaling $A_\mu\rightarrow gA_\mu$, the Lagrangian is sometimes rewritten as $\mathcal{L}^\text{QCD}=-\fr{1}{4g^2}F_{\mu\nu}^a F_{\mu\nu}^a+ \bar{\psi}(i\gamma_\mu(\partial_\mu-iA_\mu)+M)\psi$.} can thus be divided to two parts according to 
\ba \la{la}
\mathcal{L}^\text{QCD}&=& \mathcal{L}^{g}+\mathcal{L}^q \;, \\
\mathcal{L}^{g} &=& \fr14 F_{\mu\nu}^a F_{\mu\nu}^a \;, \\
\mathcal{L}^{q} &=& \bar{\psi}(i\slashed{D}+M)\psi \;,
\ea
where $\mathcal{L}^{g}$ is the Lagrangian of Yang-Mills theory. In the fermionic part, we have denoted $\bar\psi\equiv\psi^{\dag}\gamma_0$, $\slashed{D}=\gamma_\mu D_\mu$, $D_\mu=\partial_\mu-igA_\mu$ with $A_\mu \equiv A_\mu^a T^a$, $\gamma_\mu$ are the Dirac matrices, and $F_{\mu\nu}^a=\partial_\mu A_\nu^a-\partial_\nu A_\mu^a+gf^{abc}A_\mu^b A_\nu^c$.\footnote{\la{ft:LEM}Following the notation of \cite{Laine:basis2013}, here $\gamma_\mu$ are the Euclidean Dirac matrices, which are related to the Minkowskian ones through $(\gamma_0)_E=(\gamma^0)_M, (\gamma_k)_E=-i(\gamma^k)_M$. The relation between the two Lagrangians on the other hand reads $\mathcal{L}_E=-\mathcal{L}_M(\tau\rightarrow it, (\gamma_\mu)_E\rightarrow (\gamma_\mu)_M, (A_0)_E\rightarrow i(A_0)_M)$.} In these expressions, $T^a$ are the $N_c^2-1=8$ generators of the fundamental representation of the SU($N_c=3$) group, while $f^{abc}$ stand for the corresponding structure constants, satisfying the relation $[T_a,T_b]=if_{abc}T^c$. The $T_a$ are normalized such that $\tr T^a T^b=\fr12 \delta^{ab}$, while they also satisfy the Jacobi identity
\ba
\lk T^a,\lk T^b,T^c \rk\rk+\lk T^b,\lk T^c,T^a \rk\rk+\lk T^c,\lk T^a,T^b \rk\rk=0 \;.
\ea

As indicated above, in the gluon field $A_\mu^a$, the color index $a$ goes from 1 to 8, indicating that the gluon transforms in the adjoint representation. At the same time, the quark field $\psi$ in fact denotes a six-component vector in flavor space, $\psi\equiv(\psi_u, \psi_d, \psi_s, \psi_c, \psi_b, \psi_t)^{T}$, where each $\psi_k$ stands for a triplet of Dirac spinors in color space 
\ba
\psi_k=\left(
\begin{array}{c}
\psi_k^\text{red}\\
\psi_k^\text{green} \\
\psi_k^\text{blue}
\end{array}
\right) \;,\; k=u, d, s, c, b, t,
\ea
transforming in the fundamental representation. The quark mass matrix can finally be assumed to take the diagonal form $M=\text{diag}(m_u, m_d, m_s, m_c, m_b, m_t)$, as the coupling of the quarks to the Weak gauge bosons is not essential for our treatment work. The masslessness of the gluons finally follows from the gauge invariance of the theory forbidding terms such as $A_\mu A^\mu$ from appearing in the Lagrangian.

The Lagrangian \nr{la} is invariant under gauge transformations involving the (fundamental representation) matrix
\ba
U\equiv \exp\left[-iT^a\theta^a(x) \right] \;,
\ea 
where $\theta^a(x)$ are arbitrary real functions of the coordinate $x$, and the generators $T^a$ are (for $N_c=3$) related to Gell-Mann matrices $\lambda^a$ by $T^a=\fr12\lambda^a$, ($a=1,\ldots,8$). Under this transformation, the different fields transform as
\ba
\psi\rightarrow \psi^{'} &=& U\psi \;, \\
A_\mu\rightarrow A_\mu^{'} &=& U A_\mu U^{-1}+\fr{i}{g}U\partial_\mu U^{-1}\;, \la{deA}\\
D_\mu \rightarrow D_\mu^{'}&=& U D_\mu U^{-1} \;, \\
F_{\mu\nu} \rightarrow F_{\mu\nu}^{'} &=& U F_{\mu\nu} U^{-1} \;,
\ea
using which the gauge invariance of the Lagrangian is easy to check.

The gauge invariant Lagrangian \nr{la} can also be split up in an alternative way,
\ba \la{la1}
\mathcal{L}^\text{QCD}&=&\mathcal{L}_0+\mathcal{L}_{I} \;, \\
\mathcal{L}_0 &=& \fr14(\partial_\mu A_\nu^a-\partial_\nu A_\mu^a)(\partial_\mu A_\nu^a-\partial_\nu A_\mu^a) + \bar{\psi}(i\gamma_\mu\slashed{\partial}_\mu+M)\psi \;, \\
\mathcal{L}_{I} &=& \fr12 g f^{abc} (\partial_\mu A_\nu^a-\partial_\nu A_\mu^a)A_\mu^b A_\nu^c +\fr14 g^2 f^{abc} f^{ade} A_\mu^b A_\nu^c A_\mu^d A_\nu^e \nn
&& + \bar{\psi} g \gamma_\mu A_\mu\psi \;.
\ea
The two terms in $\mathcal{L}_0$ describe the free gluons field and free quark fields, respectively, while the three terms in $\mathcal{L}_{I}$ denote the three-gluon interaction, four-gluon interaction, and quark-gluon interaction, respectively. The three- and four-gluon interactions are very important components in QCD, as they are responsible for asymptotic freedom. This splitting of the Lagrangian will be shown later to play a fundemental role in perturbation theory.

\mysection{Running coupling and asymptotic freedom}

The running coupling constant of QCD is typically denoted by
\ba
\alpha_s\equiv\fr{g^2}{4\pi} \;,
\ea
where $g$ is the parameter appearing in the Lagrangian of the theory. The word ``running'' refers to the fact that $\alpha_s$ is a function of the energy scale of relevance. In perturbative calculations (explained in some detail below), one typically introduces a so-called renormalization scale $\mu$, on which $\alpha_s$ but no physical (directly measurable) quantity depends. The renormalization scale independence of various expectation values,
\ba
\mu^2\fr{d}{d\mu^2} \Av{O} =0,
\ea
can then be shown to lead to the explicit form of the running of $\alpha_s$, \ie
\ba \la{beta}
\mu^2\fr{d}{d\mu^2}\alpha_s(\mu^2) \equiv \beta\(\alpha_s\) = -\fr{\beta_0}{2\pi}\alphas^2-\fr{\beta_1}{(2\pi)^2} \alpha_s^3 -\ldots \;,
\ea 
or equivalently,
\ba 
\mu\fr{d}{d\mu}g \equiv \beta\(g\) = -g^3\(b_0+b_1g^2+b_2g^4 +\ldots \)\;,
\ea
where $\beta$ is the so-called beta-function of QCD.  The coefficients appearing above can be determined using standard calculations, and the first two coefficients read\cite{vanRitbergen:1997va}
\ba
\beta_0&=&\fr{11N_c-2N_f}{6} \;, \\
\beta_1&=&\fr{17N_c^2-5N_cN_f-3C_FN_f}{6} \;, \\
b_0&=&\fr{2\beta_0}{(4\pi)^2}=\fr{11N_c-2N_f}{3(4\pi)^2} \;,\\
b_1&=&\fr{4\beta_1}{(4\pi)^4}=\fr{34N_c^2-10N_cN_f-6C_FN_f}{3(4\pi)^4} \;,
\ea
where $C_F=(N_c^2-1)/(2N_c)$ is the value of the Casimir operator for the fundamental representation, and $N_f$ is the number of quark flavours. 

Considering \eq(\ref{beta}) as a differential equation for the coupling, the leading order solution reads
\ba\la{running}
\alpha_s\(\mu^2\)=\fr{2\pi}{\beta_0\ln\(\mu^2/\Lambda^2\)} \;,
\ea
where $\Lambda$ stands for an integration constant, the so-called QCD scale. It is related to the beta-function through the relation
\ba
\Lambda\equiv\mu\(b_0g^2\)^{-b_1/2b_0^2} e^{-1/2b_0g^2}
\exp\lb -\int_0^{g} dg^{'}\lk \fr{1}{\beta(g^{'})}+\fr{1}{b_0{g^{'}}^3}-\fr{b_1}{b_0^2g^{'}}\rk\rb \;,
\ea 
while its numerical value must be determined from experiments. Asymptotic freedom can be seen from the running coupling constant \nr{running}: As the scale parameter $\mu$ is increased, $\alpha_s$ decreases accordingly. This  means that the running coupling is small for high energy scattering, where the energy scale (momentum transfer) $\mu$ is large. In other words, the interaction of quarks and gluons is weak at high energies. In particular, the coupling $\alpha_s$ goes to 0 when $\mu\rightarrow \infty$, indicating a limit where weak coupling methods may be used.

\mysection{The partition function and its path integral representation}

Let us now specialize to the equilibrium thermodynamics of QCD and consider the evaluation of the most basic quantity in statistical mechanics, the partition function. The partition function and the corresponding density matrix are defined through the Hamiltonian $H$ of the system at finite temperature $T$,
\ba\la{part}
Z(T)&\equiv&\tr\exp\left[-\beta H\right] \;, \;\; \beta\equiv \fr1T \;, \\
\rho(T) &\equiv& Z^{-1}\exp\left[-\beta H\right]
\;,
\ea
where the trace is taken over all states of the system and the density matrix $\rho(T)$ is normalized so that $\tr\rho(T) =1$. With knowledge of the partition function, many other observables, such as the free energy $F$, the entropy $S$, and the pressure $P$, can be directly evaluated via relations such as
\ba
F&=&-T\ln Z \;,\\
S&=&-\fr{\partial F}{\partial T} =\fr{\partial \(T\ln Z\)}{\partial T} \;, \\
P&=&-\fr{\partial F}{\partial V} = \fr{\partial \(T\ln Z\)}{\partial V} \;.
\ea
In addition, the expectation value, or ensamble average, of an arbitrary observable $O$ can be calculated via the relation
\ba
\Av O = \tr\lk\rho(T)O\rk \;.
\ea
One simple example of this is the energy of the thermal system
\ba
E=\Av H =\tr\lk\rho(T)H\rk \;,
\ea
with which the entropy can be expressed as
\ba
S=-\fr{F}{T}+\fr{E}{T}\;.
\ea

In the functional integral formulation of quantum field theory, the partition function \nr{part} can be expressed in terms of a path integral over fields that are either periodic or antiperiodic in the Euclidean \textit{imaginary time} variable $\tau=it$, which runs from 0 to $\beta\equiv \fr1T$. With this convention, we obtain schematically
\ba\la{pathZ}
Z(T)=\int_{antiperiodic} \mathcal{D}\psi \mathcal{D}\bar{\psi} \int_{periodic} \mathcal{D}A_\mu \exp \lk -S_E \rk \;,
\ea
where we have assigned bosonic fields periodic and fermionic fields antiperiodic boundary conditions, and the Euclidean action reads
\ba
S_E=\int_0^\beta d\tau\int d\vec{x}\; \mathcal{L}^\text{QCD}_\text{E} \;.
\ea

While the above result is rather elegant in its simplicity, it is easy to see that this formulation contains an immense overcounting of field configurations. We have namely not taken into consideration the gauge invariance of the theory, which relates many field configurations into each other via the gauge transformation of \nr{deA} and implies that we should rather be integrating over gauge orbits than individual gauge fields. The problem can be resolved in various ways, of which the one by Fadeev and Popov is by far the most widely used. Here, one inserts into the path integral a delta function implementing the Gauss rule, and then expresses this function in terms of a functional integral of so-called ghost fields, which are fermionic (\ie~represented by Grassmann variables), yet periodic in time. As this exercise is presented in great detail in many textbooks, we will here merely state the final result, expressing the \textit{gauge-fixed} Lagrangian of the theory in the form 
\ba
\mathcal{L}^\text{QCD}_{GF}=\mathcal{L}^\text{QCD}+\mathcal{L}_{gauge-fixing}+\mathcal{L}_{ghost} \;,
\ea
where
\ba
\mathcal{L}_{gauge-fixing}&=&\fr{1}{2\xi}(\6_\mu A_\mu^a)^2 \;, \nn
\mathcal{L}_{ghost}&=&\6_\mu \bar{c}^a \6_\mu c^a +g f^{abc}\6_\mu\bar{c}^a A_\mu^b\, c^c \;,
\ea
and $\xi$ is a real number parametrizing the covariant gauge we have chosen. 

Using the above Lagrangian, the path integral representation of the partition function becomes
\ba \la{zpath}
Z(T)&=&\int_{antiperiodic} \mathcal{D}\psi \mathcal{D}\bar{\psi} 
\int_{periodic} \mathcal{D}c \mathcal{D}\bar{c}
\int_{periodic} \mathcal{D}A_\mu \exp \lb - \int_0^\beta d\tau\int d\vec{x} \rd
\\
&\times&\lk \ld\fr14 F_{\mu\nu}^a F_{\mu\nu}^a   
+\fr{1}{2\xi}(\6_\mu A_\mu^a)^2 +\6_\mu \bar{c}^a \6_\mu c^a +g f^{abc}\6_\mu\bar{c}^a A_\mu^b\, c^c
+\bar{\psi}(i\slashed{D}+M)\psi
\rk \rb \;, \nonumber
\ea
which is now well defined and free of gauge redundancies. In terms of this integral formulation, the expectation value of an arbitrary physical observable can finally be written in the form
\ba
\Av O =\fr{\int_{\psi}\int_{c}\int_{A_\mu} O \exp\lk-S_E\rk}{\int_{\psi}\int_{c}\int_{A_\mu} \exp\lk-S_E\rk} \;,
\ea
where we have introduced the symbol $\int_{\psi}\int_{c}\int_{A_\mu}$ to denote functional integrals over all relevant fields with (anti)periodic boundary conditions.

\mysection{Perturbative QCD}\la{intro:pqcd}
As we can see from \eq \nr{zpath}, the QCD partition function has a complicated form, and it is unfortunately not possible to carry out the path integral in a closed form. Fortunately, the running of the strong coupling constant tells us that the color interaction becomes weak at high temperatures or very short distances, implying that weak coupling methods may become feasible in these limits. To this end, we note that the Euclidean action of the theory can be split into two parts via
\ba
S_E=S_0+S_I \;,
\ea
where $S_0$ denotes the free action including all terms quadratic in the fields, while $S_I$ is the interaction part involving all terms featuring the coupling constant $g$  (cf.~the splitting of the Lagrangian in \eq \nr{la1}). In the absence of interactions, the two-point Green's function, \ie~the bare propagator, can be extracted from the free action $S_0$. From the interaction part of the action $S_I$, one can on the other hand read off the leading order three- and four-point Green's functions, \ie~the interaction vertices of the fields. 
	
Splitting the action in the above fashion in the functional integral defining the partition function enables us to write the quantity in the form
\ba
Z&=&\int_{\psi}\int_{c}\int_{A_\mu}\exp\lk-(S_0+S_I)\rk 
=\int_{\psi}\int_{c}\int_{A_\mu}\exp\lk-S_0\rk \sum_{k=0}^{\infty}\fr{\(-S_I\)^k}{k!} \nn
&=&\int_{\psi}\int_{c}\int_{A_\mu}\exp\lk-S_0\rk \(1+\sum_{k=1}^{\infty}\fr{\(-S_I\)^k}{k!}\) \;,
\ea
where the powers of $S_I$ carry all of the dependence of the quantity on the coupling $g$. When $g$ is small, the series in powers of $S_I$ can be expected to converge at least in the sense of an asymptotic expansion, and the sum cut off at some finite $N$. Similarly, the expectation values of various physical quantities can be evaluated to a finite order in the coupling, writing 
\ba
\Av{O}=Z^{-1}\lb \int_{\psi}\int_{c}\int_{A_\mu} O \exp\lk-S_0\rk 
+ \int_{\psi}\int_{c}\int_{A_\mu} O \(-S_I\) \exp\lk-S_0\rk + \ldots \rb \;.
\ea

This schematic procedure, carried out to the desired order in $g$ or $\alpha_s$, is commonly referred to as \textit{perturbation theory}. Most of the calculations and results that will be reviewed in this thesis rely heavily on this technique; a good reference for more details, especially in the context of QCD, is \cite{Smirnov:2006}. Some examples of thermodynamical quantities that have been computed to a high order in perturbation theory include the QCD equation of state \cite{Kajantie:2002wa,Vuorinen:2003fs,Kurkela:2009gj}, quark number (and other) susceptibilities \cite{Vuorinen:2002ue,Andersen:2012wr}, as well as various types of correlation functions \cite{Laine:2003bd,Ghiglieri:2013gia}.

\mysection{Correlators and the spectral function}

In addition to the partition function considered above, a very important class of quantities, from which one can learn about a finite temperature system, are two-point correlation functions of elementary or composite operators. It is thus worthwhile to pay some attention to the definition of different correlation or Green's functions, as well as their relations and properties. In this section, we will do just this, following to some extent ref.~\cite{Meyer:2011gj} and beginning our treatment from Minkowskian correlators.

\subsection{The Wightman functions and the spectral function}

The simplest finite-temperature correlation function of two operators $A$ and $B$ reads
\ba \la{sc}
\Av {A(t)B(0)} =\tr[\rho(T) A(t)B(0)] \;,
\ea
where it is assumed that we work in the Heisenberg picture, where the operators carry the time dependence. This in particular allows us to use the relation
\ba \la{heisenberg}
O(t)\equiv e^{iHt} O(0) e^{-iHt}\
\ea 
to prove the important \textit{Kubo-Martin-Schwinger} (KMS) formula
\ba\la{KMS}
\Av{A(t)B(0)}=Z^{-1}\tr\lk e^{-\beta H} A(t)B(0) \rk =\Av{B(0)A(t+i\beta)} \;.
\ea
For the \textit{Wightman} functions, defined by
\ba\la{wightman}
G_{AB}^>(t)&\equiv& \Av{ A(t)B(0)}\;, \\
G_{AB}^<(t)&\equiv&\Av{ B(0)A(t)}\;,
\ea
this relation leads to the result
\ba
G_{AB}^>(t)=G_{AB}^<(t+i\beta)\;.
\ea

Finally, we note that the difference of two Wightman functions, 
\ba\la{GAB}
G_{AB}(t)\equiv  G_{AB}^>(t)- G_{AB}^<(t)\;,
\ea
vanishes outside the light cone as a consequence of causality. Its Fourier transform on the other hand gives the perhaps most important thermal correlator, the \textit{spectral function}, according to
\ba\la{def:rho}
\rho_{AB}(\omega)\equiv\fr12 \int_{-\infty}^{\infty}dt\, e^{i\omega t} G_{AB}(t) \;.
\ea
Note that in our notation the argument of $\rho$ is used to distinguish this quantity from the density matrix of a statistical ensemble, which was denoted by $\rho(T)$.

\subsection{Retarded and advanced Green's functions}

Next, we move on to introduce the retarded (R) and advanced (A) Green's functions, which read
\ba
G_{AB}^R(t)&\equiv& i G_{AB}(t)\theta(t)\;, \\
G_{AB}^A(t)&\equiv&-i G_{AB}(t)\theta(-t)\;,
\ea
and clearly vanish for negative (positive) values of $t$. Denoting their Fourier transforms by
\ba
G^R_{AB}(\omega)&=&i \int_0^{\infty} dt\, e^{i\omega t}G^{AB}(t)\;,\la{GRAB} \\
G^A_{AB}(\omega)&=&-i \int_{-\infty}^0 dt\, e^{i\omega t}G^{AB}(t)\;, \la{GAAB}
\ea
the spectral function can be expressed in the form 
\ba
\rho_{AB}(\omega)=\fr{1}{2i} \( G_{AB}^R(\omega)-{G_{B^{\dag}A^\dag}^R}^*(\omega)\)
=-\fr{1}{2i} \( G_{AB}^A(\omega)-{G_{B^{\dag}A^\dag}^A}^*(\omega)\) \;,
\ea
making use of the relations $G_{AB}^{<}(t)=G_{BA}^{>}(-t)$ and $G_{A^{\dag}B^{\dag}}^>(t)={G_{BA}^{>}}^*(t)$. For the special case $B=A^{\dag}$, the spectral function thus reduces to
\ba\la{rAA}
\rho_{AA^{\dag}}(\omega)=\im G_{AA^{\dag}}^R(\omega) 
=-\im G_{AA^{\dag}}^A(\omega) \;.
\ea
From these equations, it is also evident that the spectral function is an odd function of $\omega$, \ie~$\rho_{AA^{\dag}}(-\omega)=\rho_{AA^{\dag}}(\omega)$.

\subsection{Euclidean correlators and their relation to the spectral function}

While Minkowski space correlators, which are functions of a real time $t$, are perhaps more intuitive quantities, for systems close to thermal equilibrium it is more enlightening to consider Green's functions that have been analytically continued to the imaginary time variable $\tau$. These \textit{Euclidean correlators} are in addition the only types of quantities measurable in numerical lattice simulations, and thus play a major role in the analysis of any finite temperature system.

For the reasons stated above, let us now consider the imaginary time correlator
\ba
G_{AB}^E(\tau)\equiv G_{AB}^{>}(t\rightarrow -i\tau)\;,
\ea
where $\tau$ again runs from 0 to $\beta=1/T$. Due to the finite extent of the time interval, the Fourier transform of this quantity is expressed in terms of discrete \textit{Matsubara} frequencies $\omega_n$, taking the values $2\pi n T$ for bosons and $(2n+1)\pi T$ for fermions,
\ba
G_{AB}^E(\tau)&=& T\sum_{n=-\infty}^{\infty} e^{-i\omega_n \tau}G_E(\omega_n) \;, \\
G^E(\omega_n)&=& \int_0^\beta d\tau\; e^{i\omega_n \tau} G_E(\tau) \;.
\ea
In order to establish a connection between these objects and the spectral function, we will manipulate $G_{AB}^E(\tau)$ starting from its primary definition. Making use of \eq~(\ref{heisenberg}) and inserting a complete set of eigenvectors of the Hamiltonian, $\mathop{\sum}\limits_m \ld |m\right>\left< m| \rd =1$ (with $\ld H|m\right>=\ld E_m|m\right>$), it is easy to see that this function takes the form
\ba
G_{AB}^E(\tau)=Z^{-1}\sum_{m,n} e^{-\beta E_m} e^{\tau(E_m-E_n)} 
\Av{m|A(0)\ld |n\right>\left< n| \rd B(0)|m} \; ,
\ea
while its Fourier transform becomes
\ba
\la{GE_fock}
G_{AB}^E(\omega_n)&=& Z^{-1} \int_0^\beta d\tau\; e^{i\omega_n \tau} 
\sum_{m,n} e^{-\beta E_m} e^{\tau(E_m-E_n)} \Av{m|A(0)\ld |n\right>\left< n| \rd B(0)|m} \nn
&=& Z^{-1} \sum_{m,n} \fr{e^{\beta(i\omega_n-En)}-e^{-\beta E_m}} {E_m-E_n+i\omega_n}\Av{m|A(0)\ld |n\right>\left< n| \rd B(0)|m} \;.
\ea

The above results should be contrasted to a similar relation obtained from \eq \nr{def:rho},
\ba\la{rho_fock}
\rho_{AB}(\omega) &=& \fr12 Z^{-1}  \int_{-\infty}^{\infty}dt\, e^{i\omega t} 
\sum_{m,n}  e^{-\beta E_m} \lk e^{it(E_m-E_n)} - e^{it(E_n-E_m)}\rk 
\nn &&\times \Av{m|A(0)\ld |n\right>\left< n| \rd B(0)|m} \nn
&=&  Z^{-1} 
\sum_{m,n}  e^{-\beta E_m} \pi \lk \d({E_m-E_n+\omega}) - \d({E_n-E_m+\omega})\rk 
\nn && \times \Av{m|A(0)\ld |n\right>\left< n| \rd B(0)|m} 
\nn &=& Z^{-1} 
\sum_{m,n}  e^{-\beta E_n} \(e^{\beta\omega}-1\) \pi  \d({E_n-E_m+\omega}) \Av{m|A(0)\ld |n\right>\left< n| \rd B(0)|m} \nn \\
 &=& Z^{-1} 
\sum_{m,n}  e^{-\beta E_m}  \(1-e^{-\beta\omega}\) \pi \d({E_m-E_n+\omega})  \Av{m|A(0)\ld |n\right>\left< n| \rd B(0)|m} \nonumber \;,
\ea
where we have used the fact that due to the KMS relation $G_{AB}^>(\omega)=e^{\beta\omega}G_{AB}^<(\omega)$. A comparison of  \eqs \nr{GE_fock} and \nr{rho_fock} now leads to the relation
\ba\la{GEw}
G_{AB}^E(\omega_n)= \fr{1}{\pi}\int_{-\infty}^{\infty} \fr{d\omega}{\omega-i\omega_n} \rho_{AB}(\omega)\;,
\ea
as well as its inverse,
 \ba
\rho_{AB}(\omega)=\im G_{AB}^E(\omega_n\rightarrow -i(\omega-i0^+))\;.
\ea
From \eq(\ref{GEw}), it is finally straightforward to derive a highly useful relation between the imaginary time correlator and the spectral function,
\begin{equation}
 G_{AB}^E(\tau) =
 \frac{1}{\pi}\int_0^\infty
 {\rm d}\omega \rho(\omega)
 \frac{\cosh\Big[\! \left(\frac{\beta}{2} - \tau\right)\omega\Big]}
 {\sinh\frac{\beta \omega}{2}}\, .  \la{int_rel}
\end{equation}
This demonstrates the central role played by the spectral function in any analysis of finite temperature correlation functions, a fact we will often take advantage of in the following.

\clearpage
\pagestyle{plain}



\pagestyle{myheadings}
\setcounter{chapter}{1}
%
%
%
%
%
%

\mychapter{Motivation: Bulk and shear viscosities of the QGP}
\label{CHAP 2}

\mysection{Viscosities and heavy ion physics}

Results from lattice QCD indicate that when hadronic matter at small baryonic chemical potential is heated up, it undergoes a crossover-type deconfinement transition at a temperature of the order of 150-170 MeV \cite{Borsanyi:2010zi,Karsch:2000kv}. At nonzero density, the possible existence of a critical point and a line of first order phase transitions extending to the $T=0$ axis, however, remains an unsolved problem. It is hoped that ultrarelativistic heavy ion experiments carried out at RHIC of BNL, the LHC of CERN, and perhaps most importantly at FAIR of GSI will shed light on this issue by providing direct experimental information of the created QGP. 

In heavy ion experiments, two charged nuclei are accelerated to ultrarelativistic speeds, thereby appearing as Lorentz contracted disks, cf.~the ``initial state'' in \fig \ref{methcoll}. The interpenetrating nuclei are then expected to create a complicated out-of-equilibrium system, which after about 0.5$-$1 fm/c thermalizes and forms the QGP phase. The system is, however, still very much non-uniform, and the large density and temperature gradients make it expand and cool down very fast. Eventually the temperature of the system falls below the critical temperature of the deconfinement transition, leading to the production of a multitude of hadrons, photons and dileptons later detected in the experiments.

One of the major problems in the experimental study of the QGP is clearly its short lifetime. Fortunately, it is possible to gather at least some indirect evidence of the properties of the system by analyzing the distribution of the final state hadrons. In fact, various signals, such as the famous suppression of J/$\Psi$, jet quenching, and collective flow --- all measured in experiments at RHIC and the LHC --- have consistently indicated that QGP has indeed been created in both colliders. Theoretical and experimental studies further point towards the system behaving rather as a strongly interacting fluid than a gas of weakly interacting quasiparticles, coining the term ``strongly interacting QGP'' (sQGP) \cite{Rischke:2005ne}. 

\begin{figure}
\bc
\includegraphics[width=14cm]{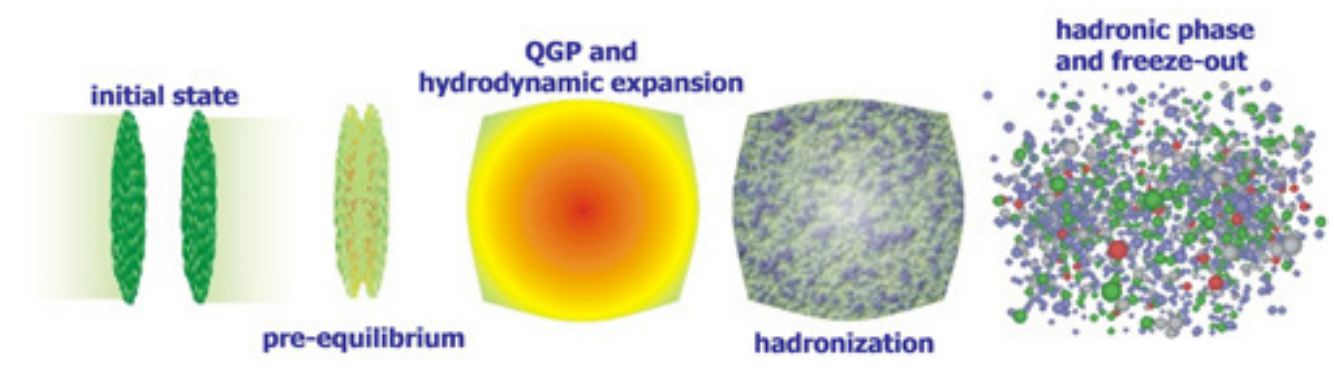}
 \caption{Evolution of the heavy ion collision process involving the creation of the QGP. The figure is taken from a presentation by S. Bass.
 \label{methcoll}}
 \ec
\end{figure}

From the point of view of this thesis, the most interesting part of the heavy ion collision process is the evolution of the (near-)thermal QGP, which can already be described in terms of relativistic hydrodynamics. In particular, the work performed in the following three chapters is intimately related to the determination of \textit{transport coefficients} appearing in the hydrodynamic equations. Most importantly, these include the shear and bulk viscosities, which have lately received wide attention in the community. A particularly famous results has been the AdS/CFT prediction of a lower bound for the ratio of the shear viscosity and entropy density in any physical system \cite{Kovtun:2004de}, 
\ba
\fr{\eta}{s}\ge \fr{1}{4\pi}\, ,
\ea
while for the bulk viscosity of the QGP, \cite{Buchel:2007mf} has proposed the relation
\ba
\fr{\zeta}{\eta}\ge 2\left(\fr{1}{3}-c_s^2\right)^2 \, ,
\ea
with $c_s$ standing for the speed of sound in the medium. At the same time, in the weakly coupled limit of QCD, the shear viscosity has been evaluated to the leading-log\cite{Arnold:2000dr} and later to the full leading order accuracy \cite{Arnold:2003zc}, and parametrically takes the form 
\ba
\fr{\eta}{s}\sim\fr{1}{g^4\ln{(1/g)}}\;,
\ea
while the leading order bulk viscosity\cite{Arnold:2006fz} reads
\ba
\fr{\zeta}{s}\sim \fr{g^4}{\ln(1/g)}\;.
\ea

There is clearly a strong discrepancy between the strong and weak coupling predictions for the viscosities. Somewhat surprisingly, it has turned out that a comparison of hydrodynamic calculations with experimental data in fact favors the former \cite{Teaney:2009qa}, pointing towards a rather low value for the $\eta/s$ ratio
\ba
\fr{\eta}{s}<0.4\;,
\ea
while the preferred value (for RHIC) appears to be $\fr{\eta}{s}\sim(1-3)\times\fr{1}{4\pi}$. In light of this, it would clearly be very important to be able to derive a nonperturbative first principles result for the transport coefficients within QCD, for which lattice calculations would in principle be the method of choice \cite{Meyer:2007ic,Meyer:2007dy}. However, as lattice QCD is restricted to the Euclidean formulation of the theory, such results necessitate a numerically extremely challenging analytic continuation from imaginary to real time to take place. Therefore, the current numerical estimates for $\eta$ and $\zeta$ still come with large systematic uncertainties.

Recently, a model independent method \cite{Burnier:2011jq} was proposed for the extraction of spectral densities and transport coefficients from Euclidean correlators measured on the lattice. A successful implementation of this idea can be found from ref.~\cite{Burnier:2012ts}, while the work performed in this thesis aims at generalizing such calculations to the bulk and shear channels of the QGP. To this end, the quantities we are after are the bulk and shear perturbative spectral functions, needed in inverting the relation 
\ba 
 G_E(\tau)=
  \int_0^\infty
 \frac{{\rm d}\omega}{\pi} \rho(\omega)
 \frac{\cosh\Big[\! \left(\frac{\beta}{2} - \tau\right)\omega\Big]}
 {\sinh\frac{\beta \omega}{2}}\, ,\quad \quad 0<\tau <\beta 
\ea
to obtain a nonperturbative spectral function from lattice results for the Euclidean correlator. Before considering these calculations in the following chapters, we will however first present a brief review of the hydrodynamic description of the QGP as well as the theoretical tools currently available for estimating the values of transport coefficients.

\pagestyle{myheadings}

\mysection{Hydrodynamic description of the QGP}
\label{sec:model}

As mentioned above, the QGP produced at RHIC and the LHC was initially expected to behave almost like an ideal gas of quasiparticles corresponding to the quark and gluon degrees of freedom. Later, studies of Au-Au collisions at RHIC \cite{Adcox:2004mh,Back:2004je,Arsene:2004fa,Adams:2005dq} and Pb-Pb collisions at the LHC \cite{Aamodt:2010pa,Aamodt:2010pb} have indeed provided strong evidences for the production of the QGP, but with rather remarkable properties indicative of a strongly interacting, almost perfect liquid with very small viscosity. An integral part in reaching this conclusion has been played by relativistic hydrodynamic simulations, into which the small but nonvanishing value of the shear viscosity has been incorporated \cite{Teaney:2009qa,Kolb:2003dz,Huovinen:2006jp,Luzum:2008cw,Luzum:2009sb,Luzum:2010ag,Song:2011hk,Song:2011qa}.

The core of a realistic hydrodynamic description of the QGP is solving the energy-momentum conservation equations
\ba
\la{eq:emc}
\partial_\mu T^{\mu\nu} = 0 \;.
\ea
For a system exhibiting only a small deviation from equilibrium, the energy-momentum tensor takes the form
\ba
T^{\mu\nu} = (e+P)u^\mu u^\nu - P g^{\mu\nu} + \Delta T^{\mu\nu} \;,
\ea
where $e$, $P$, and $u_\mu$ denote the local energy density, pressure, and flow velocity, respectively, and where we assume a (mostly minus) Minkowskian metric. For linearized viscous hydrodynamics, the function $\Delta T^{\mu\nu}$ finally reads
\ba
\la{eq:dTmn}
 \Delta T^{\mu\nu}  = \eta \left(\Delta^\mu u^\nu+\Delta^\nu u^\mu\right) + 
 \left(\fr{2}{3}\eta-\zeta\right)\Delta^{\mu\nu}\partial_\alpha u^\alpha\;,
\ea
where $\Delta_\mu=\partial_\mu-u_\mu u^\beta\partial_\beta$ and $\Delta^{\mu\nu}=u^\mu u^\nu-g^{\mu\nu}$.

In physical terms, $\Delta T^{\mu\nu}$ describes a small deviation of the energy-momentum tensor from the ideal limit due to dissipative (viscous) effects. The most important parameter in this quantity is the shear viscosity, which measures the relaxation of isotropic momentum, \ie~the resistance of the system to flow gradients. At the same time, the bulk viscosity characterizes the relaxation of non-equilibrium particle number densities, corresponding to resistance against collective expansion. 

To compare theoretical predictions for various physical quantities to the experimental heavy ion data is a complicated process, of which the hydrodynamic simulations only constitute one part. This is depicted in \fig \ref{methcoll}: First, the initial state of the system is typically described via Glauber or McLerran-Venugopalan (Color Glass Condensate) models, the early evolution of which is then studied via Monte Carlo simulations \cite{Hirano:2009ah,Hirano:2009bd}. After this, a hydrodynamic model with an (usually lattice determined) equation of state (EOS) is applied to its expansion, until the system reaches hadronization, described via fragmentation and/or recombination models. Finally, the eventual freeze-out of the hadron gas is usually modeled within Hadron Resonance Gas models \cite{Song:2008hj,Hirano:2012kj}, from which predictions for quantities such as elliptic flow are derived and compared to experimental data. How this process can be used to fix parameters such as the shear viscosity is explained e.g.~in \cite{Heinz:2013th}. More details about the practical application of hydrodynamic models can on the other hand be found e.g.~in \cite{Hirano:2008hy,Huovinen:2006jp,Hirano:2012yy}.

Finally, to understand the role of the viscosities in the hydrodynamic description of heavy ion collisions, one should note that ideal hydrodynamics has been seen to successfully describe quantities such as elliptic flow and transverse momentum spectra in the low momentum region (see, for example, \cite{Hirano:2009ah,Hirano:2009bd} and references therein). However, if one is interested in the same quantities at larger transverse momenta, such as $p_T>1.5$ GeV, one starts to see deviations between experimental data and ideal hydrodynamics calculations \cite{Teaney:2003kp}. For these quantities, a nonzero value for the shear viscosity is needed in the hydro codes to find agreement with experimental results, while the bulk viscosity is observed to play only a minor role in the dynamics \cite{Arnold:2006fz,Song:2009rh,Song:2009je}.

Next, we move one step closer to the actual topic of our thesis, namely to review the two leading methods to determine the values of the bulk and shear viscosities in deconfined QCD matter. These include first the application of linear response theory, in which the Kubo formula can be used to relate transport coefficients to the infrared (IR) limits of Green's functions, and secondly kinetic theory, in which one can derive integral representations for the same quantities. In both cases, we identify the computational tools and approximations most commonly used, and in addition review the current status of the calculations. In particular, we will identify here the role that the spectral function calculations presented in the following sections play in the process.

\mysection{Linear response theory}
\label{sec:kubo}

Transport coefficients can be viewed as describing the response of the system to infinitesimal external perturbations. Within thermal field theory, the conceptually most straightforward way of determining them is thus through linear response theory, where Kubo formulas can be used to express them as integrals over retarded Green's functions \cite{Kubo:1957mj,Hosoya:1983id,Horsley:1986dy,Zubarev}. We will now review this process in some detail.

Let us consider the Hamiltonian of a field theory system, writing it in the form
\ba
H=H_0+\int d^3\vec{x}\; J(x) O(x) \;,
\ea
where $H_0$ denotes the Hamiltonion of the unperturbed equilibrium system, $J(x)\equiv J(\vec{x},t)$ is a time-dependent external source, and $O(x)\equiv O(\vec{x},t)$ an internal variable which the source is coupled to. If $J(x)$ is small, the expectation value of $O(x)$ is slightly shifted from its local equilibrium value, and we can write
\ba
\delta\left<O(x)\right>&\equiv& \left<O(x)\right>_H-\left<O(x)\right>_{H_0} \nn
&=& -\int d^4y\; G^R(x-y) J(y)
\ea
with the retarded Green's function $G^R(x-y)=-i \theta(x^0-y^0)\left<[O(x),O(y)]\right>_{H_0}$. 

Following the linear response theory derivation of \cite{Zubarev}, the Kubo formulas for the bulk and shear viscosities can be expressed in terms of retarded correlators of the energy-momentum tensor 
\ba
\la{eq:etaret}
\eta&=& -\int \left< T_{12}(x) T_{12}(y) \right> \;,
\\
\fr43\eta+\zeta &=&-\int \left< T_{11}(x) T_{11}(y) \right> \;.\la{eq:sumret}
\ea 
Here, the angular brackets are used to denote the retarded correlators, and we have used the shorthand $\int=\lim_{\epsilon\rightarrow 0}\int d^3y \int_{-\infty}^t dt_1 e^{\epsilon(t_1-t)} \int_{-\infty}^{t_1} dt'$, $x=(t,\vec{x})$, and $y=(t',\vec{y})$. To separate the bulk viscosity from \nr{eq:sumret}, we next make use of the cubic symmetry of the spatial components of the energy-momentum tensor $T_{\mu\nu}$ (cf. \cite{Karsch:1986cq}), giving a schematic relation
\ba
\left<T_{ij}(x) T_{kl}(y)\right> = A(x,y)(\delta_{ik}\delta_{jl}+\delta_{il}\delta_{jk})+B(x,y)\delta_{ij}\delta_{kl}\;\;\; (i,j,k,l=1,2,3) \;.
\ea
This allows us to write the shear viscosity in the form
\ba
\la{eq:etaret1}
\eta &=& -\fr12 \int \left[\left<T_{11}(x)T_{11}(y)\right>-\left<T_{11}(x)T_{22}(y)\right>\right] \nn
&=&-\fr14 \int \left[\left<(T_{11}(x)-T_{22}(x))(T_{11}(y)-T_{22}(y))\right>\right] \;,
\ea
while plugging \eq \nr{eq:etaret1} into \eq \nr{eq:sumret} gives for the bulk viscosity
\ba
\la{eq:zetaret}
\zeta=-\fr19\sum_{i,j=1}^3 \int  \left< T_{ii}(x) T_{jj}(y) \right> \;.
\ea

Finally, we note that according to \eqs \nr{GRAB} and \nr{rAA}, the bulk and shear viscosities can also be written as zero frequency limits of the corresponding spectral functions (cf.~e.g.~\cite{Teaney:2006nc}) via 
\ba
\zeta &=& \fr{1}{9}\sum_{i.j=1}^3 \lim_{\omega\rightarrow 0} \fr{\rho_{ii,jj}(\omega,\vec{0})}{\omega} \;, \la{eq:zeta}
\\
\eta &=& \lim_{\omega\rightarrow 0} \fr{\rho_{12,12}(\omega,\vec{0})}{\omega}\;. \la{eq:eta}
\ea
This is in fact the most frequently used way of determining these parameters. Below, we will introduce three computational schemes used in the evaluation of the necessary spectral densities: Perturbation theory, lattice QCD, and the AdS/CFT conjecture.

\subsection{Perturbation theory}\la{se:pqcd}

The symmetrized gauge invariant energy-momentum tensor of QCD, derived from the Lagrangian of \eq \nr{la}, reads \cite{Caracciolo:1989bu} 
\ba
T_{\mu\nu} &=& T_{\mu\nu}^g + T_{\mu\nu}^q\;, \\ \la{Tg}
T_{\mu\nu}^g &=& \fr14\delta_{\mu\nu}F_{\alpha\beta}^aF_{\alpha\beta}^a-F_{\mu\alpha}^a F_{\nu\alpha}^a\;, \\
T_{\mu\nu}^q &=& \fr14 \( \bar{\psi}i\overset{\leftrightarrow}{D}_\mu\gamma_\nu\psi + \bar{\psi}i\overset{\leftrightarrow}{D}_\nu\gamma_\mu\psi \) +\fr12\delta_{\mu\nu}\bar{\psi}i\overset{\leftrightarrow}{D}_\alpha\gamma_\alpha\psi  + \delta_{\mu\nu}\bar{\psi}M\psi\;,
\ea
where $\overset{\leftrightarrow}{D}_\mu\equiv\overset{\rightarrow}{D}_\mu-\overset{\leftarrow}{D}_\mu$. Starting from this expression, it might seem straightforward to perturbatively determine the bulk and shear viscosities from the Kubo formulas (\ref{eq:etaret}) and (\ref{eq:zetaret}), or equivalently from \nr{eq:zeta} and \nr{eq:eta}. This using is, however, in practice impossible due to the extremely complicated IR structure of the spectral function, necessitating complicated resummations to be carried out before one can proceed to the region $\omega\sim gT$, let alone to the formal $\omega\to 0$ limit. 

Some work in the above direction has nevertheless been performed e.g.~in \cite{Aarts:2002cc}, but in practice this is not the most efficient way to perturbatively determine the transport coefficients. It turns out to be much more convenient to use the machinery of kinetic theory in this context (see below for more details). Using this in principle equivalent formulation of the problem, the bulk and shear viscosities of QCD (as well as certain other transport coefficients) have indeed been calculated first to the leading logarithmic order \cite{Arnold:2000dr,Arnold:2006fz} and later to the full leading order \cite{Arnold:2003zc,Arnold:2002zm}. In scalar $\phi^4$ theory, even a next to leading order result of the shear viscosity exists \cite{Moore:2007ib}. Unfortunately, it has turned out that the convergence of these expansions is rather poor, and the results are thus of limited use.

\subsection{Lattice QCD}

Whenever applicable, lattice Monte Carlo simulations should always be the method of choice in QCD calculations, as they constitute the only nonperturbative first principles tool available. In the context of transport coefficients, the earliest lattice calculations date back to the 1980s and the seminal work of Karsch and Wyld \cite{Karsch:1986cq}, after which a lot effort has been devoted to the extraction of quantities such as bulk and shear viscosities on the lattice (cf.~\cite{Meyer:2011gj} for a very nice review). Unfortunately, none of these works has been able to achieve good numerical accuracy even with the current computational resources. This is related to the fact that lattice QCD is formulated in Euclidean space, while the spectral function is a fundamentally Minkowskian quantity.

To see how the lattice determination of transport coefficients proceeds, let us recall the Green's function relation from \eq \nr{rho_fock}
\ba
G^>_{AB}(\omega)&=&2 (1+\nB(\omega))\rho_{AB}(\omega)\;, \\
G^<_{AB}(\omega)&=&2 \nB(\omega) \rho_{AB}(\omega)\;.
\ea
Setting $A=B$, we see that Euclidean correlators can be related to the spectral function through
\ba\la{eq:intrel}
 G(\tau)& =&
 \int_{-\infty}^\infty \frac{{\rm d}\omega}{2\pi} e^{-\tau\omega} G^>(\omega)
 \nn &=&
  \int_{-\infty}^0 \frac{{\rm d}\omega}{2\pi} e^{-\tau\omega} G^>(\omega)
  +\int_{0}^\infty \frac{{\rm d}\omega}{2\pi} e^{-\tau\omega} G^>(\omega)
 \nn &=&
  \int_{0}^\infty \frac{{\rm d}\omega}{2\pi} e^{\tau\omega} G^<(\omega)
  +\int_{0}^\infty \frac{{\rm d}\omega}{2\pi} e^{-\tau\omega} G^>(\omega)
 \nn &=&
  \int_{0}^\infty \frac{{\rm d}\omega}{\pi} \rho(\omega)
 \lk e^{\tau\omega} \nB(\omega)
  + e^{-\tau\omega} (1+\nB(\omega)) \rk \nn &=&
  \int_0^\infty
 \frac{{\rm d}\omega}{\pi} \rho(\omega)
 \frac{\cosh\Big[\! \left(\frac{\beta}{2} - \tau\right)\omega\Big]}
 {\sinh\frac{\beta \omega}{2}}\, ,\quad \quad 0<\tau <\beta \;, 
\ea
with $\nB{}(\omega)\equiv 1/(e^{\beta \omega}-1)$. To extract transport coefficients from the zero frequency limits of the corresponding spectral functions, one thus has to first collect very precise lattice data on the Euclidean correlators and then invert the relation in \eq \nr{eq:intrel} to obtain the spectral function. This inversion process amounts to extracting a continuous function from a finite number of discrete data points (with error bars), constituting a numerically ill-posed problem \cite{Meyer:2007ic}. Even though various methods (cf. \cite{Meyer:2007ic} and references therein) have been introduced to tackle the problem, the current results still contain sizable systematic and statistical uncertainties (see \fig 3 of \cite{Meyer:2007dy} for the bulk spectral function and \fig 3 of \cite{Meyer:2007ic} for the shear spectral function).

The only way to make substantial progress on the lattice front appears to be through first obtaining some analytic information on the spectral functions, either in the form of an educated ansatz or from perturbative calculations. The latter approach has been advocated in \cite{Burnier:2012ts}, and has been seen to produce promising results for a number of (simpler) transport coefficients. This is indeed our main motivation for our study of the bulk and shear spectral densities in the following chapters of this thesis.

\subsection{Gauge/gravity duality}

Finally, let us briefly comment on a third method, which has in recent years gained quite some popularity in the determination of transport coefficients in strongly coupled gauge theory: The gauge/gravity duality. Its roots are in the AdS/CFT correspondence, a conjectured equivalence between infinitely strongly coupled $\mathcal{N}=4$ SU($N_c$) supersymmetric Yang-Mills (SYM) theory in the $N_c=\infty$ limit and type IIB supergravity living in $AdS_5 \times S^5$ spacetime \cite{Maldacena:1997re}. This relation is extremely useful, because it enables the translation of complicated field theoretical problems (at strong coupling) to the language of General Relativity, thus in many cases enabling progress in problems, for which no standard field theory tool would be applicable.

The ${\mathcal N}=4$ SYM theory is a conformal gauge theory with a vanishing beta function to all orders in the coupling constant \cite{Avdeev:1980bh,Grisaru:1980nk,Caswell:1980ru,Sohnius:1981sn}, which makes it very different from QCD at zero temperature. In particular, this theory has no particle-like states, whereas we know that QCD predicts the existence of a host of glueballs, mesons and baryons. However, as discussed e.g.~in \cite{Liu:2006he}, at finite temperature both the conformal invariance and supersymmetry of the SYM theory are broken, and the system thus becomes much more like QCD in its deconfined phase (which at $T\gg T_c$ is known to approach a conformal limit). For this reason, combined with the observation that the plasma produced at RHIC appears to behave like a strongly coupled liquid, using the AdS/CFT conjecture even in phenomenological studies of the QGP has gained quite some popularity. Nevertheless, it should be recalled that as of today no known gravity dual to three-color QCD exists, and therefore from the point of view of the physical heavy ion system, all results derived using the gauge/gravity duality should be considered merely qualitative indicators. Detailed reviews of the use of the gauge/gravity duality in QCD research can be found in e.g.~\cite{Mateos:2007ay, CasalderreySolana:2011us}.

With the above words of caution in mind, let us briefly inspect the determination of transport coefficients using the gauge/gravity duality. In this dual language, the energy-momentum tensor $T_{\mu\nu}$ is mapped to small deviations of the metric $h_{\mu\nu}$ at the boundary, writing $g_{\mu\nu}=\eta_{\mu\nu}+h_{\mu\nu}$, where $\eta_{\mu\nu}$ stands for the 4-dimensional Minkowski metric. In particular the shear operator $T_{12}$ is found to be dual to $h_{12}$, and thus the shear viscosity on the field theory side can be found by studying the behavior of metric perturbations in the five-dimensional bulk. Up to order $1/\lambda^{3/2}$ in a strong coupling expansion, where $\lambda\equiv g^2 N_c$ is the 't Hooft coupling, the shear viscosity of $\mathcal{N}=4$ SYM plasma reads \cite{Buchel:2007mf,Buchel:2004di,Buchel:2008ac,Buchel:2008sh,Myers:2008yi}
\ba
\fr{\eta}{s}=\fr{1}{4\pi}\left(1+\fr{15\zeta(3)}{\lambda^{3/2}}+\dots\right) \;,
\ea 
where $\zeta(3)\approx 1.202$. At the same time, the bulk viscosity of  $\mathcal{N}=4$ SYM theory vanishes due to the conformality of the theory. It has, however, been studied in a variety of non-conformal models \cite{Buchel:2007mf,Benincasa:2005iv,Buchel:2005cv,Mas:2007ng,Buchel:2008uu}, in which the Kubo formula is again applied to the corresponding viscosity correlator, leading to a conjectured relation between the shear and bulk viscosities,
\ba
\fr{\zeta}{\eta}=2\(\fr13-c_s^2\)^2\;,
\ea
where $c_s$ is  the speed of sound.
These results seem to be in surprisingly good agreement with the RHIC and LHC data.

\mysection{Kinetic theory}
\label{sec:kinetic}

As discussed above, a separate, yet in certain limits equivalent way of formulating the determination of transport coefficients is through the use of kinetic theory \cite{Jeon:1994if,Jeon:1995zm,ValleBasagoiti:2002ir,Aarts:2002tn,Gagnon:2006hi,Gagnon:2007qt}. It is based on the assumption that the near-thermal system may be characterized by a phase space distribution function $f(x,p)$ of the on-shell components of the (quasi-)particles forming the fluid. Here, $x$ and $p$ are the space-time coordinate and 4-momentum of the particle, respectively. Clearly, such a description assumes that the system is in some sense weakly coupled.

In kinetic theory, the evolution of the system is governed by the Boltzmann equation \cite{Boltzmann}
\ba
\la{eq:bolt}
p^\mu\partial_\mu f=-\text{C}[f] \;,
\ea
where $\text{C}[f]$ is the so-called collision term which is a function of $f$ and depends on the interactions of all components of the fluid. This equation describes the interaction between the particles and yields as its solution the particle distribution function $f$. For an equilibrium system, $f$ is stationary, \ie~$f(x,p)=f_\text{eq}(p)$, implying that the collision term must vanish in this limit. 

The energy-momentum tensor in kinetic theory is defined as
\ba \la{fT}
T^{\mu\nu}\equiv\int p^\mu p^\nu f \fr{dp}{p^0} \;,
\ea
while the detailed evaluation of viscosities in kinetic theory can be found in \cite{Groot:1980} and references therein. In the following, we simply list the main steps of this procedure in two commonly used approximation schemes, the Chapman-Enskog and relaxation time approximations. 

\subsection{Chapman-Enskog approximation}

The Chapman-Enskog approximation is one of the most widely used approximation schemes within kinetic theory, with transport coefficients have
been determined for various systems of weakly interacting classical particles \cite{Leeuwen:1973}. In this section, we follow refs. \cite{Leeuwen:1973,Wiranata:2012br} to present the main idea behind it as well as the most important computational details encountered in the determination of the bulk and shear viscosities. 

To start with, we must specify the collision term appearing in the Boltzmann equation; for a non-degenerate system, this can be assumed to read
\ba
\text{C}[f]=\int (f f_1-f^{'}f_1^{'})\sigma (p+p_1)^2
\delta^{4}(p+p_1-p^{'}-p_1^{'}) dp_1 dp^{'} dp_1^{'} \;,
\ea
where $f\equiv f(x,p)$, $f_1\equiv f(x,p_1)$, $f^{'}\equiv f(x,p^{'})$, $f_1^{'}\equiv f(x,p_1^{'})$, $dp_i=d^3p_i/p_i^0$, and $\sigma$ denotes the cross section of the scattering process $p+p_1 \leftrightarrow p^{'}+p_1^{'}$ in the center of mass frame. The local distribution function appearing in these formulas can be written in the form
\ba
\la{eq:disdiv}
f=f^0(1+\phi)\;,
\ea
where $f^0$ stands for the equilibrium distribution, and $\phi\ll 1$ is introduced to denote a (small) departure from local equilibrium. For a system composed of massive (quasi)particles satisfying $p_\mu p^\mu=m^2$, the local distribution function in equilibrium reads
\ba
\la{eq:f0}
f^0=\fr{\rho }{4\pi m^3 T K_2(z)} e^{-p_\mu u^\mu/T}\;
\ea
with $\rho\equiv\rho(x)=\int f dp$ denoting the particle number density, $T\equiv T(x)$ the local temperature, $u^\mu\equiv u^\mu(x)$ the flow velocity, and $z\equiv m/T$. $K_2(z)$ is finally a modified Bessel function of the second kind, defined by
\ba
K_n(x)\equiv \fr{x^n}{(2n-1)!!} \int_0^{\infty} e^{-x\cosh y} \sinh^{2n}y\, dy \;.
\ea

In the Chapman-Enskog approximation, the particle distribution function is expanded around its equilibrium value in powers of space-time gradients, \ie
\ba
f=f^0+\d f^{(1)}+\d f^{(2)}+\ldots \;,
\ea
where $\d f^{(n)}$ denotes the $n$th order in the gradient expansion. To this end, in the first Chapman-Enskog approximation the small departure function $\phi=\d f^{(1)}/f^0$ in \eq \nr{eq:disdiv} must satisfy the linearized transport equation
\ba
\la{eq:boltCE}
p_\mu\partial^\mu f^0&=&-f^0 \int \int \int f_1^0 (\phi+\phi_1-\phi^{'}-\phi_1^{'}) \sigma (p+p_1)^2
\nn && \times
\delta^{4}(p+p_1-p^{'}-p_1^{'}) dp_1 dp^{'} dp_1^{'} \nn
&\equiv&-f^0L[\phi]\;.
\ea
A general solution of this differential equation (within the first Chapman-Enskog approximation) can be written in the form
\ba
\la{eq:phi}
\phi=-A\partial_\mu u^\mu-B_\mu\Delta^{\mu\nu}(\partial_\nu T/T+Du_\nu) +
C_{\mu\nu}\left<\partial^\mu u^\nu\right> \;,
\ea 
where we have defined $D=-u^\mu\partial_\mu$, and the angular brackets read $\left<t_{\mu\nu}\right>\equiv \Delta_{\mu\nu\alpha\beta}t^{\alpha\beta}$ with $\Delta_{\mu\nu\alpha\beta}\equiv \fr12(\Delta_{\mu\alpha}\Delta_{\nu\beta}+\Delta_{\mu\beta}\Delta_{\nu\alpha} ) - \fr13\Delta_{\mu\nu}\Delta_{\alpha\beta}$. The unknown functions $A$, $B_\mu$, and $C_{\mu\nu}$ are finally seen to satisfy the integral equations
\ba
\la{eq:A}
L[A] &=& -Q/T \;, \\
\la{eq:B}
L[B_\mu] &=& -(-p_\alpha u^\alpha+mh)\Delta_{\mu\nu}p^\nu /T \;, \\
\la{eq:C}
L[C_{\mu\nu}] &=& -\left< p_\mu p_\nu \right> /T \;,
\ea
making use of \eqs (\ref{eq:f0}) and (\ref{eq:phi}) in (\ref{eq:boltCE}). 
Here, we have also defined
\ba\la{eq:Q}
Q=-\fr{1}{3}m^2-p_\mu u^\mu [(1-\gamma)mh+\gamma T] 
+(\fr{4}{3}-\gamma) (p_\mu u^\mu)^2 \;,
\ea
with $h=K_3(z)/K_2(z)$ denoting the enthalpy, and $\gamma\equiv c_p/c_v$ being the ratio of the specific heats $c_p\equiv(\partial h/\partial T)_p$ and $c_v\equiv(\partial e/\partial T)_v$.

Expressed in terms of the quantities defined above, the tensor of \eq(\ref{eq:dTmn}) takes the form
\ba
\Delta T^{\mu\nu} &=& \Delta^{\mu\alpha}T_{\alpha\beta}\Delta^{\beta\nu}-p\Delta^{\mu\nu} 
\nn &=&-2\eta\left<\partial_\mu u^\nu\right> - \zeta \Delta^{\mu\nu} \partial_\gamma u^\gamma \;, \label{preten}
\ea
while we may write the quantities $B_\mu$ and $C_{\mu\nu}$ in the forms $B_\mu=B\Delta_{\mu\nu}p^\nu$ and $C_{\mu\nu}=C\left<p_\mu p_\nu\right>$, with $B$ and $C$ standing for unknown coefficients. Inserting the general solution of the Boltzmann equation \nr{eq:phi}--\nr{eq:Q} to \nr{fT} and plugging the resulting energy-momentum tensor to \eq(\ref{preten}), we then obtain for the viscosities 
\ba
\la{eq:CEbulk}
\zeta &=& -\int A Q f^0 dp \;,
\\
\la{eq:CEshear}
\eta &=& -\fr{1}{10}\int C \left<p_\mu p_\nu\right>\left<p^\mu p^\nu\right> f^0 dp \;.
\ea
At the same time, the coefficient $B$ is seen to be related to the heat conductivity, which is beyond the scope of our interest (cf.~\cite{Leeuwen:1973} for details). As will be shown below, the unknown functions $A$ and $C$ in \eqs (\ref{eq:CEbulk}) and (\ref{eq:CEshear}) can furthermore be expanded in terms of the Laguerre functions 
\ba
L_n^\alpha(\tau)\equiv\sum_{k=0}^n\left(\begin{array}{c} n+\alpha \\ n-k\end{array}\right) \fr{(-\tau)^k}{k!}\, ,
\ea
where $\tau=-(p_\mu u^\mu+m^2)/T$. 

To determine the value of $A$, we first multiply \eq (\ref{eq:A}) by the factor $f^0 L_n^{\fr12}(\tau)$, and consequently carry out the integration of $dp=d^3p/p^0$ on both sides of the equation. The result can be written in a closed form
\ba
\la{eq:AL}
[A,L_n^{\fr12}(\tau)]=\fr{m}{\rho}\alpha_n\;\;\; (n=0,1,\dots)\;,
\ea
where we have defined
\ba
\la{eq:FG}
[F,G] &\equiv& \fr14\fr{m^2}{\rho}\int\int\int\int f^0 f_1^0 \Delta(F)\Delta(G) dp \,dp_1 dp^{'} dp_1^{'}\;, \\
\Delta{\{F(p)\}} &\equiv& F(p)+F(p_1)-F(p^{'})-F(p_1^{'}) \;, \\
\la{eq:alfan}
\alpha_n &\equiv& -\fr{m}{\rho T} \int f^0 L_n^{\fr12}(\tau) Q dp\;,
\ea
Next, we expand $A$ as a linear combination of Laguerre 
polynomials
\ba\la{expA}
A=\sum_{m=0}^\infty a_m L_m^{\fr12}(\tau) \;,
\ea 
finding that the $\alpha_n$ must satisfy
\ba
\sum_{m=0}^\infty a_m a_{mn}=\fr{m}{\rho}\alpha_n\;\;\; (m,n=0,1,\dots)\;,
\ea
with $a_{mn}\equiv[L_m^{\fr12}(\tau),L_n^{\fr12}(\tau)]$. An approximate result $a_m^{(r)}$ for the first $r$ coefficients $a_m$ is obtained by constraining the upper limit of the sum to a finite number $r$, 
\ba\la{limit_am}
\sum_{m=2}^{r+1} a_m^{(r)} a_{mn}=\fr{m}{\rho}\alpha_n\;\;\; (m,n=2,3,\dots,r+1)\;.
\ea
Inserting \eqs \nr{expA} and \nr{eq:alfan} into \nr{eq:CEbulk}, the bulk viscosity is finally expressed in the form
\ba
\zeta=\fr{\rho T}{m} \sum_{n=2}^\infty a_m\alpha_m\;,
\ea
which can be calculated order by order using \eq \nr{limit_am}.

Let us finally move on to the case of the shear viscosity, involving the determination of the coefficient $C$. Similarly to the above treatment, we multiply \eq (\ref{eq:C}) by a factor of $f^0 L_n^{\fr52}(\tau)\left<p^\mu p^\nu\right>$ and integrate both sides of the equation. Inserting the series $C[\tau]=\sum_{m=0}^\infty c_m L_m^{\fr52}(\tau)$ (valid for a massive quasiparticle)\footnote{In the massless case, we would instead have $C[\tau]=\sum_{m=0}^\infty c_m L_m^{5}(\tau)$.} and employing the abbreviations
\ba
\gamma_n&\equiv&-\fr{1}{\rho T^2}\int f^0 L_n^{\fr52}(\tau)\left<p_\mu p_\nu\right> \left<p^\mu p^\nu\right>\;, \\
c_{mn} &\equiv& \fr{1}{T^{2}}[L_m^{\fr52}(\tau)\left<p_\mu p_\nu\right>,L_n^{\fr52}(\tau)\left<p^\mu p^\nu\right>]\;,
\ea
the shear viscosity becomes
\ba
\eta=\fr{1}{10}\rho T^2 \sum_{m=0}^\infty c_m\gamma_m\;,
\ea
which can be computed to the $r$th order through the identity
\ba
\sum_{m=0}^{r-1} c_m^{(r)}c_{mn}=\fr{1}{\rho T}\gamma_n \;\;\; (n=0,1,\ldots,r-1)\;.
\ea
The result of this procedure for a pion gas is plotted in \fig \ref{methfig3}.

\subsection{Relaxation time approximation}

\begin{figure}[t]
\bc
\includegraphics[width=12cm]{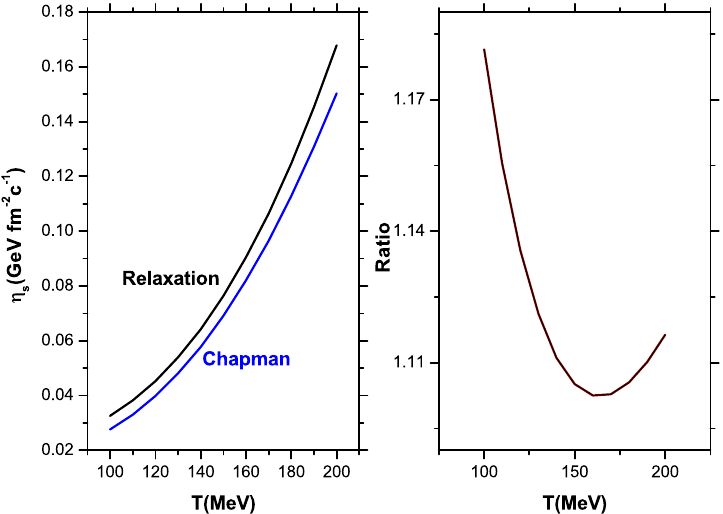}
 \caption{As an example of the kinetic theory results, we show the shear viscosity of a pion gas computed using the relaxation time approximation and the first order Chapman-Enskog approximation (left), and the ratio of the results (right). The figures are taken from \cite{Wiranata:2012br}.
 \label{methfig3}}
 \ec
\end{figure}

In physical systems, which are in some sense very close to equilibrium, another approximation scheme of kinetic theory, dubbed the Relaxation time approximation, is often useful. There, the phase space evolution is parameterized by a relaxation time $\tau$, which is defined as the inverse of the interaction rate \cite{Wiranata:2012br,Gavin:1985ph,Chakraborty:2010fr}. The collision term in the Boltzmann equation (\ref{eq:bolt}) is written in the form
\ba
C[f]=\frac{f-f^0}{\tau} \;,
\ea
while the relaxation time obtains the form (cf.~ \cite{Chakraborty:2010fr} for more details) 
\ba
\tau^{-1}=\fr12\int f_1^0 \fr{(2\pi)^4 \delta^4(p+p_1-p^{'}-p_1^{'})}{2p^{'0}\,2p_1^{'0}} |M|^2 \fr{d^3p_1}{(2\pi)^3}\fr{d^3p^{'}}{(2\pi)^3}\fr{d^3p_1^{'}}{(2\pi)^3}\;.
\ea 
Here, $|M|^2$ is the scattering amplitude for the process $p+p_1\rightarrow p^{'}+p_1^{'}$, while $f_1^0$ is the phase space distribution function of a particle with momentum $p_1$ in equilibrium.

This time the small deviation of the distribution function, cf.~ \eq (\ref{eq:disdiv}), takes the form
\ba
\la{eq:phirex}
\phi=-A\partial_\rho u^\rho +C_{\mu\nu}\left(\Delta^\mu u^\nu+\Delta^\nu u^\mu+\fr{2}{3}\Delta^{\mu\nu}\partial_\rho u^\rho\right) \;,
\ea
where $A$ and $C_{\mu\nu}$ are related to the shear and bulk viscosities, respectively. For the equilibrium distribution function $f^0=\exp[-p_\mu u^\mu/T]$, the solution of the Boltzmann equation in the rest frame of the system then reads
\ba\la{atau}
A&=&\fr{\tau}{3T p^0}\left[\left(1-3v_s^2\right)(p^0)^2-m^2\right] \;,
\\ \la{ctau}
C_{\mu\nu} &=& \fr{\tau}{2T p^0}p^\mu p^\nu \;,
\ea
where $v_s^2=\partial P/\partial e$ is the square of the velocity of sound, with $v_s^2=\fr13$ for an ideal fluid. Inserting the above solutions, \eqs \nr{eq:phirex}--\nr{ctau}, to the viscous part of the energy-momentum tensor
\ba
\Delta T^{\mu\nu}=\int \fr{d^3p}{(2\pi)^3} \fr{p^{\mu} p^\nu}{p^0} f^0 \phi\;,
\ea
and comparing with its general form in (\ref{eq:dTmn}), the bulk and shear viscosities in the local rest frame of the fluid finally take the simple forms \cite{Chakraborty:2010fr}
\ba
\zeta &=& \fr{1}{9T} \int\fr{d^3p}{(2\pi)^3} \fr{\tau}{(p^0)^2}\left[\left(1-3v_s^2\right)(p^0)^2-m^2\right]^2 f^0 \;,
\\
\eta &=& \fr{1}{15T} \int\fr{d^3p}{(2\pi)^3} \fr{\tau}{(p^0)^2} |\vec{p}|^4 f^0 \;.
\ea

%

\setcounter{chapter}{2}
%
%
%
%
%
%

\mychapter{Bulk and shear channel correlators in Yang-Mills theory}
\label{CHAP 3}

Next, we move on to the first truly novel part of the thesis, \ie~the calculations presented in the articles \cite{Laine:2011xm,Schroder:2011ht,Zhu:2012be}. We begin our treatment by first defining the bulk and shear channel correlation functions in the present chapter, as well as explaining, how it is possible to extract information about their UV behavior without having to perform complicated calculations. This section follows to a large extent the treatment of \cite{Schroder:2011ht}, as well as the corresponding bulk channel calculation performed in \cite{Laine:2010tc}.

It should be recalled that the UV limits of Green's functions are interesting not only as test cases of the eventual full computation (cf.~the following two chapters for the case of the spectral density), but also in their own right. It is namely exactly the UV limit, in which perturbation theory can be expected to converge fastest, and where lattice measurements of Euclidean correlators suffer the most from the existence of physical divergences. Indeed, while the perturbative extraction of spectral functions is important for all $\omega\sim T$, it is a very reasonable starting point to first consider the limit of $\omega\gg T$. This is most conveniently done with the help of the so-called Operator Product Expansion (OPE), which reduces the UV limit of the correlator to a sum of terms consisting of finite-temperature operators multiplying temperature independent Wilson coefficients, containing the entire dependence of the quantity on the spatial coordinates (or external momenta) \cite{Wilson:1972ee}. In our NLO calculation, we determine the values of several Wilson coefficients to unprecedented accuracy, in some cases confirming earlier results of \cite{Meyer:2008dt,CaronHuot:2009ns}. The results can in addition be compared to similar lattice and gauge/gravity calculations; see e.g.~\cite{Meyer:2010ii,Springer:2010mw,Iqbal:2009xz,Meyer:2007fc,Meyer:2010gu,Romatschke:2009ng,Huebner:2008as,
Springer:2010mf,Kajantie:2010nx,Kajantie:2011nx,Kajantie:2013gab}.

Finally, a somewhat independent motivation for our UV computations originates from so-called sum rules, \ie~integral relations derived for Green's functions based on very general nonperturbative arguments. For the bulk and shear channels of Yang-Mills theory (and QCD), they have been discussed in quite some length e.g.~in
\cite{Meyer:2007fc,Meyer:2010gu,Romatschke:2009ng,Meyer:2010ii}, the ultimate purpose being to constrain the form of the nonperturbative spectral density in different channels. As discussed in these references, the sum rules often contain a constant term, which is most conveniently evaluated using perturbative methods. The determination of these terms will be one of the side products of our calculation.

The present chapter is organized as follows. First, we define the correlation functions we set out to calculate, and later follow the treatment of \cite{Laine:2010tc,Schroder:2011ht} in explaining, what types of integrals their NLO perturbative expressions contain. After this, we go through the technical machinery needed in performing the OPE expansion, and in the end finally display the results of the corresponding calculations.

\mysection{Setup and definitions}
\la{se:setup}

All calculations presented in this thesis are performed in pure SU($N_c$) Yang-Mills theory, the Lagrangian of which is obtained from that of QCD, cf.~section \ref{se:lag}, by simply removing the quarks from the theory. We work in the $\msbar$ renormalization scheme, and denote the corresponding renormalization scale by
$ 
 \bmu^{2} \equiv {4\pi} \Lambda^2  
 {e^{-\gammaE}}\, ,
$
suppressing from the beginning all redundant factors of $\Lambda^{2\epsilon}$, which would in any case vanish in the final renormalized results. The form of the energy-momentum tensor of the theory, discussed in section \ref{se:pqcd}, then reads
\begin{equation}\la{eq:T}
 T_{\mu\nu} = \frac{1}{4} \delta_{\mu\nu} F^a_{\alpha\beta} F^a_{\alpha\beta} -F^a_{\mu\alpha} F^a_{\nu\alpha} \, .
\end{equation}
It is the correlation functions of this quantity we will be considering at some length below.

\subsection{Correlators in the bulk channel}
In the bulk channel, we are interested in two separate correlation functions
\be
 G_\theta(x) \equiv  
 \langle\, \theta(x)\, \theta(0) \,\rangle^{ }_{T}
 \;, \quad 
 G_\chi(x) \equiv 
 \langle\, \chi(x)\, \chi(0) \,\rangle^{ }_T
\;, 
\ee
where the gauge invariant scalar and pseudoscalar operators read 
\be
 \theta \equiv c^{ }_\theta\, \gB^2 F^a_{\mu\nu}F^a_{\mu\nu}
 \;, \quad
 \chi \equiv c_\chi\, \epsilon_{\mu\nu\rho\sigma} 
 \gB^2 F^a_{\mu\nu}F^a_{\rho\sigma}
 \;, \la{ops}
\ee
with the $c_i$ denoting in principle arbitrary normalization constants. The transport coefficients that the corresponding spectral functions determine are the bulk viscosity and the rate of anomalous chirality changing transitions, respectively. The corresponding Fourier transforms to momentum space are finally defined as
\be
 \tilde G_\theta(P) \equiv \int_x e^{- i P\cdot x} G_\theta(x)
 \;, \quad
 \tilde G_\chi(P) \equiv \int_x e^{- i P\cdot x} G_\chi(x)
 \;. \la{GP}
\ee

\subsection{Correlators in the shear channel}
\la{subs:corre_shear}

In the shear channel, the object of our interest is $\langle T_{12}(x)\: T_{12}(0)\rangle$, but to evaluate it in the most straightforward way we would like to be able to take advantage of the rotational invariance of the system. To this end, we consider the tensorial decomposition of the generic correlator (cf.~\cite{Kovtun:2005ev})
\begin{equation}
 G_{\mu\nu,\alpha\beta}(x)  \equiv \langle T_{\mu\nu}(x)\: T_{\alpha\beta}(0)\rangle_c\, ,
\label{eq:corr_def_coord}
\end{equation}
as well as the corresponding momentum space expression
\begin{equation}
 \tilde G_{\mu\nu,\alpha\beta}(P) \equiv \int_x e^{-iP\cdot x} G_{\mu\nu,\alpha\beta}(x) \, .
\label{eq:corr_def_mom}
\end{equation}
In these equations, the symbol $\langle\ldots\rangle_c$ stands for the connected part of the (thermal) Euclidean correlator, while we have defined $\int_x \equiv \int_0^\beta \! {\rm d}\tau \int_\vec{x}$,
$x\equiv (\tau,\mathbf{x})$; 
$P\equiv (p_n,\mathbf{p})$; 
$p_n\equiv 2\pi T n$, 
$n\in\mathbbm{Z}$. 
It should finally be noted that the Fourier transform in \eqs \nr{GP} and \nr{eq:corr_def_mom} is to be taken in $D=4-2\epsilon$ dimensions, as $1/\e$ divergences, eventually canceled by renormalization, will be frequently encountered.

To evaluate the shear channel correlator, we now introduce the projection operator
\begin{equation} \label{eq:xproj}
    X_{\mu\nu,\alpha\beta} \equiv P_{\mu\nu}^{T}P_{\alpha\beta}^{T}
	-\frac{D-2}{2}(P_{\mu\alpha}^{T}P_{\nu\beta}^{T}+P_{\mu\beta}^{T}P_{\nu\alpha}^{T}) \,,
\end{equation}
where $P_{\mu\nu}^{T}$ is a symmetric projector orthogonal to the four-momentum as well as the four vector $U=(1,\mathbf{0})$,
\begin{align}
P_{\mu\nu}^{T} &\equiv \delta_{\mu\nu}-\frac{P_\mu P_\nu
+P^2 U_\mu U_\nu-(UP)\(U_\mu P_\nu +U_\nu P_\mu\)}{P^2-(UP)^2}\;,\\
\Rightarrow &\quad
P_\mu P_{\mu\nu}^{T} = 0 = U_\mu P_{\mu\nu}^{T}\;,\quad
P_{\mu\alpha}^{T}P_{\alpha\nu}^{T}=P_{\mu\nu}^{T}\;,\quad
P_{\mu\mu}^{T}=D-2\;.
\la{eq:P}
\end{align}
Applying $X_{\mu\nu,\alpha\beta}$ to the correlator in
Eq.~(\ref{eq:corr_def_mom}) and choosing the spatial momentum $\mathbf{p}$
along the $x_{D-1}$-direction, we obtain
\begin{equation}
    X_{\mu\nu,\alpha\beta}\: \tilde G_{\mu\nu,\alpha\beta}(P) = -D(D-2)(D-3)\: \tilde G_{12,12}(P) \,,
\end{equation}
where we have indeed exploited rotational invariance in $D-2$ dimensions. As suggested by this result, we define the Euclidean correlation function that we will evaluate below by
\begin{equation}
 G_\eta(x) \equiv 2 c_\eta^2 X_{\mu\nu,\alpha\beta}(x) \, \langle T_{\mu\nu}(x)\: T_{\alpha\beta}(0)\rangle_c \,, \label{eq:g_eta_p}
\end{equation}
where $X_{\mu\nu,\alpha\beta}(x) $ denotes the corresponding projector in
coordinate space and $c_\eta$ is yet another arbitrary normalization constant. When we set $D=4$ and choose the spatial separation along the $x_3$-direction, the correlator reduces to the simple expression
\begin{equation}
 G_\eta(x)   = -16 c_\eta^2  \: \langle T_{12}(x)\: T_{12}(0)\rangle_c\, , \label{corrdef}
\end{equation}
which is indeed the shear correlator of our interest. 

With the help of \eqs \nr{eq:T} and \nr{eq:corr_def_mom}, it is finally easy to verify that when written in terms of fields, the momentum space shear correlator takes the form
\ba\la{eq:gEtaP}
\tilde G_\eta(P)=2c_\eta^2\, X_{\mu\nu,\alpha\beta} \int_x e^{-iP\cdot x}
\,\langle F_{\mu\rho}^a(x)F_{\nu\rho}^a(x)\,F_{\alpha\sigma}^b(0)F_{\beta\sigma}^b(0)\rangle_c\;.
\ea

\mysection{Correlators to next-to-leading order}
\la{se:NLO}

\begin{figure}
\bc
\includegraphics[width=10cm]{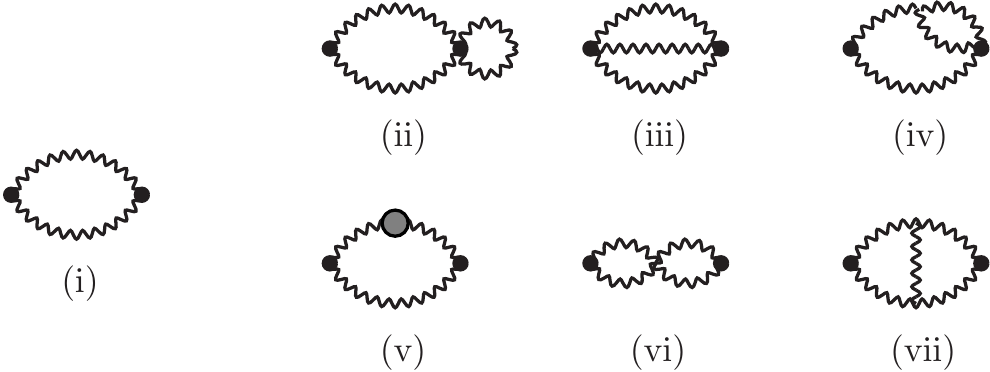}
\caption{\small 
The LO and NLO Feynman graphs contributing to the correlators of the energy momentum tensor. The figure is taken from \cite{Laine:2010tc}.}
\la{fig:graphs}
\ec
\end{figure}
%

In expanding the correlators of \eqs \nr{eq:corr_def_mom} and \nr{eq:gEtaP} to NLO in perturbation theory, we follow standard procedures: We first expand the corresponding functional integral to the desired order in $g$, then perform all necessary Wick contractions to generate the one- and two-loop graphs of Fig.~\ref{fig:graphs} (QGRAF \cite{Nogueira:1991ex} was used to check the result), and finally perform the remaining Lorentz and color algebra. Keeping $D$ unspecified for the moment and using the gluon propagator in the covariant gauge with gauge parameter $\xi$, we thereby obtain the following NLO expressions for the bulk channel correlators
\ba
 && \hspace*{-1cm} \frac{\tilde G_\theta(P)}{4 d_A c_\theta^2 } = 
 \gB^4 (D-2) \biggl[ -\mathcal{J}^{0}_\rmi{a}
 + \fr12 \mathcal{J}^{0}_\rmi{b} \biggr]
 \nn 
 & + & \gB^6 \Nc \biggl\{ 
 2 (D-2) \biggl[ - (D-2)\mathcal{I}^{0}_\rmi{a}
 + (D-4) \mathcal{I}^{0}_\rmi{b} \biggr]
 + (D-2)^2 \biggl[ \mathcal{I}^{0}_\rmi{c} - \mathcal{I}^{0}_\rmi{d} \biggr]
 \nn & &  + \, 
 \frac{22-7D}{3} \mathcal{I}^{0}_\rmi{f}
 - \frac{(D-4)^2}{2} \mathcal{I}^{0}_\rmi{g}
 + (D-2) 
 \biggl[ 
   - 3 \mathcal{I}^{0}_\rmi{e} + 3 \mathcal{I}^{0}_\rmi{h}
 + 2\mathcal{I}^{0}_\rmi{i}
    - \mathcal{I}^{0}_\rmi{j}
 \biggr] \biggr\} 
 \;, \la{Gtheta_bare} \\
 && \hspace*{-1cm} \la{Gchi_bare} \frac{\tilde G_\chi(P)}{-16 d_A c_\chi^2 (D-3)} = 
 \gB^4 (D-2) \biggl[ -\mathcal{J}^{0}_\rmi{a}
 + \fr12 \mathcal{J}^{0}_\rmi{b} \biggr]
 \nn 
 & + & \gB^6 \Nc \biggl\{ 
 2 (D-2) \biggl[ -\mathcal{I}^{0}_\rmi{a} (D-4) \mathcal{I}^{0}_\rmi{b} \biggr]
 + (D-2)^2 \biggl[ \mathcal{I}^{0}_\rmi{c} - \mathcal{I}^{0}_\rmi{d} \biggr]
 \\ & &  - \, 
 \frac{2 D^2-17D+42}{3} \mathcal{I}^{0}_\rmi{f}
 - 2 (D-4) \mathcal{I}^{0}_\rmi{g}
 + (D-2) 
 \biggl[ 
    - 3 \mathcal{I}^{0}_\rmi{e} + 3 \mathcal{I}^{0}_\rmi{h}
 + 2\mathcal{I}^{0}_\rmi{i}
    - \mathcal{I}^{0}_\rmi{j}
 \biggr] \biggr\} 
 \;,\nonumber 
\ea 
with $g_B$ standing for the bare coupling, and the master integrals defined in appendix \ref{app:master}. An important crosscheck of our calculations so far is that these results are independent of the gauge parameter $\xi$.
Implementing the above procedure for the shear correlation function \nr{eq:gEtaP} yields similarly
\ba
&&\hspace{-1.5cm}\frac{\tilde G_\eta(P)}{4 d_A c_\eta^2\Lambda^{2\epsilon}} =
\fr{D(D-2)(D-3)}{8}\(2\mathcal{J}_\rmi{a}^0-\mathcal{J}_\rmi{b}^0\) -(D-2)(D-3)\mathcal{J}_\rmi{b}^2\nn
&+&D(D-3)\(\mathcal{J}_\rmi{a}^1-\mathcal{J}_\rmi{b}^1\) \nn
&+& \gB^2 \Nc \biggl\{
\fr{D(D-2)(D-3)}{4}\bigg(2 \mathcal{I}_\rmi{a}^0 + 4 \It{b}{1} + 2 \It{b}{2} + 4 \It{d}{1} + 12 \It{d}{2} + 2 \It{e}{0}\nn
&+& \It{e}{2} + 4 \It{e}{3}
 - 3 \It{f}{1} - 4 \It{h}{0} - 4 \It{h}{1} + \It{h}{3} - 4 \It{i}{1} - 4 \It{i}{2} - 2 \It{i}{3} + \It{j}{0}\bigg)\nn
 &+&\fr{D(D-2)}{2}\bigg(-\It{e}{4} - 2 \It{e}{5} + 4 \It{e}{6} + 2 \It{e}{7} + 2 \It{h}{4} + \It{h}{5} - 4 \It{h}{6} - 2 \It{h}{7}\bigg)
 \nn
&-&\fr{(D-2)^2(D-3)}{4}\bigg(D \It{c}{0} - D \It{d}{0} - 8 \It{d}{3}\bigg) +D(D-3)\bigg(-4\It{h}{2}+2\It{j}{1}+\It{j}{2}\bigg) \nn
&-&D(D-6)\bigg(\It{j}{5}+2\It{j}{6}\bigg)+\fr{12-16D+3D^2}{2}\bigg(2\It{j}{3}+\It{j}{4}\bigg)\nn
&+&\fr{D(D-3)(3D-10)}{4}\It{e}{1} \biggr\}\;, \label{Gmasters}
\ea
with the master integrals again listed in appendix \ref{app:master}.

\mysection{UV expansion of the correlators}
\la{se:UV}

Having now identified the ``master'' sum integrals to be computed, the next step is clearly to explicitly evaluate them. For the case of the spectral function, this process will be explained in some detail in the next chapter, so here we restrict our attention to the limit where the external momentum in the graphs is much larger than $T$, corresponding to the OPE analysis explained above. Instead of going through the sum-integrals one by one, we only consider one of them in detail. The rest of the computations can be found from \cite{Laine:2010tc,Schroder:2011ht}.

\subsection{Example: The two-loop sum-integral $\It{h}{7}$} \label{subse:uvexample}

The sum-integral we have chosen to study in detail is the most complicated one encountered in the shear channel calculation, namely
\ba
\It{h}{7} & \equiv &
 \Tint{Q,R} \frac{P^2}{Q^2R^4(Q-R)^2(R-P)^2}P_T(Q)P_T(Q-R)
 \,,
\ea
where we have denoted $P_T(Q)\equiv Q_\mu Q_\nu P^T_{\mu\nu}(P)= \mathbf{q}^2-(\mathbf{q}\cdot\hat{\mathbf{p}})^2$ with $\hat{\mathbf{p}}$ a unit vector pointing in the direction of $\mathbf{p}$.
Our first step in its evaluation will be to perform the Matsubara sums, which is conveniently achieved via the well-known relation
\ba
T\sum_{p_n}F(p_n)=\int_{-\infty}^{\infty}\fr{dp}{2\pi}F(p)+\int_{-\infty-i0^+}^{\infty-i0^+}\fr{dp}{2\pi}\lk F(p) + F(-p) \rk \nB(ip) \;,
\ea
producing after some straightforward algebra
\ba
\label{Ih7c}
\It{h}{7} & = & \int_{Q,R} \frac{P^2}{Q^2R^4(Q-R)^2(R-P)^2}P_T(Q)P_T(Q-R)\nn
&+&\int_{\mathbf{q}}\frac{\nB(q)}{q}\,\Bigg[\int_{R}
\bigg( \frac{P^2\,f_q}{R^2(Q-R)^2(Q-P)^2}- \frac{P^2(1-r_n/q_n)}{R^2(Q-R)^4(Q-P)^2}\nn
&&\hspace{8em}-\frac{P^2(1-p_n/q_n)}{R^2(Q-R)^2(Q-P)^4} \bigg)\,P_T(R)P_T(Q-R)\nn
&&\hspace{8em}+\frac{2P^2}{R^4(Q-R)^2(R-P)^2}P_T(Q)P_T(Q-R) \nn
&&\hspace{8em}+ \frac{P^2}{(Q-P)^4(Q-R)^2(R-P)^2}P_T(R)P_T(Q-R) \Bigg]_Q\nn
&+&\int_{\mathbf{q}\,\mathbf{r}}\frac{\nB(q)}{q} \frac{\nB(r)}{r}\,\Bigg[
2\bigg( \frac{P^2\,f_q}{(Q-R)^2(Q-P)^2}- \frac{P^2(1-r_n/q_n)}{(Q-R)^4(Q-P)^2}\nn
&&\hspace{8em}-\frac{P^2(1-p_n/q_n)}{(Q-R)^2(Q-P)^4} \bigg)\,P_T(R)P_T(Q-R) \nn
&&\hspace{8em}+\frac{2P^2}{(Q+R-P)^2(Q-P)^4}P_T(R)P_T(Q+R)\nn
&&\hspace{8em} + \frac{P^2}{(R-Q-P)^2(Q-R)^4}P_T(Q)P_T(R)\Bigg]_{Q,R}\,,
\ea
where we have, as usual, defined $\nB(q) \equiv 1/(e^{\beta q}-1)$ and let $\int_{Q}$ and $\int_{\mathbf{q}}$ stand for $D$ and $D-1$ dimensional integrals, respectively. Observing the above result, we see that it has been naturally split to three parts, corresponding to 0, 1 and 2 powers of the Bose distribution function, while the function $f_q$ is associated with the appearance of squared propagators via the identity
\begin{equation}
    f_q \frac{\nB(q)}{q}  =    -\frac{1}{2q}\frac{\dd}{\dd q}\left( \frac{\nB(q)}{q}\right) \,.
\end{equation}
Finally, the angular brackets appearing in \eq(\ref{Ih7c}) stand for the expressions (cf.~\cite{Laine:2010tc})
\begin{align}
    [\ldots]_Q & \equiv \frac{1}{2} \sum_{q_n=\pm iq} \{\ldots\}\,,
    &[\ldots]_{Q,R} \equiv \frac{1}{4}\sum_{q_n=\pm iq}\sum_{r_n=\pm ir} \{ \ldots\}\,.
\end{align}

The parts of Eq.~(\ref{Ih7c}) proportional to 0, 1 and 2 Bose distribution functions (not counting those inside $f_q$) are from now on referred to as the 0-, 1- and 2-cut contributions, and the methods used to deal with them are different. Of the three, the 0-cut piece corresponds to the vacuum ($T=0$) correlator, and can be handled using standard integration-by-parts identities and integral tables, conveniently collected e.g.~in the TARCER Mathematica package \cite{Mertig:1998vk}. This calculation can thus be performed in a fully automated way.

The 1-cut part, on the other hand, involves a three-dimensional thermal integral, in which the integrand is a one-loop vacuum amplitude. A crucial simplification in the evaluation of its UV limit, which we are interested in, is that we may perform an expansion in positive powers of the ``on-shell'' momentum $Q$,
\ba
\label{exp}
\left[\frac{1}{(Q-R)^2}\right]_Q=\left[\frac{1}{R^2}+\frac{2Q\cdot R}{R^4}+\frac{4(Q\cdot R)^2}{R^6}+\cdots\right]_Q\,,
\ea
and similarly with $R\leftrightarrow P$. This relies on the fact that $Q$ is on shell, \ie~$Q^2=0$ inside the square brackets, while the corresponding three-momentum is cut off by the Bose distribution function. Making use of the identity $P_T(Q-R)=P_T(Q)+P_T(R)-2Q_{\mu}R_{\nu}P_{\mu\nu}^T$, the $R$ integrals can then be reduced to basic vacuum integrals easily handled within dimensional regularization, such as
\ba
\int_{R}\fr{1}{(R^2)^m[(R-P)^2]^n} 
=\fr{(P^2)^{\fr{D}{2}-m-n}}{(4\pi)^{\fr{D}{2}}} \fr{\Gamma(m+n-\fr{D}{2}) \Gamma(\fr{D}{2}-m) \Gamma(\fr{D}{2}-n)}{\Gamma(D-m-m)\Gamma(m)\Gamma(n)}\;. \;\;\;
\ea
In dealing with the remaining scalar products of $Q$ and $P$, identities such as
\ba
\big[(Q\cdot P)^2\big]_Q&=&q^2(\frac{p^2}{D-1}-p_n^2)\,,\qquad (\mathbf{q\cdot p})^2=\frac{q^2p^2}{D-1}\,,
\ea
turn out to be very handy.

In the 2-cut part of the expression (\ref{Ih7c}), one may similarly expand the propagators $1/(Q-P)^2$ and $1/(R-P)^2$ in inverse powers of $P^2$. Defining $z\equiv\mathbf{q\cdot r}/(qr)$, $q_n\equiv\sigma\, iq$, $r_n\equiv\rho\, ir$, with $\sigma,\rho=\pm$, we then obtain an expression where the only propagator involving two momenta is
\ba
\frac{1}{(Q-R)^2}=\frac{1}{2qr(\rho\sigma-z)} \;,
\ea
leading to sums like
\ba
\frac{1}{4}\sum_{\rho=\pm}\sum_{\sigma=\pm}\frac{1}{\rho\sigma-z}&=&\frac{z}{1-z^2}\,,\\
\frac{1}{4}\sum_{\rho=\pm}\sum_{\sigma=\pm}\frac{\rho\sigma}{\rho\sigma-z}&=&\frac{1}{1-z^2}\,,\\
\frac{1}{4}\sum_{\rho=\pm}\sum_{\sigma=\pm}\frac{\rho}{(\rho\sigma-z)^n}&=&\frac{1}{4}\sum_{\rho=\pm}\sum_{\sigma=\pm}\frac{\sigma}{(\rho\sigma-z)^n}=0\,,
\ea
and their derivatives. 

The remaining task is then to perform the angular integrations. The terms odd in $\mathbf{q}$ and $\mathbf{r}$ vanish due to antisymmetry, and fixing the direction of $\mathbf{r}$, we can write the remaining ones in terms of $z$-averages, such as
\ba
\left<\frac{\mathbf{\hat q}\cdot\mathbf{\hat r}\; \mathbf{\hat q}}{1-z^2}\right>_\mathbf{\hat q}&=&\mathbf{\hat r}\left<\frac{z^2}{1-z^2}\right>_z\;=\;\frac{\mathbf{\hat r}}{D-4}
\,,
\ea
where we have made use of rotational invariance and the dimensionally regularized angular integration measure. The other $z$-averages encountered in the calculation are
\ba
\left<\frac{1}{1-z^2}\right>_z &=& \frac{D-3}{D-4}
\,,\\
\left<\frac{1}{(1-z^2)^2}\right>_z &=& \frac{(D-3)(D-5)}{(D-4)(D-6)}
\,,
\ea
while the averages over the angle between $\mathbf{r}$ and $\mathbf{p}$ we need read
\ba
\left<(\mathbf{\hat r}\cdot\mathbf{\hat p})^2\right>_\mathbf{\hat r}=\frac{1}{D-1}
\,,\hspace{2em} \left<(\mathbf{\hat r}\cdot\mathbf{\hat p})^4\right>_\mathbf{\hat r}=\frac{3}{D^2-1}
\,.
\ea

Finally, for clarity of presentation, we wish to remove all $\nB^2(q)$ terms from the result, which is most conveniently achieved by utilizing the integration by parts identity
\ba
\int_{\mathbf{q}} \nB(q) q^{n-1} f_q = \frac{D-3+n}{2}\int_{\mathbf{q}}\nB(q)q^{n-3}\,.
\ea
Setting $D=4-2\epsilon$, this leads us to the final result for the UV limit of the sum-integral
\ba
&&\hspace{-1cm}\It{h}{7}  = -\frac{2^{-12+ 6\epsilon} \pi ^{ -\frac{5}{2}+2 \epsilon} \left(8 \epsilon ^3-24 \epsilon ^2+8 \epsilon +13\right)  \csc (2 \pi  \epsilon ) \Gamma^2(2-\epsilon)}{(1+\epsilon) \Gamma \left(\frac{7}{2}-\epsilon \right) \Gamma \left(5-3\epsilon\right)}\(\fr{\Lambda^{2}}{P^2}\)^{2\epsilon}P^4\nn
& + & \int_{\mathbf{q}} \frac{\nB(q)}{q} \frac{4^{-3+\epsilon } \pi ^{-1+\epsilon } \csc (\pi  \epsilon ) \Gamma (3-\epsilon )}{(3-2\epsilon ) (5-2\epsilon ) \Gamma (2-2 \epsilon )}\(\fr{\Lambda^{2}}{P^2}\)^\epsilon\nn
&\times&\!\!\Bigg[P^2 -\frac{4(8 \epsilon ^4-48 \epsilon ^3+103 \epsilon ^2-95 \epsilon +31)}{ (2-\epsilon ) (3-2 \epsilon) }q^2 +2 \epsilon  (1+\epsilon ) q^2 \fr{1}{P^2} \left(\frac{p^2}{3-2\epsilon}-p_n^2\right)\Bigg] \nn
& + & \int_{\mathbf{q}\,\mathbf{r}}\frac{\nB(q)}{q} \frac{\nB(r)}{r} \frac{(1- \epsilon ) (1-2 \epsilon ) (2-\epsilon )  }{ (1+ \epsilon ) (3-2\epsilon ) (5-2\epsilon )} \left[\frac{3  r^2}{q^2} + \frac{2 \epsilon ^3-5 \epsilon ^2+\epsilon -1}{2 \epsilon  (1-2 \epsilon ) (2-\epsilon ) }\right]\nn
&=& -\frac{13}{1440}  \left(\frac{1}{\epsilon}+\frac{5407}{780}\right)\left(\frac{\bar{\Lambda}^2}{P^2}\right)^{2\epsilon}\frac{P^4}{(4\pi )^4}\nn
& + &\frac{1}{15(4\pi )^2}\int_{\mathbf{q}} \frac{\nB(q)}{q} \Bigg[ \frac{3 P^2-62 q^2}{6 \epsilon }+ \left(\frac{ P^2}{2}-\frac{31 q^2}{3}\right)\ln\frac{\bar{\Lambda}^2}{P^2}
+\frac{q^2}{P^2} \left(\frac{p^2}{3}-p_n^2\right)\nn
&+&\frac{47 P^2}{60}+\frac{154 q^2}{45} \Bigg]
 -  \int_{\mathbf{q}\,\mathbf{r}}\frac{\nB(q)}{q} \frac{\nB(r)}{r} \Bigg[\frac{1}{30 \epsilon}- \left(\frac{29}{450}+\frac{2 r^2}{5 q^2}\right) \Bigg]+{\mathcal O}(\epsilon)\,,
\ea
where we have in the last stage performed an expansion in $\epsilon$ and again chosen the $\msbar$ scheme. 

Results for the other sum-integrals, as well as for the full correlator of \eqs \nr{Gtheta_bare}--(\ref{Gmasters}), can be obtained in a fully analogous way.

\subsection{The results}

Implementing the strategies explained above to all of the integrals appearing in the bulk and shear correlators in \eqs \nr{Gtheta_bare}-(\ref{Gmasters}), their UV expansions become
\ba
&&\hspace{-1.0cm} \frac{\tilde G_\theta(P)}{4 d_A c_\theta^2 g^4 \Lambda^{2\e}} =\nn &&\frac{P^4}{(4\pi )^2}\Bigg\{ \left(\frac{\bar{\Lambda }}{P}\right)^{2\epsilon }\left[\frac{1}{\epsilon }+1\right] \(1-\frac{g^2N_c}{(4\pi )^2}\fr{22}{3\e} \) +\frac{g^2N_c}{(4\pi )^2}\left(\frac{\bar{\Lambda }}{P}\right)^{4\epsilon }\left[\frac{11}{3\epsilon^2}+\frac{95}{6\e}\right]\Bigg\}\nn
&+&\frac{8}{P^2}\left(\frac{p^2}{3}-p_n^2\right)\Bigg[1+\frac{g^2N_c}{(4\pi)^2}\Bigg(\fr{22}{3}\ln\frac{\bar{\Lambda }^2}{P^2} +\frac{203}{18}\Bigg)\Bigg]
  \int_{\mathbf{q}}\nB(q)q \nn
&-&4g^2N_c\Bigg[\frac{3}{P^2}\left(\frac{p^2}{3}-p_n^2\right)+1\Bigg]\int_{\mathbf{q}\,\mathbf{r}}\frac{\nB(q)}{q} \frac{\nB(r)}{r}+{\mathcal O}\(g^4,\fr{1}{P^2}\)\,, \la{Gtheta_uv}\\ \nn
&&\hspace{-1.0cm} \frac{\tilde G_\chi(P)}{-16 d_A c_\chi^2 g^4 \Lambda^{2\e}}=\frac{P^4}{(4\pi )^2}\Bigg\{ \left(\frac{\bar{\Lambda }}{P}\right)^{2\epsilon }\left[\frac{1}{\epsilon }-1\right] \(1-\frac{g^2N_c}{(4\pi )^2}\fr{22}{3\e} \)
+ \frac{g^2N_c}{(4\pi )^2}\left(\frac{\bar{\Lambda }}{P}\right)^{4\epsilon }\nn
&\times&\left[\frac{11}{3\epsilon^2}+\frac{25}{2\e}\right]\Bigg\} + \frac{8}{P^2}\left(\frac{p^2}{3}-p_n^2\right)\Bigg[1+\frac{g^2N_c}{(4\pi)^2}\Bigg(\fr{22}{3}\ln\frac{\bar{\Lambda }^2}{P^2} +\frac{347}{18}\Bigg)\Bigg]
  \int_{\mathbf{q}}\nB(q)q \nn
&-&4g^2N_c\Bigg[\frac{3}{P^2}\left(\frac{p^2}{3}-p_n^2\right)-1\Bigg]\int_{\mathbf{q}\,\mathbf{r}}\frac{\nB(q)}{q} \frac{\nB(r)}{r}+{\mathcal O}\(g^4,\fr{1}{P^2}\)\,, \la{Gchi_uv} \\ \nn
&&\hspace{-1.0cm} \frac{\tilde G_\eta(P)}{4 d_A c_\eta^2}=\frac{P^4}{(4\pi )^2}\Bigg\{-\frac{2 }{5 } \left(\frac{\bar{\Lambda }}{P}\right)^{2\epsilon }\left[\frac{1}{\epsilon }-\frac{13 }{5 }\right] +\frac{g^2N_c}{(4\pi )^2}\left(\frac{\bar{\Lambda }}{P}\right)^{4\epsilon }\left[\frac{4}{9\epsilon}-\frac{206}{135}+\frac{24 \zeta(3)}{5}\right]\Bigg\}\nn
&-&\Bigg[\frac{8}{3} +\frac{8}{P^2}\left(\frac{p^2}{3}-p_n^2\right)
  -\fr{8}{9}\frac{g^2N_c}{(4\pi)^2}\Bigg(22 +\frac{41}{P^2}\left(\frac{p^2}{3}-p_n^2\right)\Bigg)\Bigg]
  \int_{\mathbf{q}}\nB(q)q \nn
&+&g^2N_c\Bigg[\fr{20}{3}+\frac{12}{P^2}\left(\frac{p^2}{3}-p_n^2\right)\Bigg]\int_{\mathbf{q}\,\mathbf{r}}\frac{\nB(q)}{q} \frac{\nB(r)}{r}+{\mathcal O}\(g^4,\fr{1}{P^2}\)\,, \label{Geta_uv}
\ea
where we have switched from the bare to the physical coupling using the relation 
\ba
g_B^2=g^2\Lambda^{2\e}\lk 1-\fr{g^2b_0}{\e}+\ldots\rk\;. \la{gB}
\ea

There are several interesting things to note about the above results. First, their $T=0$ parts agree with the previous results of \cite{Kataev:1981gr,Pivovarov:1999mr}. Furthermore, while the bulk channel results require renormalization of the coupling constant in order for their finite-temperature pieces to become finite, this is not true for the shear correlator, where all $1/\epsilon$ divergences automatically cancel. This in particular implies that there are no logs of $P^2$ in the thermal terms, suggesting that the UV expansion of the corresponding thermal spectral function only starts at a relatively high order in a $1/\omega$ expansion.

\mysection{OPE expansion}
\la{se:OPE}

As suggested above, the form of our results for the UV correlators allows for an interpretation of their $T$-dependent parts, denoted here by $\Delta\tilde G_x(P)$, in terms of an operator product expansion (OPE). Utilizing standard thermodynamics formulae for quantities such as the pressure and energy density, this leads us to the bulk channel results
\ba
\frac{\Delta\tilde G_\theta(P)}{4 c_\theta^2 g^4}&=&\frac{3}{P^2}\left(\frac{p^2}{3}-p_n^2\right)\Bigg[1 +  \frac{g^2N_c}{(4\pi)^2}\Bigg(\fr{22}{3}\ln\fr{\bar{\Lambda}^2}{P^2}+\fr{203}{18})\Bigg)\Bigg]
  (e+p)(T) \nn
&-&\fr{2}{g^2b_0}\Big[1+g^2b_0\ln\fr{\bar{\Lambda}^2}{\zeta_{\theta}P^2}\Big](e-3p)(T)+{\mathcal O}\(g^4,\fr{1}{P^2}\)\,, \la{theres2} \\ 
\frac{\Delta\tilde G_\chi(P)}{-16 c_\chi^2 g^4}&=&\frac{3}{P^2}\left(\frac{p^2}{3}-p_n^2\right)\Bigg[1 +
\frac{g^2N_c}{(4\pi)^2}\Bigg(\fr{22}{3}\ln\fr{\bar{\Lambda}^2}{P^2}+\fr{347}{18})\Bigg)\Bigg]
  (e+p)(T) \nn
&+&\fr{2}{g^2b_0}\Big[1+g^2b_0\ln\fr{\bar{\Lambda}^2}{\zeta_{\chi}P^2}\Big](e-3p)(T)+{\mathcal O}\(g^4,\fr{1}{P^2}\)\,. \la{chires2}
\ea
Here, we have used the thermodynamic identities
\ba
(e+p)(T)&=&\fr{8d_A}{3}\Bigg[\int_{\mathbf{q}}\nB(q)q-\fr{3g^2N_c}{2}\int_{\mathbf{q}\,\mathbf{r}}\frac{\nB(q)}{q} \frac{\nB(r)}{r}\Bigg]\,,\\
(e-3p)(T)&=&2d_A g^4 b_0 N_c \int_{\mathbf{q}\,\mathbf{r}}\frac{\nB(q)}{q} \frac{\nB(r)}{r} \,.
\ea
Likewise, the shear correlator in \nr{Geta_uv} can be written in the same fashion, reading
\ba
&&\hspace{-1.2cm}\frac{\Delta\tilde G_\eta(P)}{4 c_\eta^2}=-\Bigg\{1 +\frac{3}{P^2}\left(\frac{p^2}{3}-p_n^2\right)
  -\fr{1}{3}\frac{g^2N_c}{(4\pi)^2}\Bigg(22+\frac{41}{P^2}\left(\frac{p^2}{3}-p_n^2\right)\Bigg)\Bigg\}
  (e+p)(T) \nn
&+&\fr{4}{3g^2b_0}(e-3p)(T)\Big\{1-g^2b_0\ln\zeta_{12}\Big\}+{\mathcal O}\(g^4,\fr{1}{P^2}\)\,. \label{Gres2}
\ea
The values of the $T$-independent constants $\zeta_\theta$ and $\zeta_\chi$ in \eqs \nr{theres2} and \nr{chires2} as well as the $\zeta_{12}$ appearing in Eq.~(\ref{Gres2}) can be fixed with NNLO (3-loop) calculations.

The results given above can be used in several applications, such as determining the UV limits of the corresponding spectral densities and Euclidean correlators, as well as for checking sum rules. We will, however, skip these phenomenological considerations for now, and rather return to them only after having derived results for the entire spectral functions in the following chapter.

%

\setcounter{chapter}{3}
%
%
%
%
%
%

\mychapter{Thermal spectral functions: Methods}
\label{CHAP 4}
 
We have already learned that spectral functions play an important role in the analysis of the QGP, in particular because various transport coefficients can be determined from their zero frequency limits. Even though a lot of effort has been put into the lattice determination of these quantities, (see e.g.~the ``Maximum Entropy Method'' of \cite{Asakawa:2000tr}), large model dependent uncertainties still remain in the current lattice estimates of e.g.~ the bulk and shear viscosities \cite{Meyer:2007ic,Meyer:2007dy}. To this end, any progress in the extraction of spectral functions from the corresponding Euclidean correlators would clearly be highly welcome. 

A promising step in the above direction was recently taken in\cite{Burnier:2011jq}, where a model independent method was proposed for performing the analytic continuation necessary to obtain the spectral function using \eq~(\ref{eq:intrel}). In particular, it was shown that after an analytical subtraction of the UV divergent part of the quantity, the spectral function should be obtainable without any extra assumptions on its shape. The method was later applied to the determination of the flavor diffusion coefficient of QCD in \cite{Burnier:2012ts}, where it was additionally pointed out that for the practical feasibility of the strategy, it is important to have as much prior analytical understanding of the short distance behavior of the spectral density as possible. Such insights are clearly most easily accessible through perturbative calculations.

Having the ultimate goal of a nonperturbative extraction of the bulk and shear viscosities in mind, we will now set out to consider the perturbative evaluation of the corresponding spectral densities, hoping that the results will eventually find use in the analytic continuation of lattice data to Minkowskian signature. As in the previous chapter, we will work to NLO in perturbation theory, complementing earlier works (in other channels) such as \cite{Baier:1988xv,Gabellini:1989yk,Altherr:1989jc,Lu:2011df,Laine:2011is,He:2011yi,Ghosh:2011bw,Abreu:2011ic,Salvio:2011sf,Burnier:2008ia,Burnier:2010rp}. While many of these references have made important technical advances, our goal will also be to systematize the calculation of NLO spectral functions to as large an extent as possible. We will begin this from the bulk channel of pure Yang-Mills theory (cf.~\cite{Laine:2011xm}), after which we continue to the somewhat more complicated shear channel, considered in \cite{Zhu:2012be}. All of our calculations will be performed assuming the external three-momentum to vanish.

\mysection{Bulk channel}
\la{se:methodbulk}

In the bulk channel, the determination of the spectral densities reduces to the evaluation of the sum-integrals appearing in \eqs\nr{Gtheta_bare}--\nr{Gchi_bare}, of which we must take the imaginary parts according to
\ba
\rho(\omega)\equiv \im  \Bigl[ \widetilde{G}(P) 
 \Bigr]_{P \to (-i[\omega + i 0^+],\vec{0})}
 \;. \label{rho_general}
\ea
In order to explain our method as clearly as possible, we again take just one integral under consideration, choosing as a representative example $\mathcal{I}^{0}_\rmii{j}$ (cf.~also appendix A of ref.~\cite{Laine:2011xm}),
\ba
 \mathcal{I}^{0}_\rmii{j}(P) & \equiv & 
 \Tint{Q,R} \frac{P^6}{Q^2R^2[(Q-R)^2+\lambda^2](Q-P)^2(R-P)^2}
 \;. \la{Ij}
\ea
The parameter $\lambda$ appearing here has been introduced to regulate unphysical IR divergences encountered upon splitting the integral to several parts; when adding these terms together at the end of the computation, the limit $\lambda\to 0$ can finally be taken.

%
\subsection{Matsubara sums and discontinuities}

The very first step in the evaluation of the integral is again to carry out the Matsubara sums appearing in \eq\nr{Ij}. This is most conveniently achieved via the use of the relation
\ba\la{matsum}
T\sum_{p_n}\fr{e^{\pm p_n\sigma}}{p_n^2+p^2}=\fr{n_p}{2p}
\lk e^{(\beta-|\sigma\!\!\!\!\mod 2\beta|)p}+e^{|\sigma\!\!\!\!\mod 2\beta|p}\rk \;, 
-\beta\le\sigma\!\!\!\!\mod 2\beta\le\beta\;,
\ea
where we have introduced the notation
\be
 n_E \equiv \nB{}(E)
 \;.
\ee
Making next use of the formula
\ba\la{dint}
\beta\d_{p_n}=\int_0^{\beta}d\sigma e^{ip_n\sigma}\;,
\ea
and performing the resulting integrations leads to a result composed of products of terms of the form
\ba
\fr{1}{i p_n+\sum_k\sigma_k E_k}\;, \sigma_k=\pm1\;, E_k\in\{E_\vec{q},E_\vec{r},E_\vec{q-p},E_\vec{r-p},E_\vec{q-r}\}\;.
\ea
This makes it easy to perform the analytic continuation
$P \to (-i[\omega + i 0^+],\vec{0})$ and take the imaginary part of the result using the standard relation
\ba
\fr{1}{\omega\pm i 0^+} =\mathbb{P}\(\fr{1}{\omega}\) \mp i\pi \delta(\omega)\, ,
\ea
with $\mathbb{P}$ standing for the principal value.

Following along the above lines, we obtain for the ``master spectral function'' corresponding to the master integral $\mathcal{I}^{0}_\rmii{j}$
{\small
\ba
 & & \hspace*{-1cm} \rho^{ }_{\mathcal{I}^{0}_\rmii{j}}(\omega) = 
 \int_{\vec{q,r}} 
 \frac{\omega^6 \pi }{4 q r E_{qr}} \biggl\{ 
 \la{Ij_T} \\
 & & \!\!
 \frac{1}{8q^2} 
 \Bigl[\delta(\omega - 2 q) - \delta(\omega+2 q) \Bigr]
 \times 
 \nn & & \times \biggl[
 \biggl( 
 \frac{1}{(q+r-E_{qr})(q+r)} -  
 \frac{1}{(q-r+E_{qr})(q-r)} 
 \biggr)
 (1 + 2 n_q) (n_{qr}-n_r)
 \nn & & \;\; 
 +
 \biggl(
 \frac{1}{(q+r+E_{qr})(q+r)} -  
 \frac{1}{(q-r-E_{qr})(q-r)} 
 \biggr)
 (1 + 2 n_q) (1 + n_{qr}+n_r)
 \biggr]
 \nn & + & \!\!
 \frac{1}{8r^2} 
 \Bigl[\delta(\omega - 2 r) - \delta(\omega+2 r) \Bigr]
 \times 
 \nn & & \times \biggl[
 \biggl( 
 \frac{1}{(q+r-E_{qr})(q+r)} -  
 \frac{1}{(q-r-E_{qr})(q-r)} 
 \biggr)
 (1+2n_r)(n_{qr}-n_q)
 \nn & & \;\; 
 +
 \biggl(
 \frac{1}{(q+r+E_{qr})(q+r)} -  
 \frac{1}{(q-r+E_{qr})(q-r)} 
 \biggr)
 (1+2n_r)(1 + n_{qr}+n_q)
 \biggr]
 \nn & + & \!\!
 \Bigl[\delta(\omega - q - r -E_{qr}) - \delta(\omega+q+r+E_{qr}) \Bigr]
 \frac{(1+n_{qr})(1+n_q+n_r)+n_q n_r}
      {(q+r+E_{qr})^2(q-r+E_{qr})(q-r-E_{qr})}
 \nn & + &  \!\!
 \Bigl[\delta(\omega-q-r+E_{qr}) - \delta(\omega + q + r -E_{qr}) \Bigr]
 \frac{n_{qr}(1+n_q + n_r ) - n_qn_r}
      {(q+r-E_{qr})^2(q-r+E_{qr})(q-r-E_{qr})}
 \nn & + &  \!\!
 \Bigl[\delta(\omega - q + r -E_{qr}) - \delta(\omega+q-r+E_{qr}) \Bigr]
 \frac{n_r(1+n_q+n_{qr})-n_q n_{qr}}
      {(q-r+E_{qr})^2(q+r+E_{qr})(q+r-E_{qr})}
 \nn & + &  \!\!
 \Bigl[\delta(\omega + q - r -E_{qr}) - \delta(\omega-q+r+E_{qr}) \Bigr]
 \frac{n_q(1+n_r+n_{qr})-n_r n_{qr}}
      {(q-r-E_{qr})^2(q+r+E_{qr})(q+r-E_{qr})}
 \biggr\} \;. \nonumber
\ea
}
Here, we have denoted 
\be
 E_q \equiv q \;, \quad E_r \equiv r \;, \quad 
 E_{qr} \equiv \sqrt{(\vec{q}-\vec{r})^2 + \lambda^2}
 \;,
\ee
and neglected all terms proportional to $\omega^n\delta(\omega)$ as uninteresting for our current purposes.

With the above expression at hand, we are able to divide the spectral function to three distinct parts based on the physical origin of the  different contributions. The first two structures of (\ref{Ij_T}), with simple $\delta$-function constraints, will be referred to as ``factorized'' (fz) integrals, corresponding physically to virtual corrections. The latter four structures, with more complicated $\delta$-constraints, are on the other hand labeled ``phase space'' (ps) integrals, corresponding to real processes as depicted in \fig\ref{fig:processes}. The factorized integrals are finally further divided to ``powerlike'' (fz,p) integrals without Boltzmann suppression at large momenta, and ``exponential'' (fz,e) integrals with associated Boltzmann factors. 

\begin{figure}[t]
\centerline{%
\includegraphics[width=10cm]{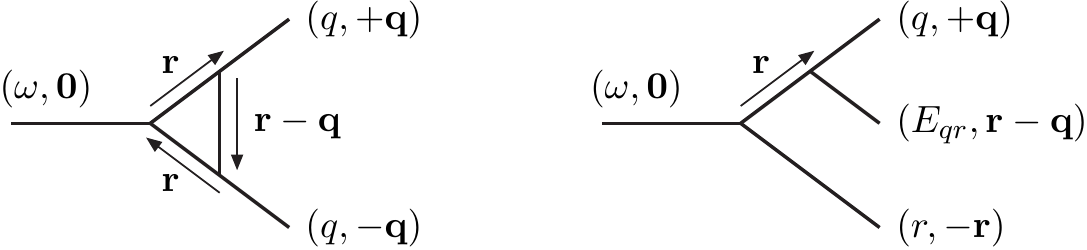}%
}
\caption[a]{\small 
Examples of the physical amplitudes contained in \eq\nr{Ij_T}. 
Left: ``virtual corrections'' or ``factorized integrals''. 
Right:  ``real corrections'' or ``phase space integrals''. The figure is taken from \cite{Laine:2011xm}.
} 
\la{fig:processes}
\end{figure}

The factorized and phase space integrals are both separately divergent (or ill-defined) in the limit $\lambda\to 0$ due to the appearance of ``soft'', ``collinear'' and  ``thermal infrared'' divergences (cf.~\cite{Salvio:2011sf}). Below, we will evaluate each of them keeping $\lambda\neq 0$, after which we sum the results together and take the limit $\lambda\to 0$.

%
\subsection{Factorized powerlike integrals}
\la{ss:fz_p}

Starting with the factorized integrals, it is clear that the symmetry $q\leftrightarrow r$ allows us to combine the first two terms of \eq\nr{Ij_T}.  Picking $\delta(\omega - 2r)$ as a representative and using the fact that $\omega > 0$, we first use the trivial relation 
\be
 \int_{\vec{r}} \pi \delta(\omega - 2 r) = \frac{\omega^2}{16\pi}
 \;,
\ee
and later deal with the $\vec{q}$-integral writing
\be
 \int_{\vec{q}} \frac{1}{2q E_{qr}}
 = \frac{1}{4\pi^2\omega} \int_0^\infty \! {\rm d}q 
 \int_{E_{qr}^-}^{E_{qr}^+} \!\! {\rm d}E_{qr}
 \;, \quad
 E_{qr}^\pm \equiv \sqrt{\Bigl(q\pm\frac{\omega}{2}\Bigr)^2 + \lambda^2}
 \;. \la{fz_measure0}
\ee
Using these variables, the factorized powerlike integral can be expressed in the form
\ba
 && \hspace*{-1cm} \rho_{\mathcal{I}^{0}_\rmii{j}}^{(\rmi{fz,p})}(\omega)
  \equiv  
  \frac{\omega^4}{(4\pi)^3}  (1 + 2 n_{\frac{\omega}{2}})
  \int_0^\infty \! {\rm d}q 
  \int_{E_{qr}^-}^{E_{qr}^+} \!\! {\rm d}E_{qr} \biggl\{ 
  \nn & & \; \; \;
 \mathbbm{P} \, \biggl[
 \frac{1}{(q+\frac{\omega}{2}+E_{qr})(q+\frac{\omega}{2})} -  
 \frac{1}{(q-\frac{\omega}{2}+E_{qr})(q-\frac{\omega}{2})} 
 \biggr]
  \biggr\}  
  \;. \la{Ij_fzp_1}  
\ea

While we could in principle immediately integrate over the variable $E_{qr}$, we have learned that it is easier to take the limit $\lambda\to 0$ if we change the order of integrations through 
\be
 \int_0^\infty \! {\rm d}q 
 \int_{ 
   \sqrt{(q - \frac{\omega}{2})^2 + \lambda^2}
  }^{ 
   \sqrt{(q + \frac{\omega}{2})^2 + \lambda^2}
  } \! {\rm d}E_{qr} 
 = 
 \int_{\lambda}^{\infty} \! {\rm d}E_{qr} 
 \int_{ 
  |\frac{\omega}{2} - \sqrt{E_{qr}^2-\lambda^2}|
 }^{
  \frac{\omega}{2} + \sqrt{E_{qr}^2-\lambda^2}
 } \!\! {\rm d}q
 \;. \la{fz_order0}
\ee
Denoting the inner integration variable as $r$ and the outer one as $q$, 
\eq\nr{Ij_fzp_1} can be easily fractioned to read
\ba
 \rho_{\mathcal{I}^{0}_\rmii{j}}^{(\rmi{fz,p})}(\omega) & = & 
  \frac{\omega^4}{(4\pi)^3}   (1 + 2 n_{\frac{\omega}{2}})
 \int_{\lambda}^{\infty} \! \frac{{\rm d}q}{q} 
 \int_{ 
  |\frac{\omega}{2} - \sqrt{ q^2-\lambda^2}|
 }^{
  \frac{\omega}{2} + \sqrt{ q^2-\lambda^2}
 }
  \!\! {\rm d}r \, \biggl\{ 
  \nn & & \; \; \;
  \mathbbm{P} \, \biggl[ 
  \frac{1}{r+\frac{\omega}{2}} - 
  \frac{1}{q+r+\frac{\omega}{2}} -   
  \frac{1}{r-\frac{\omega}{2}} +
  \frac{1}{q+r-\frac{\omega}{2}} 
  \biggr] 
  \biggr\}
  \;. \hspace*{5mm} \la{Ij_fzp_2}
\ea
This leads us to the result
\ba\la{Ij_fzp_3}
 && \hspace*{-1cm} \rho_{\mathcal{I}^{0}_\rmii{j}}^{(\rmi{fz,p})}(\omega) =  
  \frac{\omega^4}{(4\pi)^3} (1+2 n_{\frac{\omega}{2}}) \biggl\{  
  \nn && 
  \int_\lambda^{\sqrt{(\frac{\omega}{2})^2+\lambda^2}} 
  \! \frac{{\rm d}q}{q}  
      \ln\biggl| 
            \frac{q +\sqrt{q^2-\lambda^2}}
                 {q -\sqrt{q^2-\lambda^2}}
         \biggr|
         \biggl| 
            \frac{\omega+\sqrt{q^2-\lambda^2}}
                 {\omega-\sqrt{q^2-\lambda^2}}
         \biggr|
         \biggl| 
            \frac{\omega + q - \sqrt{q^2-\lambda^2}}
                 {\omega + q + \sqrt{q^2-\lambda^2}}
         \biggr|
  \\ & + & 
  \int_{\sqrt{(\frac{\omega}{2})^2+\lambda^2}}^\infty 
  \! \frac{{\rm d}q}{q} 
      \ln\biggl| 
            \frac{(q +\sqrt{q^2-\lambda^2})^2}
                 {{q^2-\lambda^2}}
         \biggr|
         \biggl| 
            \frac{\sqrt{q^2-\lambda^2} + \omega}
                 {q + \sqrt{q^2-\lambda^2} + \omega}
         \biggr|
         \biggl| 
            \frac{\sqrt{q^2-\lambda^2} - \omega}
                 {q + \sqrt{q^2-\lambda^2} - \omega}
         \biggr|
 \biggr\}
  , \nonumber
\ea
where $\sqrt{(\frac{\omega}{2})^2+\lambda^2}$ is a solution of $|\frac{\omega}{2} - \sqrt{ q^2-\lambda^2}|$=0. Sending finally $\lambda\to 0$ everywhere, where this doesn't lead to divergences, we obtain as the final result of the (fz,p) contribution
\be
 \rho_{\mathcal{I}^{0}_\rmii{j}}^{(\rmi{fz,p})}(\omega) \approx 
  \frac{\omega^4}{(4\pi)^3} (1 + 2 n_{\frac{\omega}{2}} )
  \int_\lambda^{ \frac{\omega}{2} } 
  \! \frac{{\rm d}q}{q} 
      \ln\biggl| 
            \frac{q +\sqrt{q^2-\lambda^2}}
                 {q -\sqrt{q^2-\lambda^2}}
         \biggr|
 \;. 
 \la{Ij_fzp_6} 
\ee

%
\subsection{Factorized exponential integrals}
\la{ss:fei}

The factorized exponential integrals are dealt with in a fashion similar to the (fz,p) contributions. The same change of ordering as in \eq\nr{fz_order0} is performed to keep all Bose distribution functions in the outer integral. After renaming the integral variables as $q$ and $r$, the relevant part of \eq\nr{Ij_T} becomes 
\ba
 \rho_{\mathcal{I}^{0}_\rmii{j}}^{(\rmi{fz,e})}(\omega) & \equiv & 
  \frac{\omega^4}{(4\pi)^3} (1 + 2 n_{\frac{\omega}{2}})
  \int_0^\infty \! {\rm d}q \, n_q
  \int_{ 
    \sqrt{(q - \frac{\omega}{2})^2 + \lambda^2}
   }^{ 
    \sqrt{(q + \frac{\omega}{2})^2 + \lambda^2}
   }
   \!\! {\rm d}r \, \biggl\{ 
  \nn & & \; \; \;
  \mathbbm{P} \, \biggl[ 
  \frac{1}{(q+r+\frac{\omega}{2})(q+\frac{\omega}{2})} -  
  \frac{1}{(q+r-\frac{\omega}{2})(q-\frac{\omega}{2})}
  \nn & & \hspace*{1cm} + \, 
  \frac{1}{(q-r-\frac{\omega}{2})(q-\frac{\omega}{2})} -
  \frac{1}{(q-r+\frac{\omega}{2})(q+\frac{\omega}{2})} 
  \biggr] 
  \biggr\}
  \nn & + &
  \frac{\omega^4}{(4\pi)^3} (1 + 2 n_{\frac{\omega}{2}})
 \int_{\lambda}^{\infty} \! {\rm d}q \, n_q 
 \int_{ 
  |\frac{\omega}{2} - \sqrt{ q^2-\lambda^2}|
 }^{
  \frac{\omega}{2} + \sqrt{ q^2-\lambda^2}
 }
  \!\! {\rm d}r \, \biggl\{ 
  \nn & & \; \; \;
  \mathbbm{P} \, \biggl[ 
  \frac{1}{(q+r+\frac{\omega}{2})(r+\frac{\omega}{2})} -  
  \frac{1}{(q+r-\frac{\omega}{2})(r-\frac{\omega}{2})}
  \nn & & \hspace*{1cm} + \, 
  \frac{1}{(q-r+\frac{\omega}{2})(r-\frac{\omega}{2})} -
  \frac{1}{(q-r-\frac{\omega}{2})(r+\frac{\omega}{2})} 
  \biggr] 
 \biggr\}
  \;. \hspace*{5mm} \la{Ij_fz_2}
\ea
The latter term is again partially fractioned to have a form similar to \eq\nr{Ij_fzp_2}. Integrating over $r$, this quickly leads to the result
\ba
 && \hspace*{-1cm} \rho_{\mathcal{I}^{0}_\rmii{j}}^{(\rmi{fz,e})}(\omega) =  
  \frac{\omega^4}{(4\pi)^3} (1 + 2 n_{\frac{\omega}{2}}) \biggl\{  
  \\ && 
  \int_0^\infty \! {\rm d}q \, n_q \, \mathbbm{P} \, \biggl[ 
      \frac{1}{q+\frac{\omega}{2}}
      \ln\biggl| \frac{\lambda^2}{2q\omega - \lambda^2} \biggr|
      +  \frac{1}{q-\frac{\omega}{2}} 
      \ln\biggl| \frac{\lambda^2}{2q\omega + \lambda^2} \biggr|
   \; \biggr]
  \nn & + & 
  \int_{\lambda}^\infty \! {\rm d}q \, n_q  \biggl[ 
    \frac{1}{q}
     \ln
         \biggl| 
            \frac{q + \frac{\lambda^2}{\omega}+\sqrt{q^2-\lambda^2}}
                 {q + \frac{\lambda^2}{\omega}-\sqrt{q^2-\lambda^2}}
         \biggr|
    + 
    \frac{1}{q}
     \ln
         \biggl| 
            \frac{q - \frac{\lambda^2}{\omega}+\sqrt{q^2-\lambda^2}}
                 {q - \frac{\lambda^2}{\omega}-\sqrt{q^2-\lambda^2}}
         \biggr|
  \; \biggr] \biggr\}
  \;. \la{Ij_fz_3}\nonumber
\ea

%
\subsection{Phase space integrals}\la{sse:ps_bulk}
\begin{figure}[t]

\centerline{%
\includegraphics[width=4.5cm]{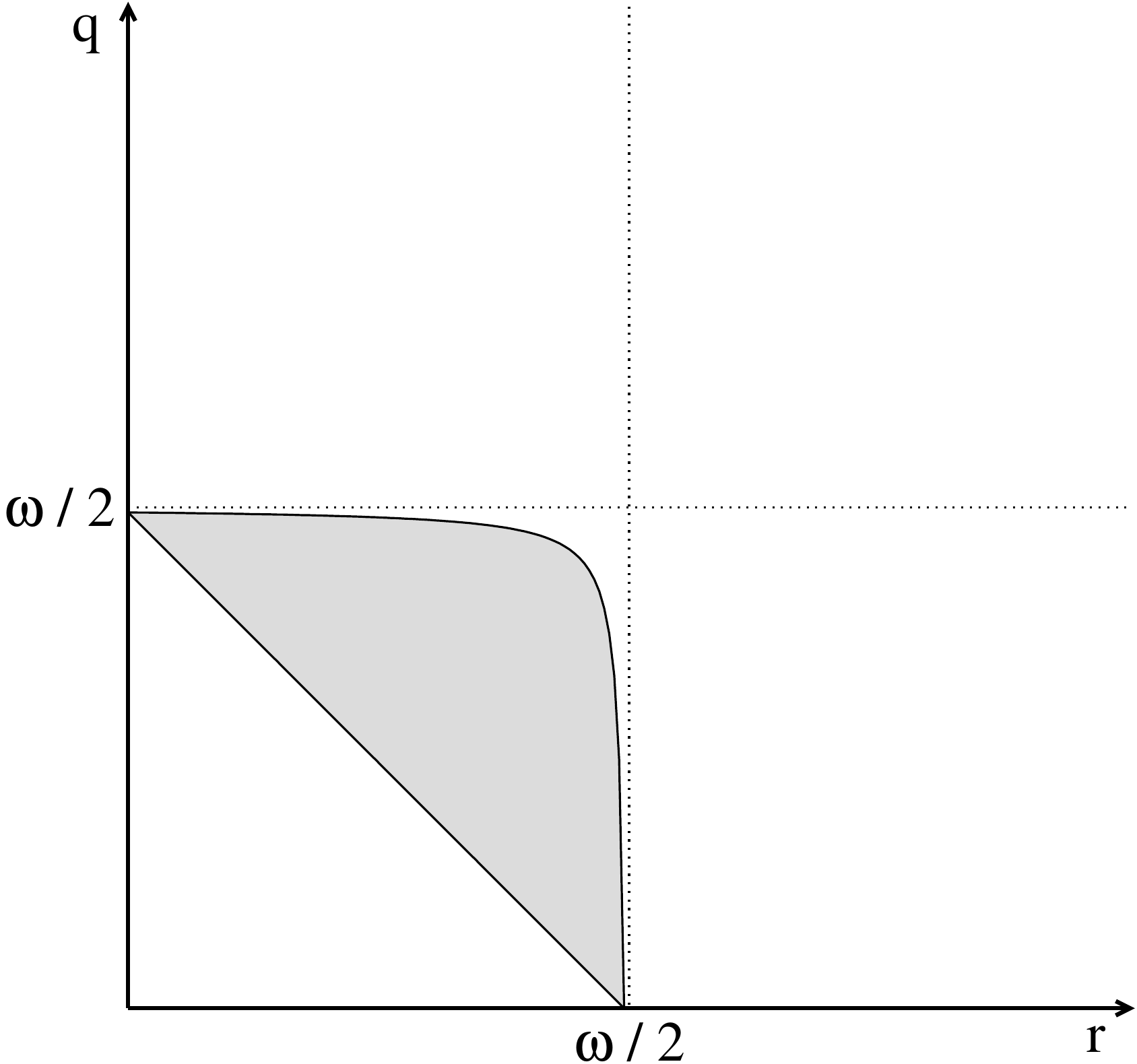}%
~~\includegraphics[width=4.5cm]{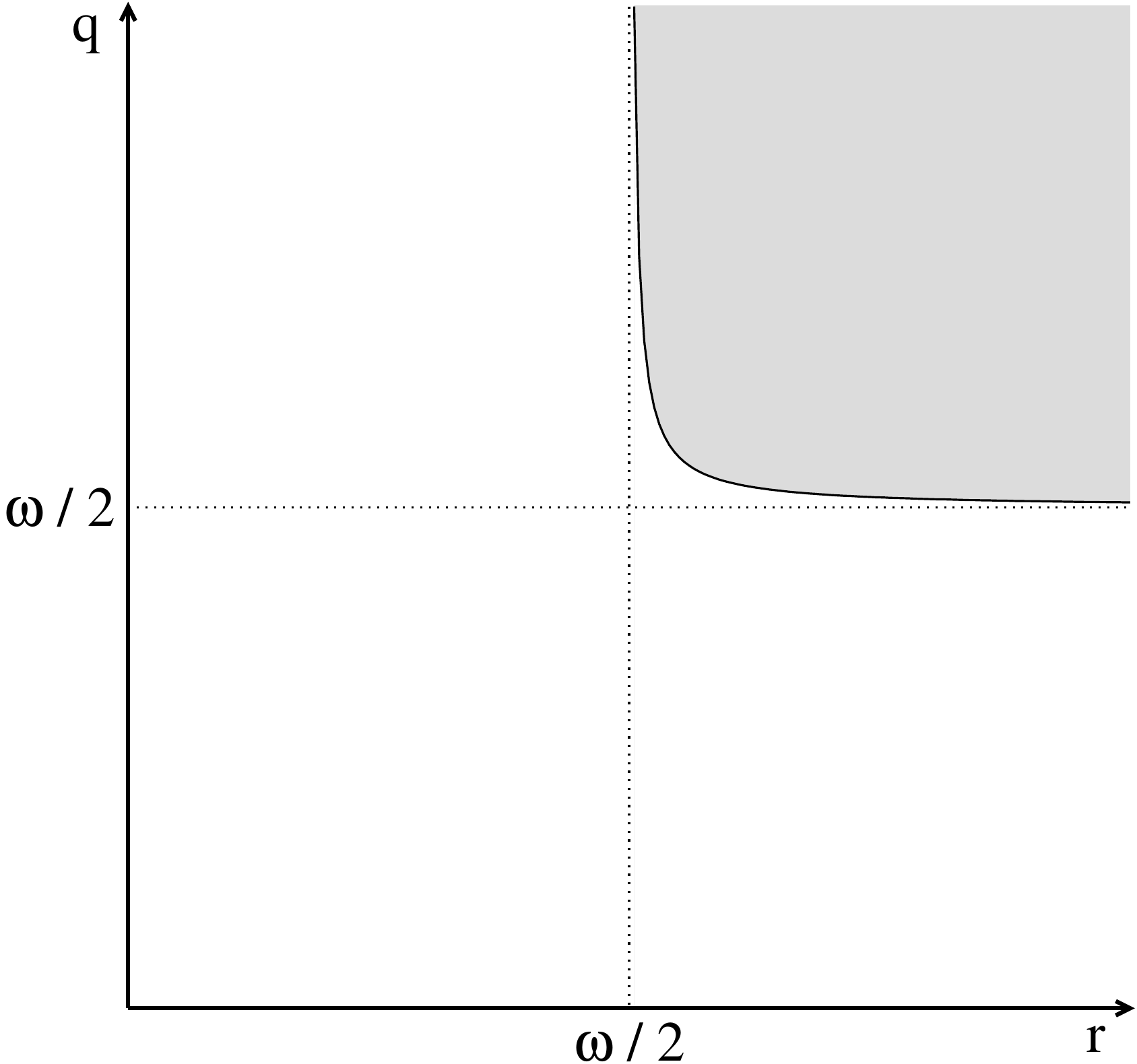}%
~~\includegraphics[width=4.5cm]{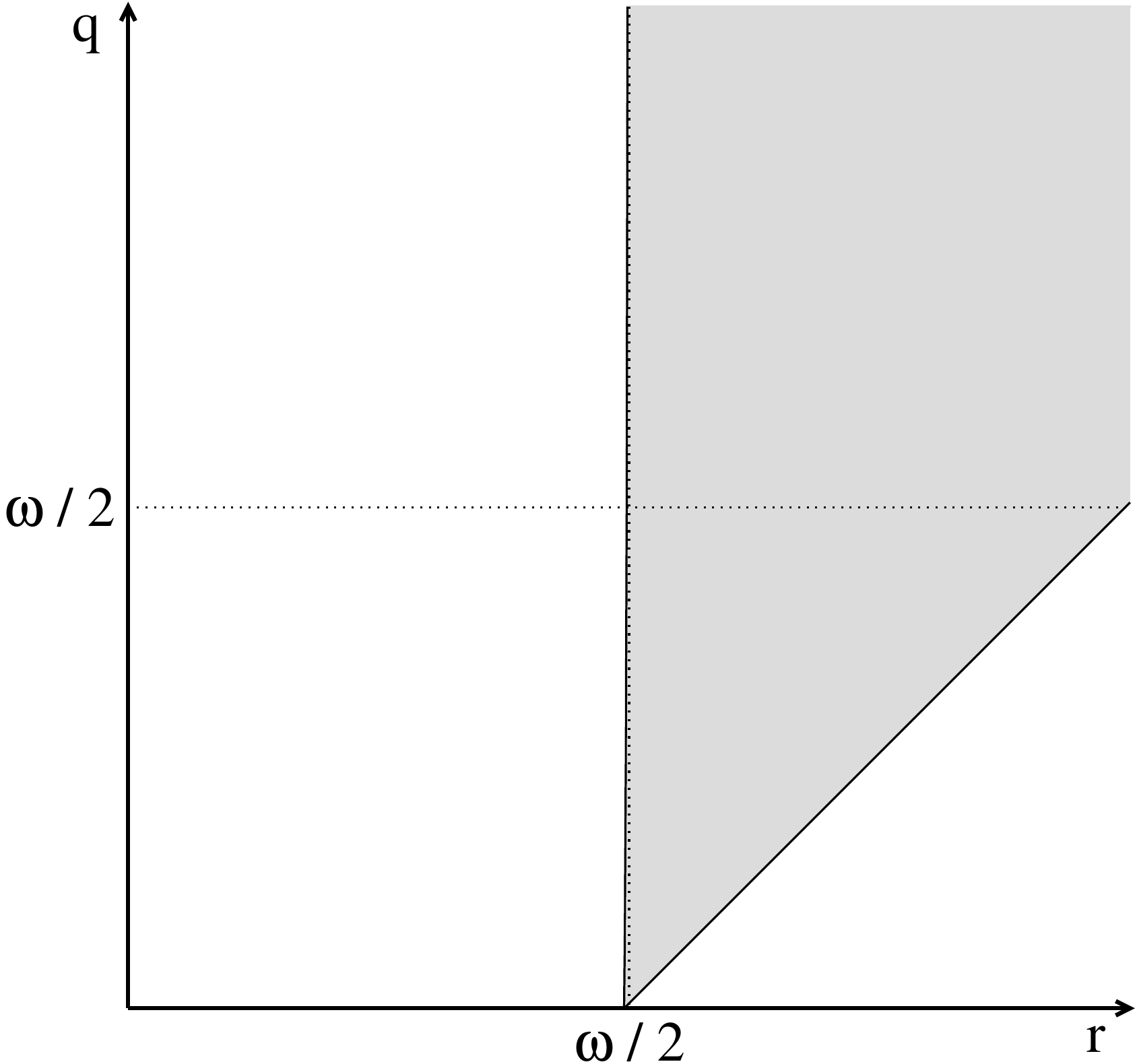}%
}

\hspace*{2.5cm} (\i) \hspace*{4.5cm} (\v) \hspace*{4.2cm} (\iv) 

\caption[a]{\small 
The original integration ranges for the phase space integrals, 
evaluated for $\lambda = \omega/10$ (\eqs\nr{range1}--\nr{range3}). The figure is taken from \cite{Laine:2011xm}.
} 
\la{fig:ranges_old}
\end{figure}
%

In the last four structures of \eq\nr{Ij_T}, we again rewrite the integration measure by substituting the angle between $\vec{q}$ and $\vec{r}$ for $E_{qr}$:
\be \la{ps_sub}
 \int_{\vec{q},\vec{r}} \frac{\pi}{4qrE_{qr}} 
 = \frac{2}{(4\pi)^3} \int_0^\infty \! {\rm d}q \int_0^\infty \! {\rm d}r
 \int_{E_{qr}^-}^{E_{qr}^+} \! {\rm d}E_{qr}
 \;, \quad
 E_{qr}^\pm \equiv \sqrt{(q\pm r)^2 + \lambda^2}
 \;.
\ee
It is also convenient to factor out a common $n_q n_r n_{qr}$ from the Bose distributions, yielding
\ba
 {(1+n_{qr})(1+n_q+n_r)+n_q n_r} & = &  
 n_q n_r n_{qr} \bigl( e^{q+r+E_{qr}} -1 \bigr)
 \;, \\
 {n_{qr}(1+n_q + n_r )-n_q n_r} & = & 
 n_q n_r n_{qr} \bigl( e^{q+r} - e^{E_{qr}} \bigr)
 \;, \\ 
 {n_r(1+n_q+n_{qr})-n_q n_{qr}} & = & 
 n_q n_r n_{qr} \bigl( e^{q+E_{qr}} - e^{r} \bigr)
 \;, \\ 
 {n_q(1+n_r+n_{qr})-n_r n_{qr}} & = & 
 n_q n_r n_{qr} \bigl( e^{r+E_{qr}} - e^{q} \bigr)
 \;. 
\ea
For the case of $0 < \lambda < \omega$, only four of the $\delta$-constraints get realized, and we can rewrite the integrals in the form 
\ba
 \rho_{\mathcal{I}^{0}_\rmii{j}}^{(\rmi{ps})}(\omega) & \equiv & 
 \frac{2\omega^4}{(4\pi)^3} 
 \int_0^\infty \! {\rm d}q \int_0^\infty \! {\rm d}r
 \int_{E_{qr}^-}^{E_{qr}^+} \! {\rm d}E_{qr}
 \, n_q n_r n_{qr} \, \biggl\{ 
 \nn
 & \mbox{(\i)} 
  & \;\;\; 
 \frac{\delta(\omega-q-r-E_{qr})}{(2r-\omega)(2q-\omega)}
 \Bigl(1 - e^{q+r+E_{qr}} \Bigr) 
 \nn 
 & \mbox{(\v)} 
  & + \, 
 \frac{\delta(\omega-q-r+E_{qr})}{(2r-\omega)(2q-\omega)}
 \Bigl(e^{E_{qr}} - e^{q+r}  \Bigr) 
 \nn 
 & \mbox{(\iv)} 
  & + \, 
 \frac{\delta(\omega+q-r-E_{qr})}{(2r-\omega)(2q+\omega)}
 \Bigl(e^{r+E_{qr}} - e^q \Bigr) 
 \nn 
 & \mbox{(\ii)} 
  & + \, 
 \frac{\delta(\omega-q+r-E_{qr})}{(2r+\omega)(2q-\omega)}
 \Bigl(e^{q+E_{qr}} - e^r \Bigr) 
 \biggr\} \;, \la{Ij_ps_1}
\ea
where the denominators can be simplified using the constraints given by the $\delta$-functions. It is also easy to see that the terms labeled by (\iv) and (\ii) are in fact equivalent and can be combined.

\begin{figure}[t]

\centerline{%
\includegraphics[width=4.5cm]{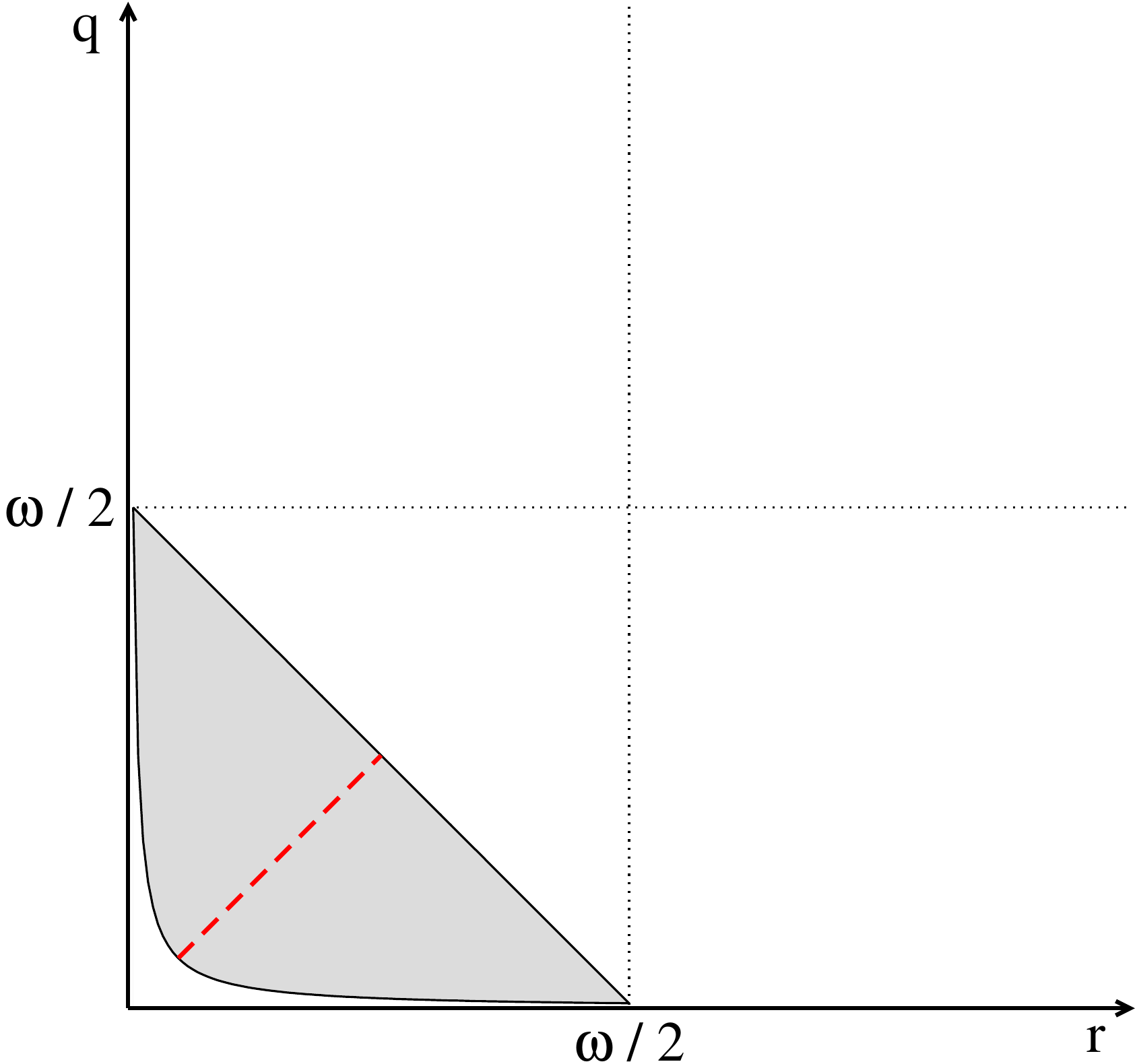}%
~~\includegraphics[width=4.5cm]{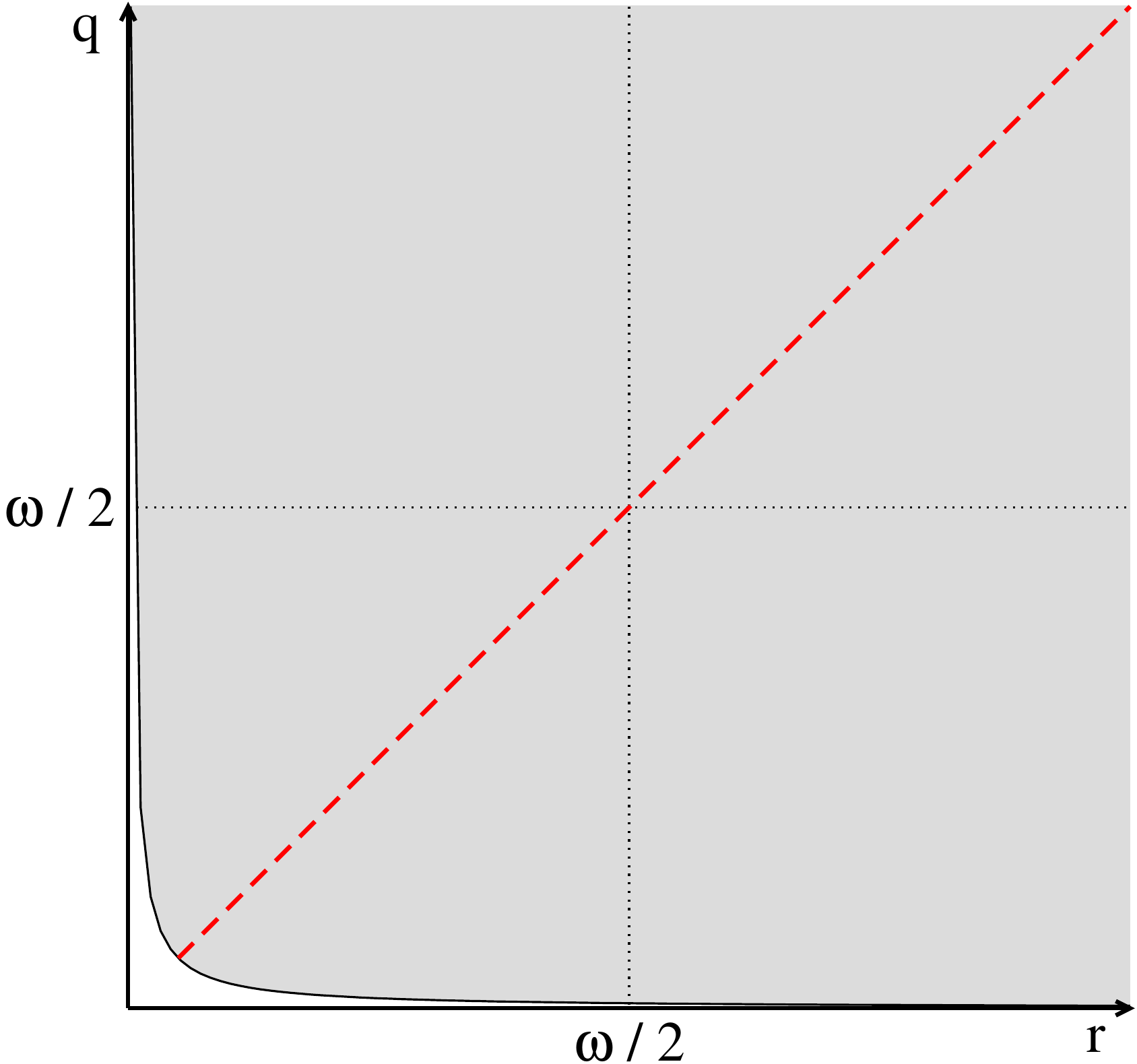}%
~~\includegraphics[width=4.5cm]{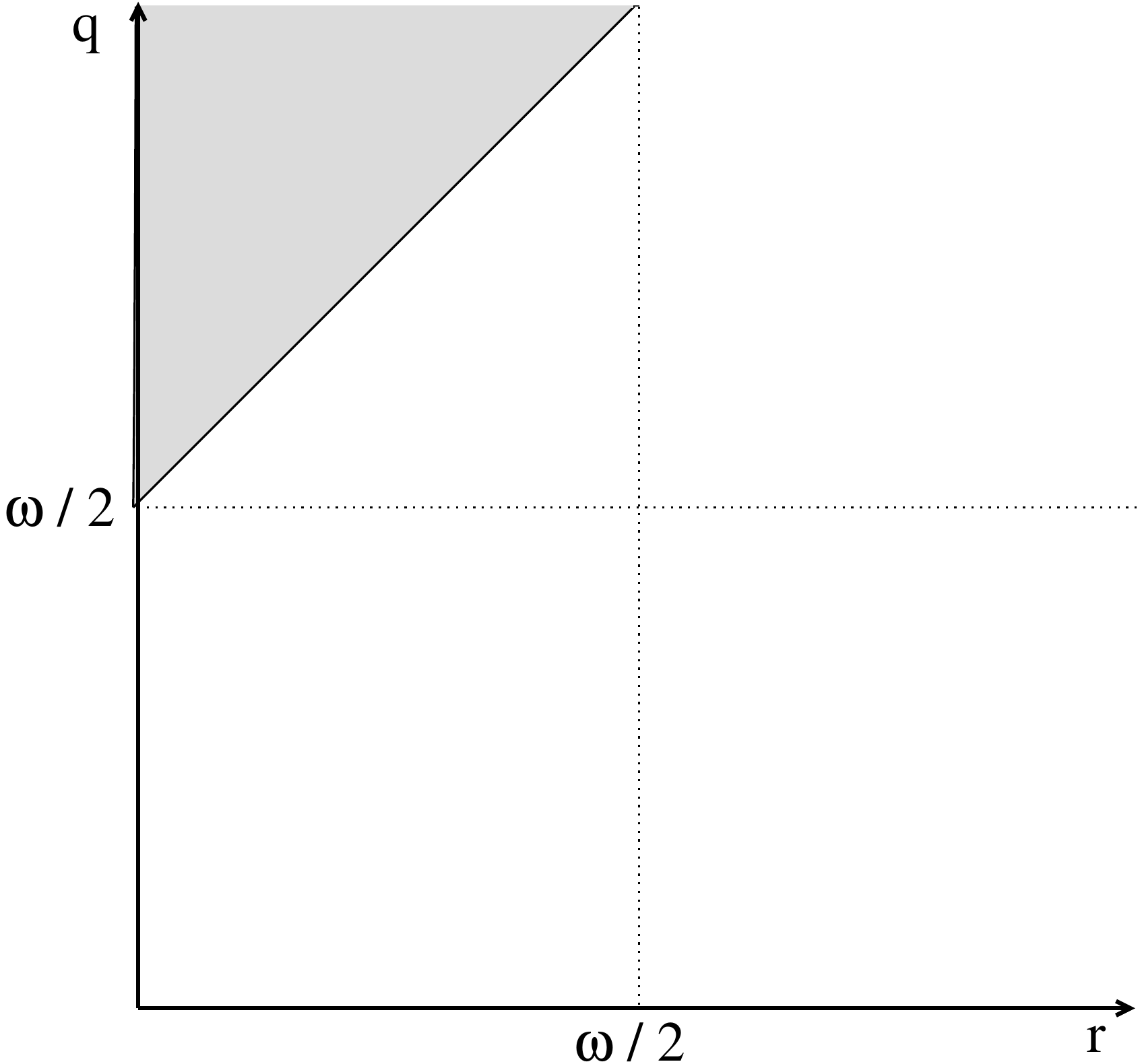}%
}
\hspace*{2.5cm} (\i) \hspace*{4.5cm} (\v) \hspace*{4.2cm} (\iv) 
\caption[a]{\small 
The same integration ranges after the shifts in \eqs\nr{nrange1}--\nr{nrange3} have been implemented, displayed again for $\lambda = \omega/10$. The dashed line indicates the diagonal $q=r$, which can be used to reflect all integrations to the octant $q \ge r$. Also this figure has been taken from \cite{Laine:2011xm}.
} 
\la{fig:ranges_new}
\end{figure}

Making next use of the restrictions coming from the $\d$-functions, we note that the following general relations hold as long as $\omega > \lambda$:
\ba
 \mbox{(\i)} &&
 \int_0^\infty \! {\rm d}q \int_0^\infty \! {\rm d}r
 \int_{E_{qr}^-}^{E_{qr}^+} \! {\rm d}E_{qr} \;
 \delta(\omega-q-r-E_{qr}) \, 
 \phi(q,r,E_{qr}) 
 \nn &  & \hspace*{3cm} = \,  
 \int_0^{\frac{\omega^2-\lambda^2}{2\omega}}
 \! {\rm d}q 
 \int_{ \frac{\omega(\omega-2q)-\lambda^2}{2\omega} }^
      { \frac{\omega(\omega-2q)-\lambda^2}{2(\omega-2q)} } 
 \! {\rm d}r \, 
 \phi(q,r,\omega-q-r)
 \;, \la{range1} \hspace*{5mm}\\ 
 \mbox{(\v)} &&
 \int_{0}^\infty \! {\rm d}q \int_0^\infty \! {\rm d}r
 \int_{E_{qr}^-}^{E_{qr}^+} \! {\rm d}E_{qr} \;
 \delta(\omega-q-r+E_{qr}) \, 
 \phi(q,r,E_{qr}) 
 \nn &  & \hspace*{3cm} = \,  
 \int_{\frac{\omega}{2}}^{\infty}
 \! {\rm d}q 
 \int_{ \frac{\omega(2q-\omega)+\lambda^2}{2(2q-\omega)} }^
      { \infty } 
 \! {\rm d}r \, 
 \phi(q,r,-\omega+q+r)
 \;, \\ 
 \mbox{(\iv)} &&
 \int_0^\infty \! {\rm d}q \int_0^\infty \! {\rm d}r
 \int_{E_{qr}^-}^{E_{qr}^+} \! {\rm d}E_{qr} \;
 \delta(\omega+q-r-E_{qr}) \, 
 \phi(q,r,E_{qr}) 
 \nn &  & \hspace*{3cm} = \,  
 \int_0^{\infty}
 \! {\rm d}q 
 \int_{ \frac{\omega(\omega+2q)-\lambda^2}{2(\omega+2q)} }^
      { \frac{\omega(\omega+2q)-\lambda^2}{2\omega} } 
 \! {\rm d}r \, 
 \phi(q,r,\omega+q-r)
 \;, \la{range3}
\ea
with the integration ranges illustrated in \fig\ref{fig:ranges_old}. 
Implementing here the changes of variables 
\ba
 \mbox{(\i)}: && 
   q \to \frac{\omega}{2} - q \;, \quad r \to \frac{\omega}{2} - r
 \;, \la{shift1} \\ 
 \mbox{(\v)}: && 
   q \to \frac{\omega}{2} + q \;, \quad r \to \frac{\omega}{2} + r
 \;, \\ 
 \mbox{(\iv)}: &&
   q \to -\frac{\omega}{2} +  q \;, \quad r \to \frac{\omega}{2} + r
 \;, \la{shift3}
\ea
the integration ranges appearing in \eqs\nr{range1}--\nr{range3} can be further converted into
\ba
 \mbox{(\i)} &&
 \int_0^{\frac{\omega^2-\lambda^2}{2\omega}}
 \! {\rm d}q 
 \int_{ \frac{\omega(\omega-2q)-\lambda^2}{2\omega} }^
      { \frac{\omega(\omega-2q)-\lambda^2}{2(\omega-2q)} } 
 \! {\rm d}r \, 
 \phi(q,r,\omega-q-r)
 \nn && \hspace*{3cm} = \, 
 \int_{\frac{\lambda^2}{2\omega}}^{\frac{\omega}{2}}
 \! {\rm d}q 
 \int_{ \frac{\lambda^2}{4q} }^
      { \frac{\omega(\omega-2q)+\lambda^2}{2\omega} } 
 \! {\rm d}r \, 
 \phi\Bigl(\frac{\omega}{2}-q,\frac{\omega}{2}-r,q+r\Bigr)
 \;, \la{nrange1} \\ 
 \mbox{(\v)} &&
 \int_{\frac{\omega}{2}}^{\infty}
 \! {\rm d}q 
 \int_{ \frac{\omega(2q-\omega)+\lambda^2}{2(2q-\omega)} }^
      { \infty } 
 \! {\rm d}r \, 
 \phi(q,r,-\omega+q+r)
 \nn &  & \hspace*{3cm} = \,  
 \int_{0}^{\infty}
 \! {\rm d}q 
 \int_{ \frac{\lambda^2}{4q} }^
      { \infty } 
 \! {\rm d}r \, 
 \phi\Bigl(\frac{\omega}{2} + q,\frac{\omega}{2} + r,q+r\Bigr)
 \;, \\ 
 \mbox{(\iv)} &&
 \int_0^{\infty}
 \! {\rm d}q 
 \int_{ \frac{\omega(\omega+2q)-\lambda^2}{2(\omega+2q)} }^
      { \frac{\omega(\omega+2q)-\lambda^2}{2\omega} } 
 \! {\rm d}r \, 
 \phi(q,r,\omega+q-r)
 \nn &  & \hspace*{3cm} = \,  
 \int_{\frac{\omega}{2}}^{\infty}
 \! {\rm d}q 
 \int_{ - \frac{\lambda^2}{4q} }^
      { \frac{\omega(2q-\omega)-\lambda^2}{2\omega} } 
 \! {\rm d}r \, 
 \phi\Bigl(q-\frac{\omega}{2},\frac{\omega}{2}+r,q-r\Bigr)
 \;, \la{nrange3} \hspace*{8mm}
\ea
illustrated in \fig\ref{fig:ranges_new}. This allows us to simplify 
\eq\nr{Ij_ps_1} to the form 
\ba
 \rho_{\mathcal{I}^{0}_\rmii{j}}^{(\rmi{ps})}(\omega) & = & 
 \frac{\omega^4}{2(4\pi)^3} (e^\omega - 1)  \biggl\{ 
 \nn 
 & \mbox{(\i)}  &  - \,
 \int_{\frac{\lambda^2}{2\omega}}^{\frac{\omega}{2}}
 \! {\rm d}q
 \int_{\frac{\lambda^2}{4q}}^{\frac{\omega(\omega-2q)+\lambda^2}{2\omega}} 
 \!\!\!\! {\rm d}r \; \mathbbm{P}
 \biggl( \frac{ 
 1 } {qr} \biggr)\;
 n_{\fr{\omega}2-q} \, 
 n_{\fr{\omega}2-r} \, 
 n_{q+r}
 \nn
 & \mbox{(\v)}  & - \, 
 \int_{ 0 }^{\infty}
 \! {\rm d}q
 \int_{\frac{\lambda^2}{4q}}^{ \infty }
 \! {\rm d}r \; \mathbbm{P}
 \biggl( \frac{ 1
 } {qr} \biggr)\;
 n_{\fr{\omega}2+q} \, 
 n_{\fr{\omega}2+r} \, 
 n_{q+r}
 e^{q+r} 
 \nn
 & \mbox{(\iv)}  & + \, 
 \int_{ \frac{\omega}{2} }^{\infty}
 \! {\rm d}q
 \int_{ - \frac{\lambda^2}{4q}}^{
    \frac{\omega(2q-\omega)-\lambda^2}{2\omega}}
 \!\!\!\! {\rm d}r \; \mathbbm{P}
 \biggl( \frac{ 2
  } {qr} \biggr)\;
 n_{q-\fr{\omega}2} \, 
 n_{\fr{\omega}2+r} \, 
 n_{q-r} \,
 e^{q - \frac{\omega}{2}}
 \biggr\} \;, \la{Ij_ps_2} \;\;\;\;\;
\ea
where in the cases (\i) and (\v), we can finally reflect the ranges $q < r$ to $q > r$, as indicated in \fig\ref{fig:ranges_new}. After these manipulations, all integrations only start at a nonzero value of $q$, and explicit IR divergences are averted.

So far we have made no approximations concerning the value of $\lambda$  in the phase space integrals (apart from assuming that $\lambda<\omega$). 
For practical reasons, it is, however, useful to rearrange the result somewhat at this point, separating from it a simple part that contains all the terms divergent in the $\lambda\to 0$ limit and allowing to set the parameter to zero in the rest. The divergences that need to be subtracted have two separate origins: they either come from a prefactor $1/r$ or from an IR divergent Bose distribution function $n_{q+r}$ (or $n_{q-\frac{\omega}{2}}$ in the case (\iv)). To handle the latter, we subtract from the integrand a term of the form $\alpha\, n_{q+r}/(qr)$, where $\alpha$ is the ``residue'' of this structure at the origin; the subtracted term can then be integrated analytically by changing variables from $(q,r)$ to $(x,y)\equiv (q+r,q-r)$ and carrying out the subsequent integration over $y$. To handle the remaining $1/r$-divergences, we again subtract a term $\beta(q)/(qr)$, where $\beta(q)$ is an $r=0$ ``residue''. After these subtractions, the remaining integral is observed to remain finite upon taking the limit $\lambda\to 0$.

Following the recipe described above, we now proceed to separate the divergent parts of the integrals using the identities
\ba
 && \hspace{-2cm}n_{\fr{\omega}2-q} \, 
 n_{\fr{\omega}2-r} \, 
 n_{q+r} (1-e^{\omega})
 \nn &=& 
 - \bigl(1 + 2 n_{\frac{\omega}{2}} \bigr)
 \biggl[ 1 + n_{q+r} +  n_{\fr{\omega}2-q}
  +(1+n_{\fr{\omega}2-r})  
  \frac{n_{q+r} n_{\fr{\omega}2-q}}{n_r^2}
 \biggr]
 \;, \la{ps_simp_3}
 \\
  &&\hspace{-2cm} n_{\fr{\omega}2+q} \, 
 n_{\fr{\omega}2+r} \, 
 n_{q+r} \, 
 e^{q+r}
 (1  - e^{\omega}) 
 \nn &=& 
 (1 + 2 n_{\frac{\omega}{2}} )
 \biggl[
   -n_{q+r} + n_{q+\frac{\omega}{2}} - 
    (1 + n_{q+\frac{\omega}{2}}) 
    \frac{n_{q+r}n_{r+\frac{\omega}{2}}}{n_r^2}
 \biggr] 
 \;, \la{ps_simp_1}
 \\
  && \hspace*{-2cm}
 n_{q-\fr{\omega}2} \, 
 n_{\fr{\omega}2+r} \, 
 n_{q-r} \, 
 e^{q-\frac{\omega}{2}} (e^{{\omega}} - 1 )
 \nn &=& 
 ( 1 + 2 n_{\frac{\omega}{2}} )
 \biggl[
    n_{q-\frac{\omega}{2}} - n_{q} 
   - n_{q-\frac{\omega}{2}} 
    \frac{(1+n_{q-r})(n_q - n_{r+\frac{\omega}{2}})}
    {n_r n_{-\frac{\omega}{2}}}
 \biggr]
 \;  \la{ps_simp_2}
\ea
in the cases $(\i)$, $(\v)$, and $(\iv)$, respectively. In each of these cases, the IR divergent parts turn out to be contained in the terms consisting of only one distribution function (or just unity) in the square brackets, which are quite straightforward to handle.

\begin{figure}[t]
\centerline{%
~~\includegraphics[width=6.5cm]{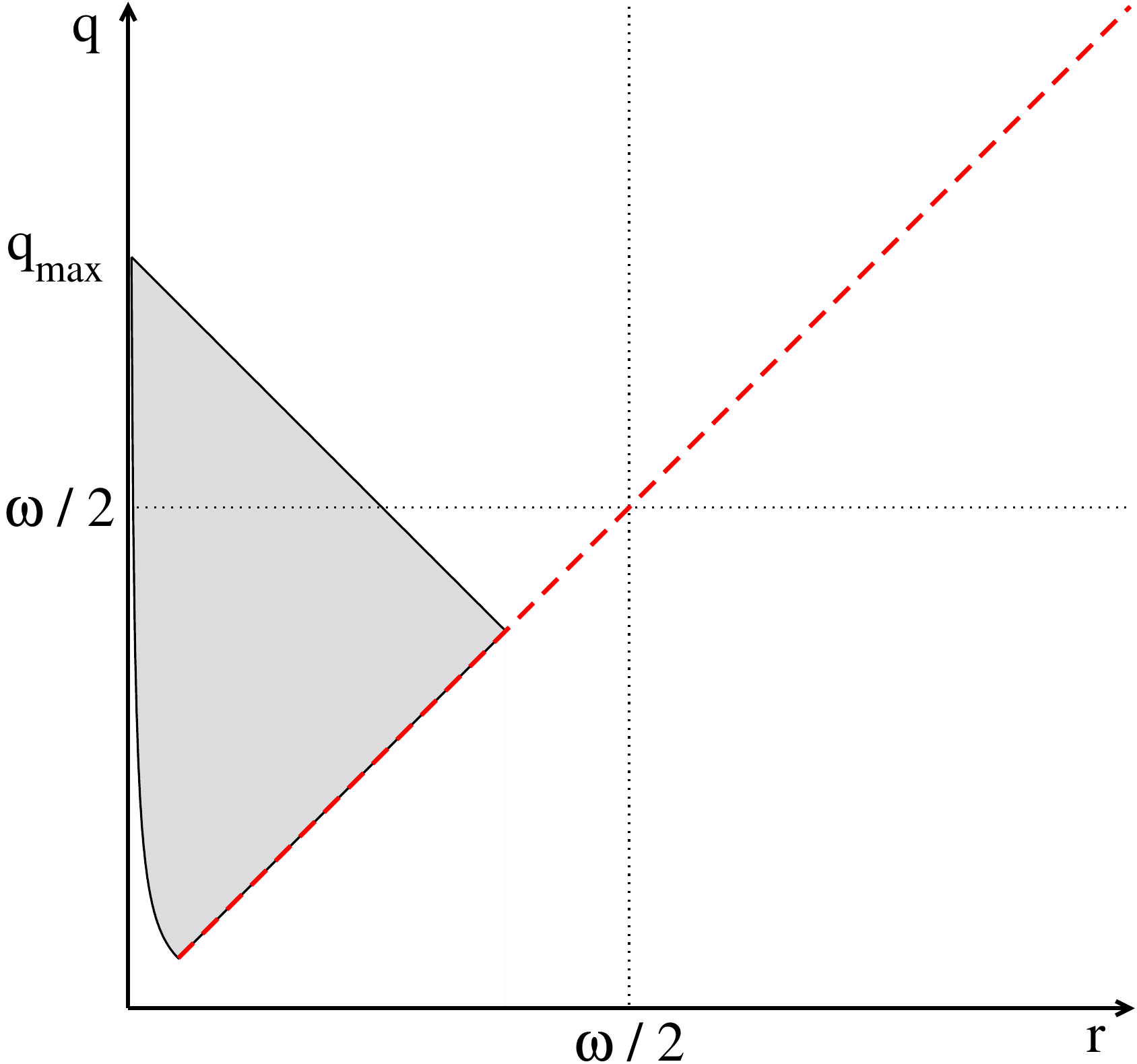}%
}
\caption[a]{\small 
An integration range for which a substitution of the type 
in \eq\nr{xy} can be carried out. The figure is taken from \cite{Laine:2011xm}.
} 
\la{fig:ranges_xy}
\end{figure}

The next step is clearly to carry out the divergent integrals. The terms coming with a factor $n_{q+r}$ are dealt with using the variables $(x,y)=(q+r,q-r)$, producing
\ba
 \mathcal{I} & \equiv & 
 \int_{\frac{\lambda}{2}}^{q_\rmii{max}} 
 \! \frac{{\rm d}q}{q}
 \int_{\frac{\lambda^2}{4q}}^{r_\rmii{max}(q)}
 \! \frac{{\rm d}r}{r}
 \, \phi(q+r,q-r)
 \\ 
 & = &  
 \int_{\lambda}^{x_\rmii{max}}
 \! \frac{{\rm d}x}{x}
 \int_{0}^{\sqrt{x^2-\lambda^2}} \!\!\!
 {\rm d}y \, 
 \biggl( \frac{1}{x+y} + \frac{1}{x-y} \biggr) 
 \phi(x,y)
 \left. 
 \right|_{x_\rmii{max} = q_\rmii{max}+ r_\rmii{max}(q_\rmii{max})}
 \;, \la{xy}\nonumber
\ea
as illustrated in \fig\ref{fig:ranges_xy}. In the case under consideration, the function $\phi$ is independent of the variable $y$, \ie~$\phi(x,y)\equiv\phi(x)$, implying that the integral over $y$ can be carried out trivially. This leads to the result 
\be
 \mathcal{I} = \int_{\lambda}^{x_\rmii{max}} 
 \! \frac{{\rm d}x}{x}
 \ln \left| \frac{x+\sqrt{x^2-\lambda^2}}{x-\sqrt{x^2-\lambda^2}} \right|
 \phi(x)
 \;, \la{xy_trick}
\ee
which finds direct use in cases ($\i$) and ($\v$). Dealing with the $1/r$ terms using the relations
\ba
 && 
 \int_{ \frac{\lambda^2}{4q} }^{ q} 
 \! {\rm d}r \; \mathbbm{P} \Bigl( \frac{1}{r} \Bigr)
 = 
 \ln\biggl| 
       \frac{4q^2}{\lambda^2}
    \biggr|
 \;, \\ 
 &&
 \int_{ \frac{\lambda^2}{4q}}^{
 \frac{\omega(\omega-2q)+\lambda^2}{2\omega}
 }
 \! {\rm d}r \; \mathbbm{P} \Bigl( \frac{1}{r} \Bigr)
 = \ln\biggl| 
      \frac{2q[\omega(\omega-2q)+\lambda^2]}{\omega\lambda^2}
    \biggr| 
 \;.
\ea
and taking $\lambda\to 0$ whenever possible, the three phase space integrals of \eq (\ref{Ij_ps_2}) then become
\ba
 \rho_{\mathcal{I}^{0}_\rmii{j},\rmi{(\i)}}^{(\rmi{ps})}(\omega)  &\approx & 
 \frac{\omega^4}{(4\pi)^3}  
  \Bigl(1+ 2 n_{\frac{\omega}{2}} \Bigr)
 \biggl\{ 
 \int_{ {\lambda} }^{ \frac{\omega}{2} }
 \! \frac{{\rm d}q}{q}
 (1+n_q) \ln \biggl| 
      \frac{q - \sqrt{q^2-\lambda^2}}{q + \sqrt{q^2-\lambda^2}}
    \biggr|
 \nn & +& 
 \int_{ \frac{\lambda}{2} }^{ \frac{\omega}{4} }
 \! \frac{{\rm d}q}{q} n_{\frac{\omega}{2} - q}
   \ln\biggl| 
       \frac{\lambda^2}{4q^2}
    \biggr|
 + 
 \int_{ \frac{\omega}{4} }^{ \frac{\omega}{2} }
 \! \frac{{\rm d}q}{q}
   n_{\frac{\omega}{2} - q} 
  \ln\biggl| 
      \frac{\omega\lambda^2}{2q[\omega(\omega-2q)+\lambda^2]}
    \biggr| 
 \nn &- &   
 \int_0^{ \frac{\omega}{2} } \! \frac{ {\rm d}q }{q} 
  n_{\frac{\omega}{2} - q} 
 \int_0^{ \frac{\omega}{4} - |q-\frac{\omega}{4}| } \! \frac{ {\rm d}r }{r}  
  \frac{n_{q+r} (1+n_{\fr{\omega}2-r})}{n_r^2}
 \biggr\} \;, \hspace*{5mm} \la{Ij_ps_i}
 \\ 
 \rho_{\mathcal{I}^{0}_\rmii{j},\rmi{(\v)}}^{(\rmi{ps})}(\omega) & \approx & 
 \frac{\omega^4}{(4\pi)^3}  (1 + 2 n_{\frac{\omega}{2}} )
 \biggl\{ 
 \int_{ {\lambda} }^{\infty}
 \! \frac{{\rm d}q}{q} \, n_q \, 
 \ln\biggl| 
      \frac{q - \sqrt{q^2-\lambda^2}}{q + \sqrt{q^2-\lambda^2}}
    \biggr|
 \\ &+ & 
 \int_{ \frac{\lambda}{2} }^{\infty}
 \! \frac{{\rm d}q}{q} 
  n_{q+\frac{\omega}{2}}
 \ln\biggl| 
      \frac{4q^2}{\lambda^2}
    \biggr|
 - \int_0^\infty \! \frac{{\rm d}q}{q}  
 \bigl( 1 + n_{q+\frac{\omega}{2}} \bigr)
 \int_{0}^{ q }
 \! \frac{ {\rm d}r }{r} 
 \frac{n_{q+r}n_{r+\frac{\omega}{2}}}{n_r^2}
 \biggr\} \;, \hspace*{5mm} \la{Ij_ps_v}\nn 
 \rho_{\mathcal{I}^{0}_\rmii{j},\rmi{(\iv)}}^{(\rmi{ps})}(\omega) & \approx & 
 \frac{\omega^4}{(4\pi)^3}  \Bigl(1 + 2 n_{\frac{\omega}{2}} \Bigr)
 \biggl\{ 
 \int_{ \frac{\omega}{2} }^{\infty}
 \! \frac{{\rm d}q}{q} 
 \Bigl(n_{q-\frac{\omega}{2}} - n_q \Bigr)
 \ln\biggl| 
      \frac{2q[\omega(\omega - 2q)+\lambda^2]}{\omega\lambda^2}
    \biggr|
 \nn &- & 
 \int_{ \frac{\omega}{2} }^{\infty}
 \! \frac{{\rm d}q}{q}
  n_{q-\frac{\omega}{2}} 
 \int_{0}^{ q - \frac{\omega}{2} }
 \! \frac{ {\rm d}r }{r} 
    \frac{(1+n_{q-r})(n_q - n_{r+\frac{\omega}{2}})}
    {n_r n_{-\frac{\omega}{2}}}
 \biggr\} \;. \la{Ij_ps_iv}
\ea
Here, we have renamed $x$ as $q$ after having applied the relations \nr{xy} and \nr{xy_trick}, and used ``$\approx$'' as a reminder of having set $\lambda\to 0$ wherever possible.

\subsection{Final result}
\la{ss:collect}
Adding up the (fz) and (ps) parts from above, we see that the result for the entire master sum-integral can be divided to 0-, 1-, and 2-dimensional integrals. Performing some minor reshufflings of the terms, and changing variables so that in the 1d integrals all Bose distribution functions have $q$ as their argument, all of the $\lambda\to 0$ divergences can be seen to cancel. The spectral function thereby boils down to the relatively simple expression
\ba
 && \hspace*{-2cm} 
 \frac{(4\pi)^3\rho^{ }_{\mathcal{I}^{0}_\rmii{j}}(\omega)}
 {\omega^4 (1+2 n_{\frac{\omega}{2}})}  = 
 \nn 
  & & 
 \int_0^{ \frac{\omega}{4} } \! { {\rm d}q } \, n_{q} \; 
   \biggl[  
     \Bigl( \frac{1}{q-\frac{\omega}{2}} - \frac{1}{q}  \Bigr)
     \ln\Bigl( 1 - \frac{2q}{\omega} \Bigr)
   - \frac{\frac{\omega}{2}}{q(q+\frac{\omega}{2})}
     \ln\Bigl( 1 + \frac{2q}{\omega} \Bigr)
   \biggr]  
 \nn 
 & + & 
 \int_{ \frac{\omega}{4} }^{ \frac{\omega}{2} } 
 \! { {\rm d}q } \, n_{q} \; 
   \biggl[  
     \Bigl( \frac{2}{q-\frac{\omega}{2}} - \frac{1}{q}  \Bigr)
     \ln\Bigl( 1 - \frac{2q}{\omega} \Bigr)
   - \frac{\frac{\omega}{2}}{q(q+\frac{\omega}{2})}
     \ln\Bigl( 1 + \frac{2q}{\omega} \Bigr)
  \nn &&- 
     \frac{1}{q-\frac{\omega}{2}} 
     \ln\Bigl(\frac{2q}{\omega} \Bigr)
   \biggr]  
%
 +
 \int_{ \frac{\omega}{2} }^{ \infty } 
 \! { {\rm d}q } \, n_{q} \; 
   \biggl[  
     \Bigl( \frac{2}{q-\frac{\omega}{2}} - \frac{2}{q}  \Bigr)
     \ln\Bigl( \frac{2q}{\omega} -1 \Bigr)
 \nn &&
   - \frac{\frac{\omega}{2}}{q(q+\frac{\omega}{2})}
     \ln\Bigl( 1 + \frac{2q}{\omega} \Bigr)
     - \frac{\frac{\omega}{2}}{q(q-\frac{\omega}{2})}
     \ln\Bigl(\frac{2q}{\omega} \Bigr)
   \biggr]  
%
 \nn & + & 
 \int_0^{ \frac{\omega}{2} } \! { {\rm d}q }  
 \int_0^{ \frac{\omega}{4} - |q-\frac{\omega}{4}| } 
 \! { {\rm d}r }  \;
 \biggl( - \frac{1}{qr }\biggr)\;
  \frac{ n_{\frac{\omega}{2} - q}\, n_{q+r} (1+n_{\fr{\omega}2-r})}{n_r^2}
 \nn & + & 
 \int_{ \frac{\omega}{2} }^{\infty}
 \! {{\rm d}q}
 \int_{0}^{ q - \frac{\omega}{2} }
 \! { {\rm d}r } \;
 \biggl( - \frac{1}{qr }\biggr) \;
    \frac{  n_{q-\frac{\omega}{2}} 
   (1+n_{q-r})(n_q - n_{r+\frac{\omega}{2}})}
    {n_r n_{-\frac{\omega}{2}}}
 \nn & + & 
 \int_0^\infty \! {{\rm d}q}  
 \int_{0}^{ q }
 \! { {\rm d}r } \;
 \biggl( - \frac{1}{qr }\biggr) \;
 \frac{ ( 1 + n_{q+\frac{\omega}{2}} )
 n_{q+r}n_{r+\frac{\omega}{2}}}{n_r^2}
 \; + \; \rmO(\lambda\ln\lambda)
 \;, \hspace*{5mm} \la{Ij_final} 
\ea
from which the 0d parts have furthermore altogether vanished. 

Results for all other bulk channel master integrals follow in a highly analogous fashion, and can be found from \cite{Laine:2011xm}.

\mysection{Shear channel}
In the shear channel, we are faced with the challenge of evaluating the master integrals appearing in \eq\nr{Gmasters}. While the techniques we use here are to a large extent rather straightforward generalizations of the ones already introduced in the above bulk channel calculation, there are in addition some new types of problems that need to be addressed. These are in most cases related to the appearance of the functions $P_T(Q)\equiv Q_\mu Q_\nu P^T_{\mu\nu}(P) = \mathbf{q}^2-(\mathbf{q}\cdot\hat{\mathbf{p}})^2$ as well as squared propagators in the integrands, as is evident from the list of master integrals given in appendix \ref{app:master}. 

A major simplification originates from the fact that we have chosen to work in the limit of vanishing external three-momenta. This implies that we can take advantage of rotational invariance in $D-1$ dimensions and use the results
\ba
\left<\mathbf{\hat p}_i\mathbf{\hat p}_j\right>_\mathbf{\hat p}&=&\frac{1}{D-1}\delta_{ij}\, , \\
\left<\mathbf{\hat p}_i\mathbf{\hat p}_j\mathbf{\hat p}_k\mathbf{\hat p}_l\right>_\mathbf{\hat p}&=&\frac{1}{D^2-1}\left(\delta_{ij}\delta_{kl} +\delta_{ik}\delta_{jl} +\delta_{il}\delta_{jk}\right)\, 
\ea
to make the following substitutions in the integrands of all master integrals:
\ba
P_T(Q)\rightarrow\frac{D-2}{D-1}q^2\,,\hspace{2em} P^2_T(Q)\rightarrow\frac{D(D-2)}{D^2-1}q^4\,,\nn
P_T(Q)P_T(R)\rightarrow\frac{D^2-2 D-2}{D^2-1}q^2r^2+\frac{2}{D^2-1}(\mathbf{q}\cdot\mathbf{r})^2\, , \label{PTrels}
\ea
This ``scalarization'' turns out to be extremely useful in our considered computations.

Next, we introduce the notation 
\ba
 \rho^{ }_{\mathcal{J}^{n}_\rmii{x}}(\omega) \equiv 
 \int_{\vec{q}}f_{\mathcal{J}^{n}_\rmii{x}}\;, 
 \hspace{3em}
 \rho^{ }_{\mathcal{I}^{n}_\rmii{x}}(\omega) \equiv 
 \int_{\vec{q,r}}f_{\mathcal{I}^{n}_\rmii{x}}\; , \label{fdef}
\ea
listing the integrands $f_{\mathcal{J}^{0}_\rmii{x}}$ and $f_{\mathcal{I}^{0}_\rmii{x}}$, corresponding to the bulk channel integrals, in appendix \ref{fs}. Making use of the replacements of \eq (\ref{PTrels}), we then immediately obtain the results
\ba
 f_{\mathcal{J}^{1}_\rmii{b}} =
 -\frac{D-2}{D-1}\frac{q^2}{\omega^2}f_{\mathcal{J}^{0}_\rmii{b}}\;,
 &\hspace{2em}&
 f_{\mathcal{J}^{2}_\rmii{b}} =
 \frac{D(D-2)}{D^2-1}\frac{q^4}{\omega^4}f_{\mathcal{J}^{0}_\rmii{b}}\;, \\
 f_{\mathcal{I}^{1}_\rmii{b}} =
 -\frac{D-2}{D-1}\frac{q^2}{\omega^2}f_{\mathcal{I}^{0}_\rmii{b}}\;,
 &\hspace{2em}&
 f_{\mathcal{I}^{2}_\rmii{b}} =
 -\frac{D-2}{D-1}\frac{r^2}{\omega^2}f_{\mathcal{I}^{0}_\rmii{b}}\;,\\
 f_{\mathcal{I}^{1}_\rmii{d}} =
 -\frac{D-2}{D-1}\frac{q^2}{\omega^2}f_{\mathcal{I}^{0}_\rmii{d}}\;,
 &\hspace{2em}&
 f_{\mathcal{I}^{2}_\rmii{d}} =
 -\frac{D-2}{D-1}\frac{r^2}{\omega^2}f_{\mathcal{I}^{0}_\rmii{d}}\;,\\
 f_{\mathcal{I}^{3}_\rmii{d}} =
 \frac{D(D-2)}{D^2-1}\frac{r^4}{\omega^4}f_{\mathcal{I}^{0}_\rmii{d}}\;,
 &\hspace{2em}&
 f_{\mathcal{I}^{1}_\rmii{f}} =
 -\frac{D-2}{D-1}\frac{q^2}{\omega^2}f_{\mathcal{I}^{0}_\rmii{f}}\;,\\
 f_{\mathcal{I}^{1}_\rmii{h}} =
 -\frac{D-2}{D-1}\frac{q^2}{\omega^2}f_{\mathcal{I}^{0}_\rmii{h}}\;,
 &\hspace{2em}&
 f_{\mathcal{I}^{2}_\rmii{h}} =
 -\frac{D-2}{D-1}\frac{r^2}{\omega^2}f_{\mathcal{I}^{0}_\rmii{h}}\;,\\
 f_{\mathcal{I}^{1}_\rmii{j}} =
 -\frac{D-2}{D-1}\frac{q^2}{\omega^2}f_{\mathcal{I}^{0}_\rmii{j}}\;,
 &\hspace{2em}&
 f_{\mathcal{I}^{2}_\rmii{j}} =
 -\frac{D-2}{D-1}\frac{E_{qr}^2-\lambda^2}{\omega^2}f_{\mathcal{I}^{0}_\rmii{j}}\;,\\
 f_{\mathcal{I}^{3}_\rmii{j}} =
 \frac{D(D-2)}{D^2-1}\frac{q^4}{\omega^4}f_{\mathcal{I}^{0}_\rmii{j}}\;,
 &\hspace{2em}&
 f_{\mathcal{I}^{4}_\rmii{j}} =
 \frac{D(D-2)}{D^2-1}\frac{(E_{qr}^2-\lambda^2)^2}{\omega^4}f_{\mathcal{I}^{0}_\rmii{j}}\;,\\
 &&\hspace{-17em}f_{\mathcal{I}^{5}_\rmii{j}} =
 \frac{1}{\omega^4}\left(\frac{D^2-2 D-2}{D^2-1}q^2r^2+\frac{2}{D^2-1}(\mathbf{q}\cdot\mathbf{r})^2\right)f_{\mathcal{I}^{0}_\rmii{j}}\;,\\
 &&\hspace{-17em}f_{\mathcal{I}^{6}_\rmii{j}} =
 \frac{1}{\omega^4}\left(\frac{D^2-2 D-2}{D^2-1}q^2(E_{qr}^2-\lambda^2)+\frac{2}{D^2-1}(\mathbf{q}\cdot(\mathbf{q-r}))^2\right)f_{\mathcal{I}^{0}_\rmii{j}}\, ,
\ea
where $\lambda^2$ is the IR regulator introduced above. 

The remaining master integrals --- $\mathcal{I}^{3}_\rmii{h},\, \mathcal{I}^{4}_\rmii{h},\, \mathcal{I}^{5}_\rmii{h},\, \mathcal{I}^{6}_\rmii{h},\, \mathcal{I}^{7}_\rmii{h},\, \mathcal{I}^{1}_\rmii{i},\, \mathcal{I}^{2}_\rmii{i}$ and $\mathcal{I}^{3}_\rmii{i}$ --- all contain additional squared propagators, which is the reason we have not been able to trivially relate them to the bulk cases. In dealing with them, we have found it useful to employ the identity
\ba  
  \frac{1}{R^4} &=& - \lim_{m\to 0} \Bigg\{ \frac{{\rm d}}{{\rm d}m^2} 
  \frac{1}{R^2+m^2}\Biggr\} \, , \la{rmass}
\ea
which however necessitates introducing a fictitious mass parameter $m$ in the propagators to be squared. In addition to introducing some additional complications in the calculations, this procedure also makes the IR regulator $\lambda$ redundant, so we set it to zero at this point. This leads to the identities
\ba
 f_{\mathcal{I}^{3}_\rmii{h}} &=&
 -\frac{D-2}{D-1}
 \lim_{m\to 0} 
 \Bigg\{
 \frac{{\rm d}}{{\rm d}m^2}r^2
 f_{\mathcal{I}^{0}_\rmii{h}}
 \Biggr\} \;,
 \\
 f_{\mathcal{I}^{4}_\rmii{h}} &=&
 \frac{D(D-2)}{D^2-1}
 \lim_{m\to 0} 
 \Bigg\{
  \frac{{\rm d}}{{\rm d}m^2}\frac{q^4}{\omega^2}
 f_{\mathcal{I}^{0}_\rmii{h}}
 \Biggr\}\;,
 \\
 f_{\mathcal{I}^{5}_\rmii{h}} &=&
 \frac{D(D-2)}{D^2-1}
 \lim_{m\to 0} 
 \Bigg\{
  \frac{{\rm d}}{{\rm d}m^2}\frac{r^4}{\omega^2}
 f_{\mathcal{I}^{0}_\rmii{h}}
 \Biggr\}\;,
 \\
 f_{\mathcal{I}^{6}_\rmii{h}} &=&
 \lim_{m\to 0} 
 \Bigg\{
  \frac{{\rm d}}{{\rm d}m^2}
 \left(\frac{D^2-2 D-2}{D^2-1}q^2r^2+\frac{2}{D^2-1}(\mathbf{q}\cdot\mathbf{r})^2\right) 
\frac{f_{\mathcal{I}^{0}_\rmii{h}}}{\omega^2} 
 \Biggr\}\; , \\
 f_{\mathcal{I}^{7}_\rmii{h}} &=&
 \lim_{m\to 0} 
 \Bigg\{
  \frac{{\rm d}}{{\rm d}m^2}
 \left(\frac{D^2-2 D-2}{D^2-1}q^2(\mathbf{q-r})^2+\frac{2}{D^2-1}(\mathbf{q}\cdot(\mathbf{q-r}))^2\right) \frac{f_{\mathcal{I}^{0}_\rmii{h}}}{\omega^2}
 \Biggr\}\, ,\nonumber\\
\ea
in which the relations in \eq (\ref{PTrels}) have again been used. The integrand corresponding to $\mathcal{I}^{0}_\rmii{h}$, but containing $m$ instead of the $\lambda$ regulator is given in \eq\nr{fIh0}. 

For the remaining i type integrals, we again introduce the mass $m$, and subsequently decompose $\mathcal{I}^{0}_\rmii{i}$ as
\ba
 \mathcal{I}^{0}_\rmii{i} &=& \Tint{Q,R} 
 \frac{4(Q\cdot P)^2 + m^2P^2}{Q^2[R^2+m^2](Q-R)^2(R-P)^2} +
 \Tint{Q,R} 
 \frac{m^2}{Q^2[R^2+m^2](R-P)^2}\nn
 &-&
 \Tint{Q,R} 
 \frac{P^2}{Q^2(Q-R)^2(R-P)^2}
 +
 \Tint{Q,R} 
 \frac{P^2}{Q^2[R^2+m^2](Q-R)^2} \nn
 &+&\Tint{Q,R} 
 \frac{2}{Q^2[R^2+m^2]} -
 \Tint{Q,R} 
 \frac{1}{Q^2R^2}\;, \la{Ii0de}
\ea
in which the last three terms give no contributions to the spectral function. The other integrals are related to this one through the formulas
\ba
 f_{\mathcal{I}^{2}_\rmii{i}} &=&
 \lim_{m\to 0} 
 \Bigg\{
  \frac{{\rm d}}{{\rm d}m^2}\omega^2
 f_{\mathcal{I}^{1}_\rmii{i}}
 \Biggr\} \; , \\
f_{\mathcal{I}^{3}_\rmii{i}}
 &=&-\frac{D-2}{D-1}
 \lim_{m\to 0}
 \Bigg\{
  \frac{{\rm d}}{{\rm d}m^2}
 r^2
 f_{\mathcal{I}^{0}_\rmii{i}}
 \Biggr\} \nn
 &=&-\frac{D-2}{D-1}
 \lim_{m\to 0} 
 \Bigg\{
  \frac{{\rm d}}{{\rm d}m^2}
 r^2
 f_{\mathcal{I}^{}_\rmii{i'}}
 \Biggr\}
 - f_{\mathcal{I}^{2}_\rmii{h}} 
 - f_{\mathcal{I}^{2}_\rmii{b}}\; ,
\ea
while $f_{\mathcal{I}^{1}_\rmii{i}}$ is given in \eq (\ref{fIi1}).

Taking advantage of the above relations, the computational method introduced in section \ref{se:methodbulk} can in principle be immediately implemented to all of the shear master integrals. As there, however, are some subtleties arising from the squared propagators, we will below go through the evaluation of one subset of the integrals, $\mathcal{I}^{0}_\rmii{h}$--$\mathcal{I}^{3}_\rmii{h}$, in some detail. The other cases will then follow in an analogous way.

\subsection{Example calculation}\label{Ihs_exp}

We will first present a detailed treatment of the master spectral function $\rho_{\mathcal{I}^{0}_\rmii{h}}$, which has been considered in section B.10 of \cite{Laine:2011xm}, but is now generalized to the presence of the mass parameter $m$. After this, we use the result to derive similar expressions for the related spectral functions $\rho_{\mathcal{I}^{1}_\rmii{h}}$--$\rho_{\mathcal{I}^{3}_\rmii{h}}$.

\subsubsection*{$\rho^{ }_{\It{h}{0}}$}

Just as in the bulk channel example considered above, also the evaluation of $\rho^{ }_{\It{h}{0}}$ begins from the Matsubara sum, now performed in the presence of nonzero $m$. The resulting expression (\ref{fIh0}) is then once again divided into ``factorized powerlike'' (fz,p), ``factorized exponential'' (fz,e), and ``phase space'' (ps) parts, which are evaluated separately and added together at the very end.  

Beginning from the (fz,p) contribution, we obtain the UV divergent expression
\ba
 \rho^{(\rmi{fz,p})}_{\mathcal{I}^{0}_\rmii{h}}(\omega) &\equiv&
 \int_{\vec{q,r}}
 \frac{\omega^4 \pi }{16 q r E_r E_{qr}}
 \delta(\omega - E_{r} - r)
 \nn &\times&
 \biggl(
 \frac{1}{q+E_{r}+E_{qr}} +
 \frac{1}{q-E_{r}+E_{qr}}
 \biggr)
 (1+n_{E_r}+n_r)
  \;,  \la{Ih_T}
\ea
where we have denoted
\be
 E_q \equiv q \;, \quad E_r \equiv \sqrt{r^2+m^2} \;, \quad 
 E_{qr} \equiv |\vec{q}-\vec{r}|
 \;,
\ee
and the $\mathbf{r}$ integral can be easily performed using 
\ba \la{fz_intr}
 \int_{\vec{r}} \delta(\omega - E_{r} - r) & = & \left. \frac{2\;r^{D-2} E_r}{(4\pi)^\frac{D-1}{2}\Gamma(\frac{D-1}{2})\, \omega}\right|_{r=\frac{\omega^2-m^2}{2\omega}} .
\ea
The remaining integral over $\vec{q}$ then reduces to
\ba
&& \hspace{-3em}\int_{\vec{q}}
 \frac{1 }{4 q E_{qr}}
 \biggl(
 \frac{1}{q+E_{r}+E_{qr}} +
 \frac{1}{q-E_{r}+E_{qr}}
 \biggr) \nonumber 
 =
  \int_Q \frac{1}{Q^2(Q-R)^2}
 \Bigg{|}_{R=(E_r i ,r \vec{e}_r )}
  \la{vacint_h} \\*
 &=&
 \frac{\Lambda^{-2\epsilon}}{(4\pi)^2}
 \biggl(
   \frac{1}{\epsilon} + \ln \frac{\bmu^2}{m^2} + 2 + \rmO(\epsilon)
 \biggr)
 \;,
\ea
leading to the result
\ba
  \rho^{(\rmi{fz,p})}_{\mathcal{I}^{0}_\rmii{h}}(\omega)
 =
 \frac{\omega^2(\omega^2-m^2) \Lambda^{-4\epsilon} }{4(4\pi)^3} (1+2n_{\frac{\omega}{2}})
 \bigg(
   \frac{1}{\epsilon} + \ln \frac{\bmu^2}{(\omega-\fr{m^2}{\omega})^2}
   + \ln \frac{\bmu^2}{m^2} + 4
 \bigg)
 \;. \;\;
\ea
In the last stage, we have here set $n_{E_r}+n_{r}=2n_{\fr{\omega}{2}}$, owing to the identity
\ba
\lim_{m\to 0}
 \bigg\{
\frac{{\rm d}}{{\rm d}m^2}  (n_{E_r}+n_r )
 \Biggr\}  = 0 \; . 
 \ea
 
According to \eq (\ref{fIh0}), the UV-finite (fz,e) part reads for $\omega>0$
\ba
 \rho^{(\rmi{fz,e})}_{\mathcal{I}^{0}_\rmii{h}}(\omega) &=& 
 \int_{\vec{q,r}} \frac{\omega^4 \pi }{8 q r E_{qr}} \Bigg\{  
 \frac{1}{2E_r} 
 \Bigl[\delta(\omega - E_{r} - r) \Bigr]
 \times 
 \\ & & \times \biggl[
 \biggl( 
 \frac{1}{q+E_{r}-E_{qr}} +  
 \frac{1}{q-E_{r}-E_{qr}} 
 \biggr)
 (1+n_{E_r}+n_r)(n_{qr}-n_q)
 \nn & & \;\; 
 +
 \biggl(
 \frac{1}{q+E_{r}+E_{qr}} +  
 \frac{1}{q-E_{r}+E_{qr}} 
 \biggr)
 (1+n_{E_r}+n_r)(n_{qr}+n_q)
 \biggr]
 \Biggr\}
 \; . \nonumber
\ea
Integrating over $r$ according to \nr{fz_intr}, performing a change of integration variables as in \nr{fz_measure0} (with $\lambda\to 0$)
and interchanging the order of integrations in the terms including $n_{qr}$ as in \nr{fz_order0}, we then obtain
\ba
 \rho_{\mathcal{I}^{0}_\rmii{h}}^{(\rmi{fz,e})}(\omega) &\equiv&
 \Bigg[
  \frac{\omega^3}{2(4\pi)^3} (1+n_{E_r}+n_r )
  \Bigg\{
  \int_0^\infty \! {\rm d}q\, n_q
  \int_{E_{qr}^-}^{E_{qr}^+} \! {\rm d}E_{qr}   \,
  \nn &\times&   
   \mathbb{P}\left(\frac{1}{\Delta_{00}} +\frac{1}{\Delta_{10}} -\frac{1}{\Delta_{01}} -\frac{1}{\Delta_{11}}\right)
  \nn 
  &+&
  \int_{0}^\infty \! {\rm d}E_{qr}\, n_{qr}
  \int_{
  |r - E_{qr}|
 }^{
  r + E_{qr}} \! {\rm d}q  \, 
    \mathbb{P}\left(\frac{1}{\Delta_{00}} +\frac{1}{\Delta_{01}} +\frac{1}{\Delta_{10}} +\frac{1}{\Delta_{11}}\right)
       \Biggr\}
  \Bigg]_{r=\frac{\omega^2-m^2}{2\omega}} \nn
&=&  
  \frac{\omega^3}{(4\pi)^3} (1+2n_{\fr{\omega}{2}} )
  \int_0^\infty \! {\rm d}q \, n_q \,
  \ln\left|\fr{2q\,\omega-m^2}{2q\,\omega+m^2}\times\fr{2q+\omega}{2q-\omega}\right|
  \;,  \la{Ih0_fz_e_l0}
\ea
where
$\Delta_{ij}\equiv q+(-1)^{i}E_r+(-1)^{j}E_{qr}$.

For $\omega>0$, the (ps) part of the integral, corresponding to the last four rows of \eq (\ref{fIh0}), finally reads
\ba
 \rho_{\mathcal{I}^{0}_\rmii{h}}^{(\rmi{ps})}(\omega) &\equiv& 
 -\int_{\vec{q,r}}\frac{\omega^4 \pi }{8 q r E_{qr}} \Bigg\{  \\ 
 &  & \!\!
 \delta(\omega - q - r -E_{qr})
 \frac{(1+n_{qr})(1+n_q+n_r)+n_q n_r}
      {(q+E_{r}+E_{qr})(q-E_{r}+E_{qr})}
 \nn & + &  \!\!
 \delta(\omega-q-r+E_{qr}) 
 \frac{n_{qr}(1+n_q + n_r )-n_qn_r}
      {(q+E_{r}-E_{qr})(q-E_{r}-E_{qr})}
 \nn & + &  \!\!
 \delta(\omega - q + r -E_{qr})
 \frac{n_r(1+n_q+n_{qr})-n_q n_{qr}}
      {(q-E_{r}+E_{qr})(q+E_{r}+E_{qr})}
 \nn & + &  \!\!
 \delta(\omega + q - r -E_{qr})
 \frac{n_q(1+n_r+n_{qr})-n_r n_{qr}}
      {(q-E_{r}-E_{qr})(q+E_{r}-E_{qr})}
 \Biggr\}
 \;,  \nonumber
\ea
 Following the steps laid out in section \ref{sse:ps_bulk}, we reach (after some labor) the result
\ba
 \rho_{\mathcal{I}^{0}_\rmii{h}}^{(\rmi{ps})}(\omega) & = &
 \frac{\omega^4}{(4\pi)^3} (1+2 n_{\frac{\omega}{2}})  \Bigg\{
 \nn
 & \mbox{(\i)}  &  + \,
 \fr12\int_{0}^{\frac{\omega}{2}}
 \! {\rm d}q
 \int_{0}^{\omega/2-q}
 \! {\rm d}r \;
  \mathbb{P}\Bigg(\frac{F_{\mathcal{I}^{0}_\rmii{h}}(\frac{\omega}{2}-q,\frac{\omega}{2}-r,q+r)}
  {-2\omega r+m^2} \la{Ih0_ps} \\
  &&+\, \frac{F_{\mathcal{I}^{0}_\rmii{h}}(\frac{\omega}{2}-r,\frac{\omega}{2}-q,q+r)}
  {-2\omega q+m^2} \Bigg)
  \biggl[ 1 + n_{q+r} +  n_{\fr{\omega}2-q}
  \nn && 
  +(1+n_{\fr{\omega}2-r})
  \frac{n_{q+r} n_{\fr{\omega}2-q}}{n_r^2}\biggr]
 \nn
 & \mbox{(\v)}  & + \,
 \fr12\int_{ 0 }^{\infty}
 \! {\rm d}q
 \int_{0}^{ \infty }
 \! {\rm d}r \;
\mathbb{P}\Bigg(\frac{F_{\mathcal{I}^{0}_\rmii{h}}(\frac{\omega}{2}+q,\frac{\omega}{2}+r,q+r)}
  {2\omega r+m^2} \nn
 &&+\, \frac{F_{\mathcal{I}^{0}_\rmii{h}}(\frac{\omega}{2}+r,\frac{\omega}{2}+q,q+r)}
  {2\omega q+m^2} \Bigg)
  \left[ n_{q+r} - n_{q+\frac{\omega}{2}} +
    (1 + n_{q+\frac{\omega}{2}})
    \frac{n_{q+r}n_{r+\frac{\omega}{2}}}{n_r^2} \right]
 \nn
 & \mbox{(\iv)}  & + \,
 \int_{ \frac{\omega}{2} }^{\infty}
 \! {\rm d}q
 \int_{0}^{q-\omega/2}
 \! {\rm d}r \;
 \mathbb{P} \Bigg(\frac{F_{\mathcal{I}^{0}_\rmii{h}}(q-\frac{\omega}{2},\frac{\omega}{2}+r,q-r)}
  {2\omega r+m^2} \nn
  && +\, \frac{F_{\mathcal{I}^{0}_\rmii{h}}(\frac{\omega}{2}+r,q-\frac{\omega}{2},q-r)}
  {-2\omega q+m^2} \Bigg)
  \biggl[
 n_{q-\frac{\omega}{2}} - n_{q}
   \nn && - n_{q-\frac{\omega}{2}}
    \frac{(1+n_{q-r})(n_q 
    -n_{r+\frac{\omega}{2}})}
    {n_r n_{-\frac{\omega}{2}}}  \biggr]
 \nn
 &&\hspace{1em}\Bigg\} \;, \nonumber
\ea
where we have made use of the symmetries of the integrand to write
\ba
 &\mbox{(\i)}&: \frac{1}
  {-2\omega r+m^2}
  \rightarrow
  \fr12\left(\frac{1}
  {-2\omega r+m^2} + \frac{1}
  {-2\omega q+m^2} \right),
 \nn &\mbox{(\v)}&:
 \frac{1}
  {2\omega r+m^2}
  \rightarrow
  \fr12\left(\frac{1}
  {2\omega r+m^2} + \frac{1}
  {2\omega q+m^2} \right)  \la{symIh0}
\ea
in the first two parts of the expression. In addition to this, we have for further reference introduced the function $F(x,y,z)$ in all of the integrands, which in the present case obtains the value $F_{\mathcal{I}^{0}_\rmii{h}}(x,y,z)=1$ . It is straightforward to see that \eq (\ref{Ih0_ps}) reduces to the $\lambda=0$ version of \eq (B.42) of ref.~\cite{Laine:2011xm}, when the limit $m\rightarrow 0$ is taken.

Next, we must perform the integrals in the above three parts (i)--(iii). In each case, we separate the result (with the overall normalization factor $\frac{\omega^4}{(4\pi)^3} (1+2 n_{\frac{\omega}{2}})$ taken out) to three subcontributions based on the distribution functions they contain, denoting by $A$ those parts either containing one $n_{q+r}$ but no other $n$'s or no distribution functions at all; by $B$ those with distribution functions independent of $r$; and by $C$ the remaining piece, proportional to a negative power of $n_r$. In some cases, we will also find it convenient to perform a change of variables from $q$ and $r$ to $q+r\equiv x$ and $q-r \equiv y$ as explained in section \ref{sse:ps_bulk}.

\paragraph{(\i)}
Taking advantage of the symmetry of the integral, it is easy to see that the $A$ contribution can be written in the form
\ba
A^{(\i)}_{\mathcal{I}^{0}_\rmii{h}} &=&
  \fr12 \int_{0}^{\frac{\omega}{2}}
 \! {\rm d}x
 \int_{0}^{x}
 \! {\rm d}y \;
\mathbb{P} \left(\frac{1}
  {-(x-y)\omega +m^2} + \frac{1}
  {-(x+y)\omega +m^2} \right)
  \left( 1 + n_{x} \right)
  \nonumber \\* &=&
 \fr1{2\omega}\int_{0}^{\frac{\omega}{2}}
 \! {\rm d}q
 \left( 1 + n_{q} \right)
 \ln \frac{m^2}{|2q\omega -m^2|}
 \;,
\ea
where we have in the last stage renamed $x=q$ and where the ``vacuum'' contribution is clearly analytically integrable. For the $B$ and $C$ parts, we use the symmetry of the integrand to restrict the integration region to $q>r$, ending up with
\ba
B^{(\i)}_{\mathcal{I}^{0}_\rmii{h}} &=&
\fr1{2\omega}\int_{0}^{\frac{\omega}{4}}
 \! {\rm d}q  \; n_q
  \left( \ln \frac{m^2}{|2q\omega -m^2|}
  + \frac{q} {-\fr{\omega}{2}+q+\frac{m^2}{2\omega}}
 \right)\nonumber \\
&+&  \fr1{2\omega}
    \int_{\frac{\omega}{4}}^{\frac{\omega}{2}}
 \! {\rm d}q \; n_q\;
 \mathbb{P}\left( \ln \frac{m^2}{|\omega^2-2q\omega -m^2|}
  - \frac{q-\fr{\omega}{2}} {-\fr{\omega}{2}+q+\frac{m^2}{2\omega}}
 \right)
\; , \\
C^{(\i)}_{\mathcal{I}^{0}_\rmii{h}} &=&
  -\fr1{2\omega}\int_{0}^{\frac{\omega}{2}}
 \! {\rm d}q
 \int_{0}^{\fr{\omega}{4}-|q-\fr{\omega}{4}|}
 \! {\rm d}r \;
 \mathbb{P} \left(\frac{1}
  {r-\fr{m^2}{2\omega}} + \frac{1}
  {q-\fr{m^2}{2\omega}} \right)\nonumber \\
  &\times&
  (1+n_{\fr{\omega}2-r})
  \frac{n_{q+r} n_{\fr{\omega}2-q}}{n_r^2}
  \;.
\ea
The latter of these expressions must be evaluated as a two-dimensional numerical integral.

\paragraph{(\v)}
For these integrals, the exact same steps as above lead to the results
\ba
 A^{(\v)}_{\mathcal{I}^{0}_\rmii{h}}&=&
- \fr1{2\omega}\int_{0}^{\infty}
 \! {\rm d}q \; n_q
 \ln  \frac{m^2}{2q\omega + m^2}
 \;, \\
 B^{(\v)}_{\mathcal{I}^{0}_\rmii{h}}&=&
  \fr1{2\omega}
  \int_{\frac{\omega}{2}}^{\infty}
 \! {\rm d}q \; n_q
 \left( \ln \frac{m^2}{-\omega^2+2q\omega + m^2}
  - \frac{q-\fr{\omega}{2}} {-\fr{\omega}{2}+q+\frac{m^2}{2\omega}}
 \right)
 \; , \\
 C^{(\v)}_{\mathcal{I}^{0}_\rmii{h}}&=& 
 \fr1{2\omega}\int_{0}^{\infty}
 \! {\rm d}q
 \int_{0}^{q}
 \! {\rm d}r \;
 \left(\frac{1}
  { r+\fr{m^2}{2\omega}} + \frac{1}
  { q+\fr{m^2}{2\omega}} \right)
    (1 + n_{q+\frac{\omega}{2}})
    \frac{n_{q+r}n_{r+\frac{\omega}{2}}}{n_r^2}
  \;. \hspace{1.5cm}
\ea

\paragraph{(\iv)}
In this case, the $A$ contribution vanishes, while the two others produce
\ba
B^{(\iv)}_{\mathcal{I}^{0}_\rmii{h}} &=&
  -\fr1{2\omega} \int_{0 }^{\infty}
 \! {\rm d}q \; n_{q}
 \left( \ln \frac{m^2}{2q\omega + m^2}
  + \frac{q} {\fr{\omega}{2}+q -\frac{m^2}{2\omega}} \right)\nn 
&+& \fr1{2\omega} \int_{ \frac{\omega}{2} }^{\infty}
 \! {\rm d}q \;n_{q}
 \left( \ln \frac{m^2}{-\omega^2+2q\omega + m^2}
  + \frac{q-\fr{\omega}{2}} {q-\frac{m^2}{2\omega}} \right)
  \;, \\
C^{(\iv)}_{\mathcal{I}^{0}_\rmii{h}} &=&
 -\fr1{2\omega}\int_{ \frac{\omega}{2} }^{\infty}
 \! {\rm d}q
 \int_{ 0}^{
    q-\omega/2}
 \! {\rm d}r \;
 \left(\frac{1}
  { r+\fr{m^2}{2\omega}} - \frac{1}
  {q-\fr{m^2}{2\omega}} \right)\nonumber \\
  &\times &
  n_{q-\frac{\omega}{2}}
    \frac{(1+n_{q-r})(n_q - n_{r+\frac{\omega}{2}})}
    {n_r n_{-\frac{\omega}{2}}}
   \;.
\ea
This constitutes the evaluation of the (ps) contribution to the master integral.

Collecting all of the above pieces ((fz) and (ps)) together and reorganizing the terms, we finally obtain a lengthy expression for the function $\rho_{\mathcal{I}^{0}_\rmii{h}}^{ }(\omega)$
\ba
&&\hspace{-2.5em}{\Lambda}^{4\epsilon}\fr{2(4\pi)^3\rho_{\mathcal{I}^{0}_\rmii{h}}^{ }(\omega)} {\omega^3 (1+2n_{\fr{\omega}{2}} )} =
  \frac{\omega^2-m^2}{2\omega}
 \biggl(
   \frac{1}{\epsilon} + \ln \frac{\bmu^2}{(\omega-\fr{m^2}{\omega})^2}
   + \ln \frac{\bmu^2}{m^2} + 4
 \biggr)
 \nn
&+& \fr{\omega}{2}\left[
 1+\left(1-\fr{m^2}{\omega^2}\right)
 \ln\left(\fr{m^2}{\omega^2-m^2}\right)
 \right]
 \nn 
&+&\int_{0}^{\frac{\omega}{4}}
 \! {\rm d}q  \; n_q \Bigg\{
 2\ln\frac{2q+\omega}{-2q+\omega}
+\left( \fr{\omega}{2}-\frac{m^2}{2\omega} \right)
 \left( \frac{1} {q+\fr{\omega}{2}-\frac{m^2}{2\omega}}
   + \frac{1} {q-\fr{\omega}{2}+\frac{m^2}{2\omega}}\right)
 \Biggr\}
  \nn 
&+& \int_{\frac{\omega}{4}}^{\frac{\omega}{2}}
 \! {\rm d}q \; n_q \Bigg\{
 2\ln\frac{2q+\omega}{-2q+\omega}
+ \ln\fr{2q\omega-m^2}{|\omega^2-2q\omega+m^2|}
 \nn 
&-&2+\fr{\omega}{2} \frac{1} {q+\fr{\omega}{2}-\frac{m^2}{2\omega}}
 +\frac{m^2}{2\omega} \mathbb{P} \left( \frac{1} {q-\fr{\omega}{2}+\frac{m^2}{2\omega}}
   - \frac{1} {q+\fr{\omega}{2}-\frac{m^2}{2\omega}}\right)
 \Biggr\}
 \nn
&+& \int_{ \frac{\omega}{2} }^{\infty}
 \! {\rm d}q \; n_q \Bigg\{
 2\ln\frac{2q+\omega}{2q-\omega}-2\ln\frac{2q\omega-\omega^2+m^2}{2q\omega-m^2}
 + \fr{\omega}{2} \left(
  \frac{1} {q+\fr{\omega}{2}-\frac{m^2}{2\omega}}
  -\frac{1} {q-\frac{m^2}{2\omega}}\right)
 \nn 
&-& 1 -\frac{m^2}{2\omega} \left(
  \frac{1} {q+\fr{\omega}{2}-\frac{m^2}{2\omega}}
  -\frac{1} {q-\fr{\omega}{2}+\frac{m^2}{2\omega}}
  -\frac{1} {q-\frac{m^2}{2\omega}}\right)
 \Biggr\}
 \nn 
&+&\int_{0}^{\frac{\omega}{2}}
 \! {\rm d}q
 \int_{0}^{\fr{\omega}{4}-|q-\fr{\omega}{4}|}
 \! {\rm d}r \;\mathbb{P}
 \left(\frac{1}
  {- r+\fr{m^2}{2\omega}} + \frac{1}
  {- q+\fr{m^2}{2\omega}} \right)
  (1+n_{\fr{\omega}2-r})
  \frac{n_{q+r} n_{\fr{\omega}2-q}}{n_r^2}
 \nn 
&+&\int_{0}^{\infty}
 \! {\rm d}q
 \int_{0}^{q}
 \! {\rm d}r \;
 \left(\frac{1}
  { r+\fr{m^2}{2\omega}} + \frac{1}
  { q+\fr{m^2}{2\omega}} \right)
    (1 + n_{q+\frac{\omega}{2}})
    \frac{n_{q+r}n_{r+\frac{\omega}{2}}}{n_r^2}
 \nn 
&-&\int_{ \frac{\omega}{2} }^{\infty}
 \! {\rm d}q
 \int_{ 0}^{
    \frac{2q-\omega}{2}}
 \! {\rm d}r \;
 \left(\frac{1}
  { r+\fr{m^2}{2\omega}} + \frac{1}
  { -q+\fr{m^2}{2\omega}} \right)
  n_{q-\frac{\omega}{2}}
    \frac{(1+n_{q-r})(n_q - n_{r+\frac{\omega}{2}})}
    {n_r n_{-\frac{\omega}{2}}}
  \;. \la{rho_Ih0_m}
\ea
Unlike in the (fz,p), (fz,e) and (ps) parts separately, taking the $m\to 0$ limit of this expression leads to a finite result, which is seen to agree with that derived in ref.~\cite{Laine:2011xm}. We will nevertheless keep $m$ nonzero for the time being, as this will turn out to be very useful in deriving results for some of the other master spectral functions.

\subsubsection*{$\rho^{ }_{\It{h}{1}}$}
As shown in the beginning of this section, the integrand of $\rho^{ }_{\It{h}{1}}$ is related to that of $\rho^{ }_{\It{h}{0}}$ through the relation
\ba
 f_{\mathcal{I}^{1}_\rmii{h}} =
 -\frac{D-2}{D-1}\frac{q^2}{\omega^2}f_{\mathcal{I}^{0}_\rmii{h}}\;,
\ea
from which it is easy to derive a result for the (fz,p) contribution to the integral,
\ba
  \rho^{(\rmi{fz,p})}_{\mathcal{I}^{1}_\rmii{h}}(\omega)
 &=&
  -\frac{D-2}{D-1}\frac{\omega \pi \Lambda^{-2\epsilon} r^{D-3}(1+n_{E_r}+n_r)}{2(4\pi)^{\frac{D-1}{2}} \Gamma(\frac{D-1}{2})}
 \left. \int_Q \frac{q^2}{Q^2(Q-R)^2}
 \right|_{R=(E_r i ,r \vec{e}_r )}\nn
 &=&
 \frac{\omega^4\Lambda^{-4\epsilon}} {(4\pi)^3}(1+2n_{\frac{\omega}{2}})
 \Bigg\{
 \frac{1}{72} \left(-1+\frac{m^6}{\omega ^6}\right)\left(\frac{1}{\epsilon}+\ln \frac{\bmu^2}{m^2}+\ln \frac{\bmu^2}{(\omega-\fr{m^2}{\omega})^2}\right)
 \nn &&
   -\frac{23}{432}+\frac{23 m^6}{432 \omega ^6}-\frac{m^4}{144 \omega ^4}+\frac{m^2}{144 \omega ^2}
 \Biggr\} \;. \la{Ih1_fzp}
\ea
For the (fz,e) part, we similarly obtain
\ba
 \rho_{\mathcal{I}^{1}_\rmii{h}}^{(\rmi{fz,e})}(\omega) &=&
 -\frac{2}{3}\Biggl[
  \frac{\omega}{2(4\pi)^3} (1+n_{E_r}+n_r ) \nn
  &\times&  \Bigg\{
  \int_0^\infty \! {\rm d}q
  \int_{E_{qr}^-}^{E_{qr}^+} \!\! {\rm d}E_{qr}  \, n_q \, q^2
      \left(\frac{1}{\Delta_{00}} +\frac{1}{\Delta_{10}} -\frac{1}{\Delta_{01}} -\frac{1}{\Delta_{11}}\right)
  \\ &+ &
  \int_{0}^\infty \! {\rm d}E_{qr}
  \int_{
  |r - E_{qr}|
 }^{
  r + E_{qr}} \!\! {\rm d}q   \, n_{qr} q^2
    \left(\frac{1}{\Delta_{00}} +\frac{1}{\Delta_{01}} +\frac{1}{\Delta_{10}} +\frac{1}{\Delta_{11}}\right)
       \Biggr\}
  \Biggl]_{r=\frac{\omega^2-m^2}{2\omega}}
    \;.\nonumber  \la{Ih1_fz_e}
\ea
Proceeding finally to the (ps) part, we recall that in the derivation of \eq (\ref{Ih0_ps}), a change of variables of the type
\ba
 & \mbox{(\i)}  &
 \phi(q,r,E_{qr}) \rightarrow \phi\left(\frac{\omega}{2}-q,\frac{\omega}{2}-r,q+r\right)\;, \nn
 & \mbox{(\v)}  &
 \phi(q,r,E_{qr}) \rightarrow \phi\left(\frac{\omega}{2}+q,\frac{\omega}{2}+r,q+r\right)\;, \la{psshift} \\
 & \mbox{(\iv)}  &
 \phi(q,r,E_{qr}) \rightarrow \phi\left(q-\frac{\omega}{2},\frac{\omega}{2}+r,q-r\right) \nonumber 
\ea
was carried out. It is then straightforward to see that the (ps) contribution to $\It{h}{1}$ takes the form of \eq (\ref{Ih0_ps}) with 
\ba
F_{\It{h}{1}}(x,y,z)&=& -\frac{2x^2}{3\omega^2}\;.
\ea

The integration of the above expressions can be dealt with using the methods discussed previously in this section.

\subsubsection*{$\rho^{ }_{\It{h}{2}}$}
For $\rho^{ }_{\mathcal{I}^{2}_\rmii{h}}(\omega)$, the relation to $\rho^{ }_{\It{h}{0}}$ reads
\ba  
f_{\mathcal{I}^{2}_\rmii{h}} =
 -\frac{D-2}{D-1}\frac{r^2}{\omega^2}f_{\mathcal{I}^{0}_\rmii{h}}\;,
\ea
from which one straightforwardly obtains the relations
\ba
  &&\hspace{-1.5cm}\rho^{(\rmi{fz,p})}_{\mathcal{I}^{2}_\rmii{h}}(\omega)
  =
  -\frac{D-2}{D-1}\left(\fr{\omega^2-m^2}{2\omega^2}\right)^2
 \rho_{\mathcal{I}^{0}_\rmii{h}}^{(\rmi{fz,p})}(\omega)
 \\ &=& 
 \frac{\omega^4\Lambda^{-4\epsilon}} {(4\pi)^3}
 \frac{\left(m^2-\omega ^2\right)^3}{24 \omega ^6}(1+2n_{\frac{\omega}{2}})
 \Bigg\{
 \frac{1}{\epsilon}+\ln \frac{\bmu^2}{m^2}+\ln \frac{\bmu^2}{(\omega-\fr{m^2}{\omega})^2}+\fr{11}{3}
 \Biggr\} 
 \;, \la{Ih2_fzp}\nonumber\\ 
 &&\hspace{-1.5cm}\rho_{\mathcal{I}^{2}_\rmii{h}}^{(\rmi{fz,e})}(\omega) =
 -\frac{2}{3}\left(\fr{\omega^2-m^2}{2\omega^2}\right)^2
 \rho_{\mathcal{I}^{0}_\rmii{h}}^{(\rmi{fz,e})}(\omega)\la{Ih2_fz_e}
  \\ &=&  
  -
  \frac{2\omega}{3(4\pi)^3} 
  \left(\fr{\omega^2-m^2}{2\omega}\right)^2
  (1+2n_{\fr{\omega}{2}} )
  \int_0^\infty \! {\rm d}q \, n_q \, 
  \ln\left|\fr{2q\omega-m^2}{2q\omega+m^2}\times\fr{2q+\omega}{2q-\omega}\right|
    \;,  \nonumber
\ea
as well as
\ba
F_{\It{h}{2}}(x,y,z)&=& -\frac{2y^2}{3\omega^2}\;.
\ea
Again, all integrals encountered here can be carried out with methods used in $\rho^{ }_{\mathcal{I}^{0}_\rmii{h}}$.

\subsubsection*{$\rho^{ }_{\It{h}{3}}$}

With $\It{h}{3}$, we encounter a master integral containing a squared propagator. To this end, we proceed to take a mass derivative of $\rho^{}_{\mathcal{I}^{2}_\rmii{h}}(\omega)$, obtaining the relation
\ba
\rho^{ }_{\mathcal{I}^{3}_\rmii{h}}(\omega) 
 &=&\omega^2\lim_{m\to 0} 
 \Bigg\{
  \frac{{\rm d}}{{\rm d}m^2} \rho^{}_{\mathcal{I}^{2}_\rmii{h}}(\omega)
  \Biggr\}
 \;.
\ea
As we have above computed $\rho^{ }_{\mathcal{I}^{2}_\rmii{h}}(\omega)$ keeping $m$ nonzero, taking the derivative in this expression is in principle a very straightforward task. The only somewhat problematic issue is related to the IR divergences that appear in individual parts of the integral once one proceeds to set $m\to 0$ in the end. Our strategy with them is to identify those parts of the integrals that diverge in this limit, and subsequently add and subtract from the integrands terms that are analytically computable (with nonzero $m$), yet render the original integral convergent. The divergent terms are expected to cancel against each other once we assemble the result for the full master spectral function, while the remaining finite parts are dealt with using the methods described above.

In the one-dimensional integrals, originating from the $A$ and $B$ parts defined above, the divergences in the $m\to 0$ limit may in principle appear due to one of two reasons: A pole at $q=\fr{\omega}{2}$, coming from a factor $1/(q-\fr{\omega}{2}+\fr{m^2}{2\omega})$, or an explicit $\ln m^2$ term originating from the analytic $r$ integral. For ${\mathcal{I}^{3}_\rmii{h}}$, the latter does not occur, while the former can be dealt with by writing
\ba
\fr{(4\pi )^3\rho^{(\rmi{1d,div}) }_{\mathcal{I}^{3}_\rmii{h}}(\omega)} { 1+2n_{\fr{\omega}{2}}}
&=& \int_{ \frac{\omega}{2} }^{\infty}
 \! {\rm d}q \;n_{q}
 \Bigg[
 \frac{-m^4+\omega ^4}{24 \left(q+\frac{m^2}{2 \omega }-\frac{\omega }{2}\right)}+\frac{m^6-2 m^4 \omega ^2+m^2 \omega ^4}{48 \left(q+\frac{m^2}{2 \omega }-\frac{\omega }{2}\right)^2 \omega }
 \Bigg] 
 \hspace{7em}\nn
 &=&
\int_{ \frac{\omega}{2} }^{\infty}
 \! {\rm d}q \, \Bigg\{n_{q}
 \Bigg[
 \frac{-m^4+\omega ^4}{24 \left(q+\frac{m^2}{2 \omega }-\frac{\omega }{2}\right)}+\frac{m^6-2 m^4 \omega ^2+m^2 \omega ^4}{48 \left(q+\frac{m^2}{2 \omega }-\frac{\omega }{2}\right)^2 \omega }
 \Bigg]
 \nn &&
 -n_{\fr{\omega}{2}} \frac{\omega }{2q}
 \Bigg[
 \frac{-m^4+\omega ^4}{24 \left(q+\frac{m^2}{2 \omega }-\frac{\omega }{2}\right)}+\frac{m^6-2 m^4 \omega ^2+m^2 \omega ^4}{48 \left(q+\frac{m^2}{2 \omega }-\frac{\omega }{2}\right)^2 \omega }
 \Bigg]
 \Bigg\}
 \nn &&
 +\int_{ \frac{\omega}{2} }^{\infty}
 \! {\rm d}q \;n_{\fr{\omega}{2}} \frac{\omega }{2q}
 \Bigg[
 \frac{-m^4+\omega ^4}{24 \left(q+\frac{m^2}{2 \omega }-\frac{\omega }{2}\right)}+\frac{m^6-2 m^4 \omega ^2+m^2 \omega ^4}{48 \left(q+\frac{m^2}{2 \omega }-\frac{\omega }{2}\right)^2 \omega }
 \Bigg]
 \;. \nonumber
\ea
Here, the integral containing the curly brackets is seen to be finite, while the divergent term on the last row can be evaluated analytically, producing
\ba
&&\hspace{-2cm}\int_{ \frac{\omega}{2} }^{\infty}
 \! {\rm d}q \;n_{\fr{\omega}{2}} \frac{\omega }{2q}
 \Bigg[
 \frac{-m^4+\omega ^4}{24 \left(q+\frac{m^2}{2 \omega }-\frac{\omega }{2}\right)}+\frac{m^6-2 m^4 \omega ^2+m^2 \omega ^4}{48 \left(q+\frac{m^2}{2 \omega }-\frac{\omega }{2}\right)^2 \omega }
 \Bigg]
\nn& =&
 -\frac{\omega ^4}{24}  \left(\frac{m^2}{\omega^2 }-1 +  \ln\fr{m^2}{\omega ^2}\right) n_{\fr{\omega}{2}}
 \,. \hspace*{1cm}\nonumber
\ea

Turning our attention next to the two-dimensional part, we observe that the only divergence originates from the $r=0$ limit of $C^{(\iv)}_{\mathcal{I}^{3}_\rmii{h}}$. This can be regulated by subtracting from the integrand a term of the type $\alpha/(r+\frac{m^2}{2\omega})$, where $\alpha$ is a suitably chosen residue. This amounts to writing $C^{(\iv)}_{\mathcal{I}^{3}_\rmii{h}}$ in the somewhat complicated form
\ba
\fr{2}{3}C^{(\iv)}_{\mathcal{I}^{3}_\rmii{h}}&=& \frac{ \omega ^4 }{24} n_{\fr{\omega}{2}} \ln \fr{m^2}{\omega^2}  -\frac{\omega^4}{24 T} \int_{ \frac{\omega}{2} }^{\infty}
 \! {\rm d}q n_q(1+n_q) \ln\left|\fr{2 q \omega -\omega ^2}{\omega^2}\right|
\nn 
&+&
\fr{2}{3}C^{(\iv)}_{\mathcal{I}^{3}_\rmii{h}} + 
\frac{\omega^4}{24 T} \int_{ \frac{\omega}{2} }^{\infty}
 \! {\rm d}q\, n_q(1+n_q)\, 
 \int_{ 0}^{
    \frac{2q-\omega}{2}}
 \!\!\!\! {\rm d}r \;
 \frac{1}{\left(r+\fr{m^2}{2\omega} \right)} 
 \;,
\ea
where the two terms on the latter row combine to a finite integral.

Apart from the terms discussed above, all other parts of $\rho^{}_{\mathcal{I}^{3}_\rmii{h}}$ are IR finite and computable with standard methods. The very same strategy can later be applied to the other h-type integrals, and ultimately even to the other master spectral functions. As this process is explained in considerable detail in \cite{Zhu:2012be}, we will refrain from considering the other integrals here, and rather move on to analyze the final outcome of the calculation in the following chapter.

%

\setcounter{chapter}{4}
%
%
%
%
%
%

\mychapter{Thermal spectral functions: Results}
\label{CHAP 5}

Having explained the most important steps in the evaluation of NLO thermal spectral functions in the previous chapter, we are now ready to take a closer look at the results derived for the bulk and shear channels of SU($N$) Yang-Mills theory in \cite{Laine:2011xm,Zhu:2012be}. This will complement the UV analysis of chapter \ref{CHAP 3} by extending the results to all $\omega\sim T$, which hopefully will turn out to be very useful in the extraction of a nonperturbative spectral function from lattice data. In the bulk channel calculation presented in \cite{Laine:2011xm}, the results have even been extended to cover the region $\omega \sim gT$ by employing the Hard Thermal Loop (HTL) effective theory \cite{Braaten:1991gm}. The details of this calculation will not be reviewed in this thesis, but we will nevertheless display the effects of the resummation on our numerical results below.

Once the NLO spectral functions have been evaluated, a straightforward application of the results will be the derivation of Euclidean imaginary time correlators, available through \eq\nr{eq:intrel}. Below, we will perform a detailed comparison of our results for these quantities with lattice data, and in some cases even with AdS/CFT predictions. In addition, we will discuss the use of the perturbative results in confirming and refining sum rules proposed e.g.~in \cite{Meyer:2007fc,Meyer:2010gu,Romatschke:2009ng,Meyer:2010ii}.
Finally, some attention will also be paid to the analytic treatment of the IR and UV limits of the spectral functions.

\mysection{Bulk channel}
\la{se:spf_bulk}

Extracting the bulk channel spectral functions using the methods explained in chapter \ref{CHAP 4} and writing the bare gauge coupling in terms of the renormalized one to remove the remaining $1/\epsilon$ poles, we obtain a result with the schematic form
\ba
\hspace{-0.4cm} \frac{\rho^{ }_\theta(\omega)}{4 d_A c_\theta^2 } & = &  
 \frac{\pi\omega^4}{(4\pi)^2}
 \bigl( 1 + 2 n_{\frac{\omega}{2}} \bigr)
 \biggl\{
   g^4 + \frac{g^6\Nc}{(4\pi)^2}
    \biggl[
       \frac{22}{3} \ln\frac{\bmu^2}{\omega^2} + \frac{73}{3} 
     + 8\, \phi^{ }_T(\omega) 
    \biggr] 
 \biggr\}
 + \rmO(g^8)
 \;, \la{rhofinal_theta} \nonumber\\
 \\ 
\hspace{-0.4cm} \frac{-\rho^{ }_\chi(\omega)}{16 d_A c_\chi^2 } & = &  
 \frac{\pi\omega^4}{(4\pi)^2}
 \bigl( 1 + 2 n_{\frac{\omega}{2}} \bigr)
 \biggl\{
   g^4 + \frac{g^6\Nc}{(4\pi)^2}
    \biggl[
       \frac{22}{3} \ln\frac{\bmu^2}{\omega^2} + \frac{97}{3} 
     + 8\, \phi^{ }_T(\omega) 
    \biggr] 
 \biggr\}
 + \rmO(g^8)
 \;. \la{rhofinal_chi} \nonumber\\
\ea
The logarithmic and constant terms appearing here can be seen to agree with the OPE results of ref.~\cite{Laine:2010tc}, while the complicated $T-$dependent function $\phi^{ }_T$, appearing in both $\theta$ and $\chi$ channels, reads
\ba
 \phi^{ }_T(\omega)  & = &   
 \int_0^{ \frac{1}{2} } \! { {\rm d}\sigma } \, 
   \hat n_{\sigma} \; 
   \biggl\{
      \biggl[\frac{1}{\sigma} -\frac{1}{\sigma-1}  
      -2 + \sigma -\sigma^2
      \biggr]
       \ln\bigl( 1 - \sigma \bigr)
 \nn & & \hspace*{2cm} + \,
      \biggl[\frac{1}{\sigma} -\frac{1}{\sigma+1} + 
      2 + \sigma + \sigma^2
      \biggr]
       \ln\bigl( 1 + \sigma \bigr)
 \nn & & \hspace*{2cm} + \,
  \frac{11}{12} 
   \biggl[
     \frac{1}{\sigma+1} 
   + \frac{1}{\sigma-1} \biggr] 
   + \fr{5\sigma}6 
   \; \biggr\}  
 \nn 
 & + & 
 \int_{ \frac{1}{2} }^{ 1 } 
 \! { {\rm d}\sigma } \, 
     \hat n_{\sigma}  \; 
   \biggl\{
      \biggl[\frac{1}{\sigma} -\frac{2}{\sigma - 1} 
      - \fr{11}{4} + \sigma - \frac{3\sigma^2}{2}
      \biggr]
       \ln\bigl( 1 - \sigma \bigr)
 \nn & & \hspace*{2cm} + \,
      \biggl[\frac{1}{\sigma} -\frac{1}{\sigma+1}  
     + 2  + \sigma + \sigma^2
      \biggr]
       \ln\Bigl( 1 + \sigma \Bigr)
 \nn & & \hspace*{2cm} + \,
      \biggl[\frac{1}{\sigma-1}  
      + \fr34 + \frac{\sigma^2}{2}
      \biggr]
       \ln\bigl( \sigma \bigr)
 \nn & & \hspace*{2cm} + \,
  \frac{11}{12} 
   \frac{1}{\sigma+1} 
   - \fr13 - 2\sigma - \frac{\sigma^2}{3}
   \;\biggr\}  
 \nn 
 & + & 
 \int_{ 1 }^{ \infty } 
 \! { {\rm d}\sigma } \, 
    \hat n_{\sigma} \; 
   \biggl\{
      2\biggl[\frac{1}{\sigma} -\frac{1}{\sigma-1} 
      - 2 + \sigma - \sigma^2
       \biggr]
       \ln\bigl(  \sigma - 1 \bigr)
 \nn & & \hspace*{2cm} + \,
      \biggl[\frac{1}{\sigma} -\frac{1}{\sigma+1}  
      +2+\sigma + \sigma^2
      \biggr]
       \ln\bigl( 1 + \sigma \bigr)
 \nn & & \hspace*{2cm} + \,
      \biggl[\frac{1}{\sigma-1} - \frac{1}{\sigma}  
      + 2 - \sigma + \sigma^2
      \biggr]
       \ln\bigl( \sigma \bigr)
 \nn & & \hspace*{2cm} + \,
  \frac{11}{12} 
   \biggl[ \frac{1}{\sigma+1 } - \frac{1}{\sigma} \biggr] 
  + \frac{23}{12} - \frac{13\sigma}{4} - \frac{11\sigma^2}{12}
   \;\biggr\}  
 \nn & + & 
 \int_0^{ 1 } 
 \! { {\rm d}\sigma } 
 \int_0^{ \frac{1}{2} - |\sigma-\frac{1}{2}| } 
 \! { {\rm d}\tau }  \,
  \frac{  
   \hat n_{1-\sigma} \,
   \hat n_{\sigma+\tau} 
  (1+\hat n_{1-\tau}) }
  {\hat n_{ \tau }^2}
  \; \times 
 \nn & & \hspace*{2cm} \times \,
  \biggl[
  \frac{1}{\sigma\tau} 
  - \frac{5 - 4 \tau + 2\tau^2}{4\sigma}
  - \frac{5 - 4 \sigma + 2 \sigma^2}{4 \tau}
  + \frac{3}{2}
  \biggr] 
 \nn & + & 
 \int_{ 1 }^{\infty}
 \! {{\rm d}\sigma}
 \int_{0}^{ \sigma - 1 }
 \! { {\rm d}\tau } \,
    \frac{  
    \hat n_{\sigma - 1} 
    (1+\hat n_{\sigma - \tau} )
    (\hat n_{ \sigma } 
    - \hat n_{\tau + 1 } )
    }
    { \hat n_{ \tau } \hat n_{- 1} }
  \; \times 
 \nn & & \hspace*{2cm} \times \,
  \biggl[
  \frac{1}{\sigma\tau} 
  + \frac{5 + 4 \tau + 2\tau^2}{4\sigma}
  - \frac{5 - 4 \sigma + 2 \sigma^2}{4 \tau}
  - \frac{3}{2}
  \biggr] 
 \nn & + & 
 \int_0^\infty \! {{\rm d}\sigma}  
 \int_{0}^{ \sigma }
 \! { {\rm d}\tau } \,
 \frac{ 
  ( 1 + \hat n_{ \sigma + 1 } )
  \,\hat n_{ \sigma + \tau }
  \,\hat n_{ \tau + 1 } }
  {\hat n_{ \tau }^2}
  \; \times 
 \nn & & \hspace*{2cm} \times \,
  \biggl[ 
  \frac{1}{\sigma\tau} 
  + \frac{5 + 4 \tau + 2\tau^2}{4\sigma}
  + \frac{5 + 4 \sigma + 2 \sigma^2}{4 \tau}
  + \frac{3}{2}
  \biggr] 
 \;, \hspace*{5mm} \la{phi_T} 
\ea
with $\hat n_x \equiv n_{\frac{\omega x}{2}}$. While \eq\nr{phi_T} is rather lengthy, all of the integrals appearing in it are finite (assuming the principal value prescription), and can thus be carried out with numerical methods. The behavior of the function is displayed in \fig\ref{fig:phiT}.

\begin{figure}[t]
\centerline{%
 \includegraphics[width=8.5cm]{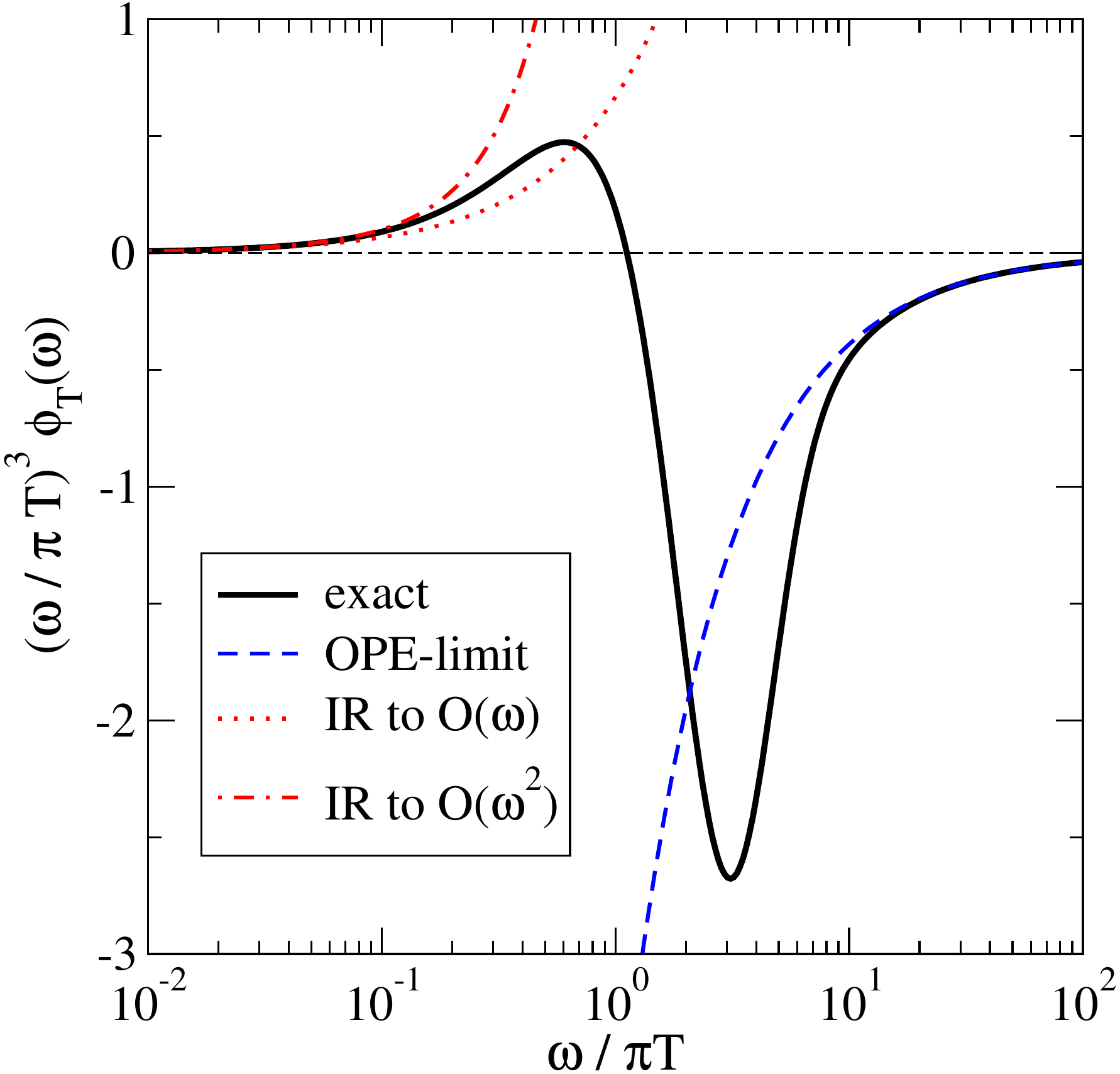}%
}
\caption[a]{\small
The function $\phi^{ }_T$ of \eq\nr{phi_T}, multiplied by $(\frac{\omega}{\pi T})^3$ and plotted as a function of $\frac{\omega}{\pi T}$. Also shown are the ultraviolet (``OPE'') limit from \eq\nr{phi_T_as} as well as the infrared (``IR'') limit 
from \eq\nr{phi_IR}. The figure is taken from \cite{Laine:2011xm}.
}
\la{fig:phiT}
\end{figure}

\subsection{Limits and numerical evaluation}
\la{ss:limits}

We begin the analysis of the above results by performing some consistency checks, first concerning the UV ($\omega \gg \pi T$) behavior of the function $\phi^{ }_T$. Setting $P=(-i[\omega+0^+],\vec{0})$ and taking the imaginary part in \eqs \nr{Gtheta_uv} and \nr{Gchi_uv}, we obtain a prediction for the UV behavior of the function $\phi^{ }_T$:
\be
 \phi^{ }_T(\omega) 
 \;\; = \;\;
 - \frac{176}{45} \frac{\pi^4 T^4}{\omega^4}
 + \rmO\Bigl( \frac{\pi^6 T^6}{\omega^6} \Bigr)
 \;. \la{phi_T_as}
\ee
As demonstrated by \fig\ref{fig:phiT}, this is indeed found to be the large-$\omega$ behavior of \eq\nr{phi_T}. In the opposite limit, \ie~$\omega \ll \pi T$, a semianalytic treatment of the integrals appearing in \eq\nr{phi_T} on the other hand produces a result of the form
\be
 \phi_T^{ }(\omega) \approx 
  \frac{2 \pi^2 T^2}{3 \omega^2} + 
 3.31612\, \frac{\pi T}{\omega} 
 + \rmO\Bigl(\ln \frac{\omega}{\pi T}\Bigr)
 \;. \la{phi_IR}
\ee
This is again seen to be in good agreement with the numerical result in \fig\ref{fig:phiT}.

Despite the simplicity of \eq(\ref{phi_IR}), this unfortunately does not represent the physical small-$\omega$ behavior of the spectral function. Indeed, it is in general not possible to continue our results all the way to the $\omega\to 0$ limit, as the complicated IR structure of the physical spectral function would require extremely tedious resummations to be carried out. Even extending the applicability of our results down to frequencies of the order of $\omega\sim g T$ necessitates a Hard Thermal Loop (HTL) \cite{Braaten:1991gm} type resummation, which we indeed have carried out in \cite{Laine:2011xm} and will now briefly explain.

Schematically, the HTL resummed spectral function can be written in the form
\be
 \rho^\rmii{QCD}_\rmii{resummed} 
 \; = \; 
 \rho^\rmii{QCD}_\rmii{resummed} 
 -  \rho^\rmii{HTL}_\rmii{resummed} 
 + \rho^\rmii{HTL}_\rmii{resummed}  
 \; \approx \;
 \rho^\rmii{QCD}_\rmii{naive} 
 -  \rho^\rmii{HTL}_\rmii{naive} 
 + \rho^\rmii{HTL}_\rmii{resummed}  
 \;, \la{master_resum}
\ee
where we have subtracted and added an HTL part. In doing this, we have noted that if the effective theory is used correctly, then the difference between the full and the effective theory computations is infrared safe, so that no resummation is needed for the difference. Having reported the result for the function $\rho^\rmii{QCD}_\rmii{naive}$ above, our task thus becomes to evaluate the two functions $\rho^\rmii{HTL}_\rmii{naive}$ and $\rho^\rmii{HTL}_\rmii{resummed}$.

The details of the HTL computations are given in appendix C of ref.~\cite{Laine:2011xm}, while here we merely quote the corresponding results. First, the naive HTL parts obtain the form 
\ba
 & & \hspace*{-1cm}
 \left. \frac{\rho_\theta^\rmii{HTL}(\omega)}{4 d_A c_\theta^2}
 \right|_\rmi{naive}
  =
 \left. \frac{-\rho_\chi^\rmii{HTL}(\omega)}{16 d_A c_\chi^2}
 \right|_\rmi{naive}
  =
 \frac{\pi g^4 \bigl( 1 + 2 n_{\frac{\omega}{2} } \bigr)}{(4\pi)^2} 
 \Bigl\{ \omega^4 + \omega^2 \mE^2 
\Bigr\} 
 + \rmO(g^8) 
 \;, \hspace*{0.5cm} \la{nlo_htl_compact}
\ea
where the Debye mass parameter is defined by 
\be
  \mE^2 \equiv \frac{ g^2 \CA  T^2}{3}   
  \;. \la{mE}
\ee
The first term of \eq\nr{nlo_htl_compact} clearly matches the leading order QCD result in \eqs\nr{rhofinal_theta} and \nr{rhofinal_chi}, 
whereas the second term exactly corresponds to the leading term of \eq\nr{phi_IR}. This in particular implies that in the difference
$
 \rho^\rmii{QCD}_\rmii{naive} 
 -  \rho^\rmii{HTL}_\rmii{naive} 
$
appearing in \eq\nr{master_resum}, the dominant IR divergence clearly drops out.

\begin{figure}[t]
\centerline{%
 \includegraphics[width=7cm]{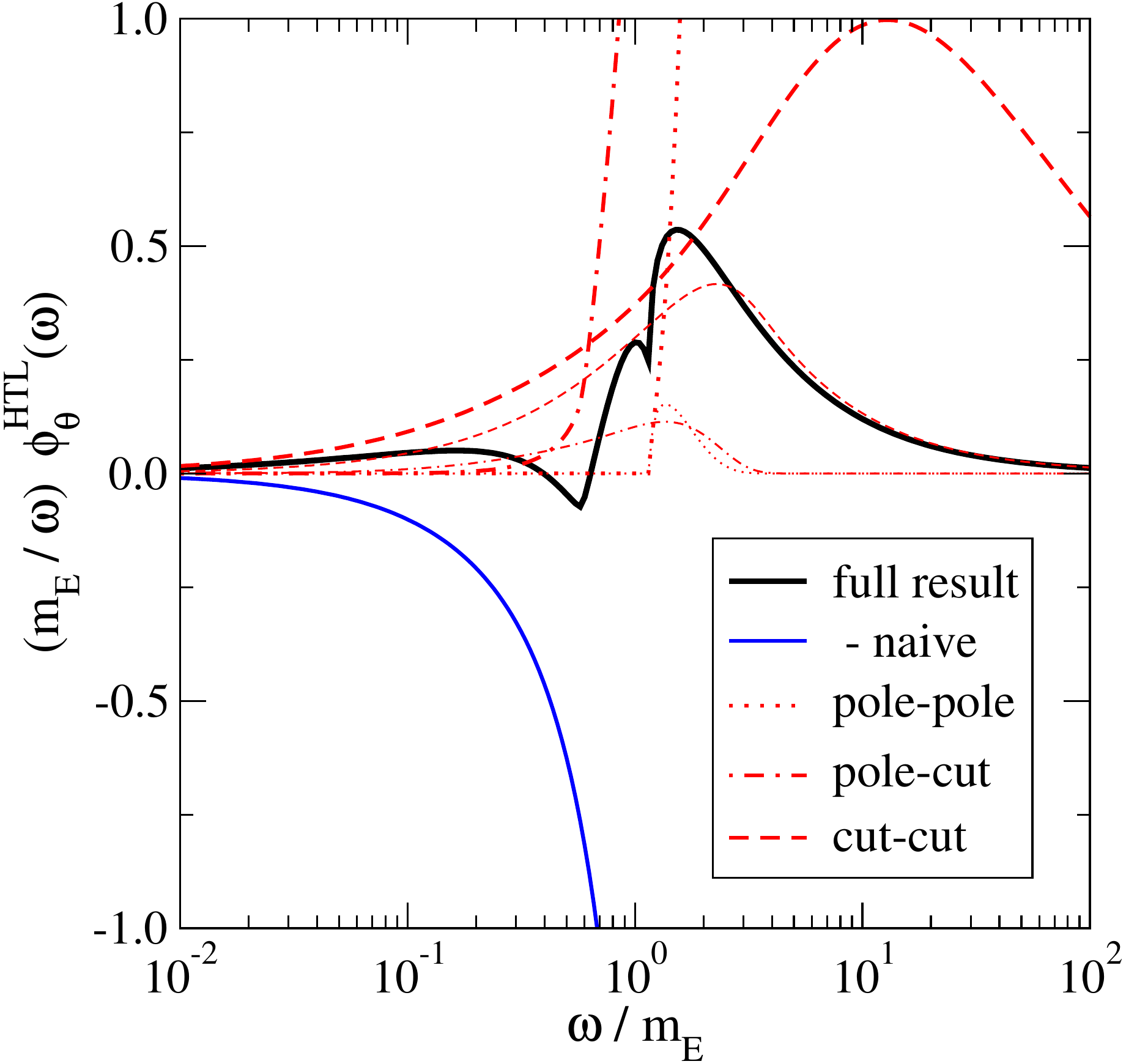}%
~~~\includegraphics[width=7cm]{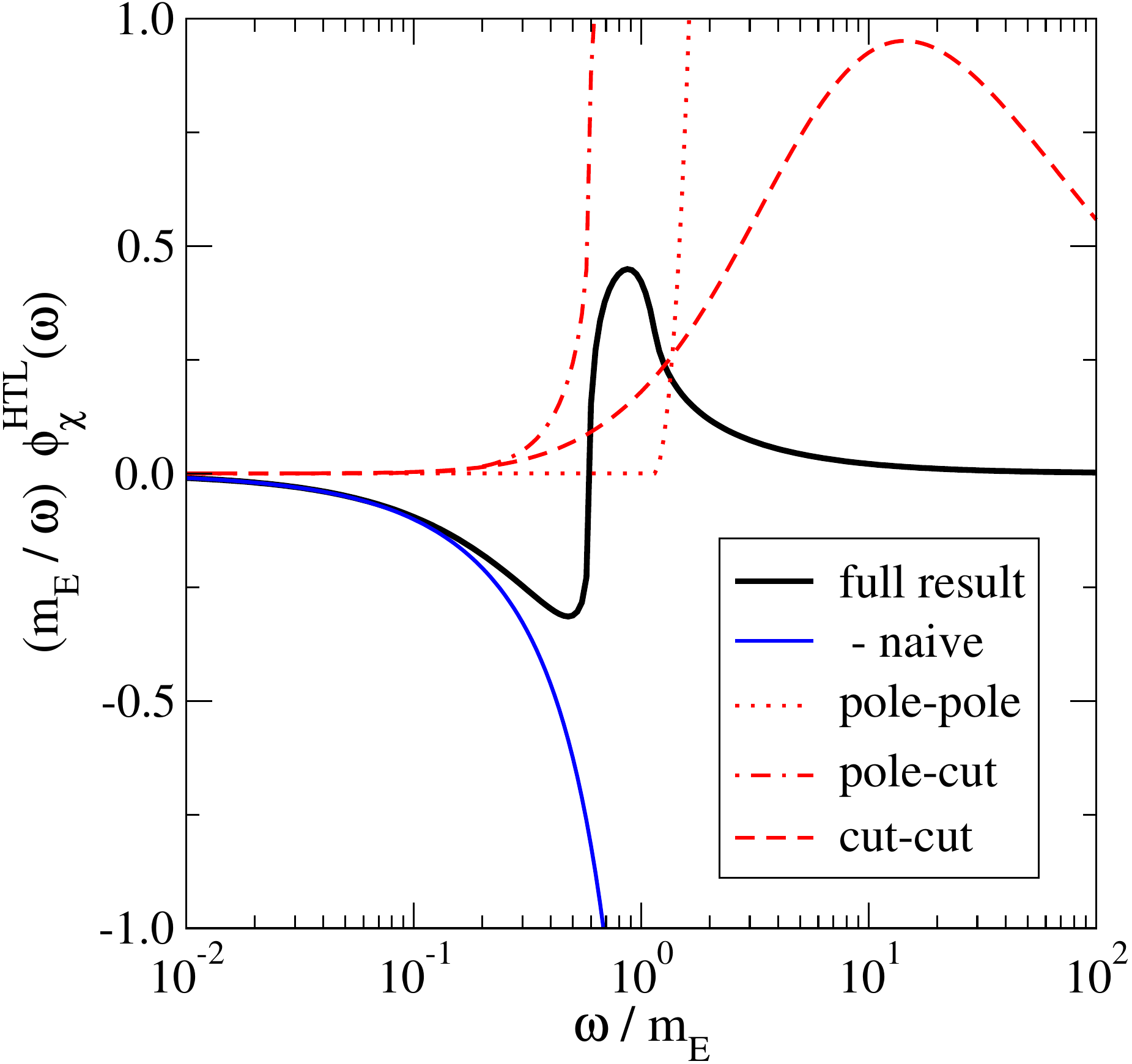}
}
\caption[a]{\small
The functions $\phi^\rmii{HTL}_{\theta,\chi}$ of \eqs\nr{HTL_resum_theta} and \nr{HTL_resum_chi}, multiplied by $\frac{\mE}{\omega}$.  Also shown are partial contributions from various sub-processes. The figure is taken from \cite{Laine:2011xm}.
}

\la{fig:phi_HTL}
\end{figure}

The resummed HTL computation on the other hand leads to the result
\ba
 \left. \frac{\rho_\theta^\rmii{HTL}(\omega)}{4 d_A c_\theta^2}
 \right|_\rmi{resummed}
 \!\!\! & \equiv & 
  \frac{\pi g^4 \bigl( 1 + 2 n_{\frac{\omega}{2} } \bigr)}{(4\pi)^2} 
  \Bigl\{ \omega^4 + \omega^2 \mE^2 +\mE^4
  \, \phi_\theta^\rmii{HTL}(\omega)
  \Bigr\}
 \;, \la{HTL_resum_theta} \\ 
 \left. \frac{-\rho_\chi^\rmii{HTL}(\omega)}{16 d_A c_\chi^2}
 \right|_\rmi{resummed}
  \!\! & \equiv & 
  \frac{\pi g^4 \bigl( 1 + 2 n_{\frac{\omega}{2} } \bigr)}{(4\pi)^2} 
  \Bigl\{ \omega^4 + \omega^2 \mE^2 +\mE^4
  \, \phi_\chi^\rmii{HTL}(\omega)
  \Bigr\}
  \;, \la{HTL_resum_chi}
\ea
where the separated terms obviously cancel against the naive version
in the difference of the two terms in \eq\nr{master_resum}. The functions $\phi^\rmii{HTL}_{\theta,\chi}$ must be evaluated numerically, and are displayed in \fig\ref{fig:phi_HTL}. As expected, one may explicitly check that those functions only modify the full spectral function in the deep IR regime.

Next, we proceed to numerically inspect the spectral function, and to this end, must assign a value for the coupling $g^2$. As we are working to NLO in perturbation theory, we use the two-loop running coupling, with the renormalization scale varied by a factor of 2 around a piecewise defined optimal value. In the regime $\omega \gg \pi T$, we apply the ``fastest apparent convergence'' criterion to \eqs\nr{rhofinal_theta} and \nr{rhofinal_chi}, obtaining
\be
 \ln(\bmu^\rmi{opt($\omega$)}_{\theta}) \equiv \ln(\omega) 
 -  \frac{73}{44}
 \;, \quad
 \ln(\bmu^\rmi{opt($\omega$)}_{\chi}) \equiv \ln(\omega) 
 -  \frac{97}{44}
 \;, \la{muopt_w}
\ee
while in the infrared regime $\omega \ll \pi T$ we follow the usual EQCD choice (cf.\ ref.~\cite{Laine:2005ai} and references therein), resulting in
\be
 \ln(\bmu^\rmi{opt($T$)}_{\theta,\chi}) \equiv 
 \ln(4\pi T) - \gammaE - \frac{1}{22}
 \;. \la{muopt_T}
\ee 
For a given intermediate $\omega$, the larger of these scales is chosen; the  switch happens at $\omega \approx 11.3\, \pi T$ for $\rho^{ }_\theta$,  and at $\omega \approx 19.5\, \pi T$ for $\rho^{ }_\chi$. Finally, we connect the critical temperature of the deconfinement transition in pure SU(3) gauge  theory to the scale parameter appearing inside $g$ using the lattice result $\Tc = 1.25 \Lambdamsbar$.

\begin{figure}[t]
\centerline{%
\includegraphics[width=7cm]{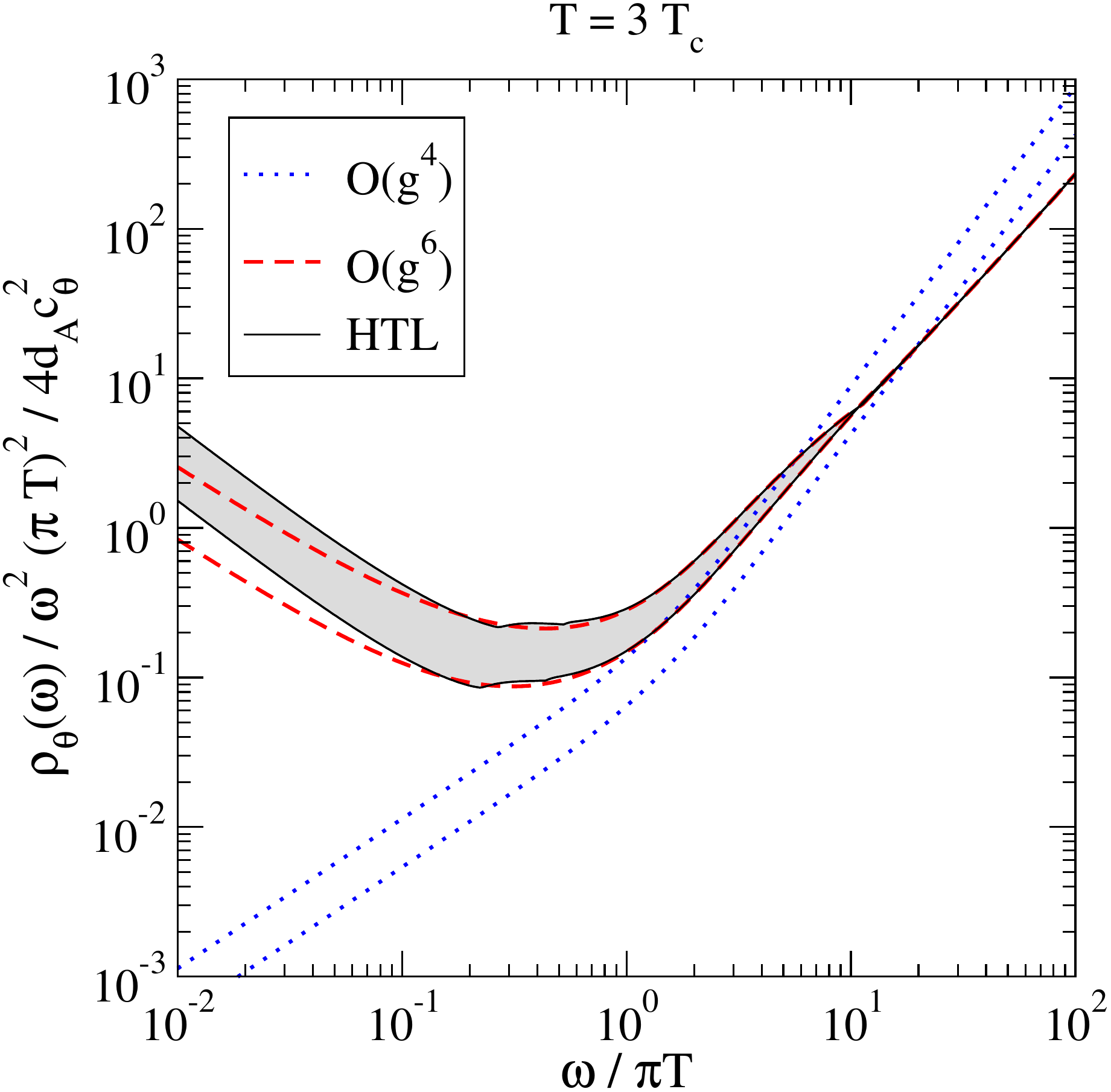}%
~~\includegraphics[width=7cm]{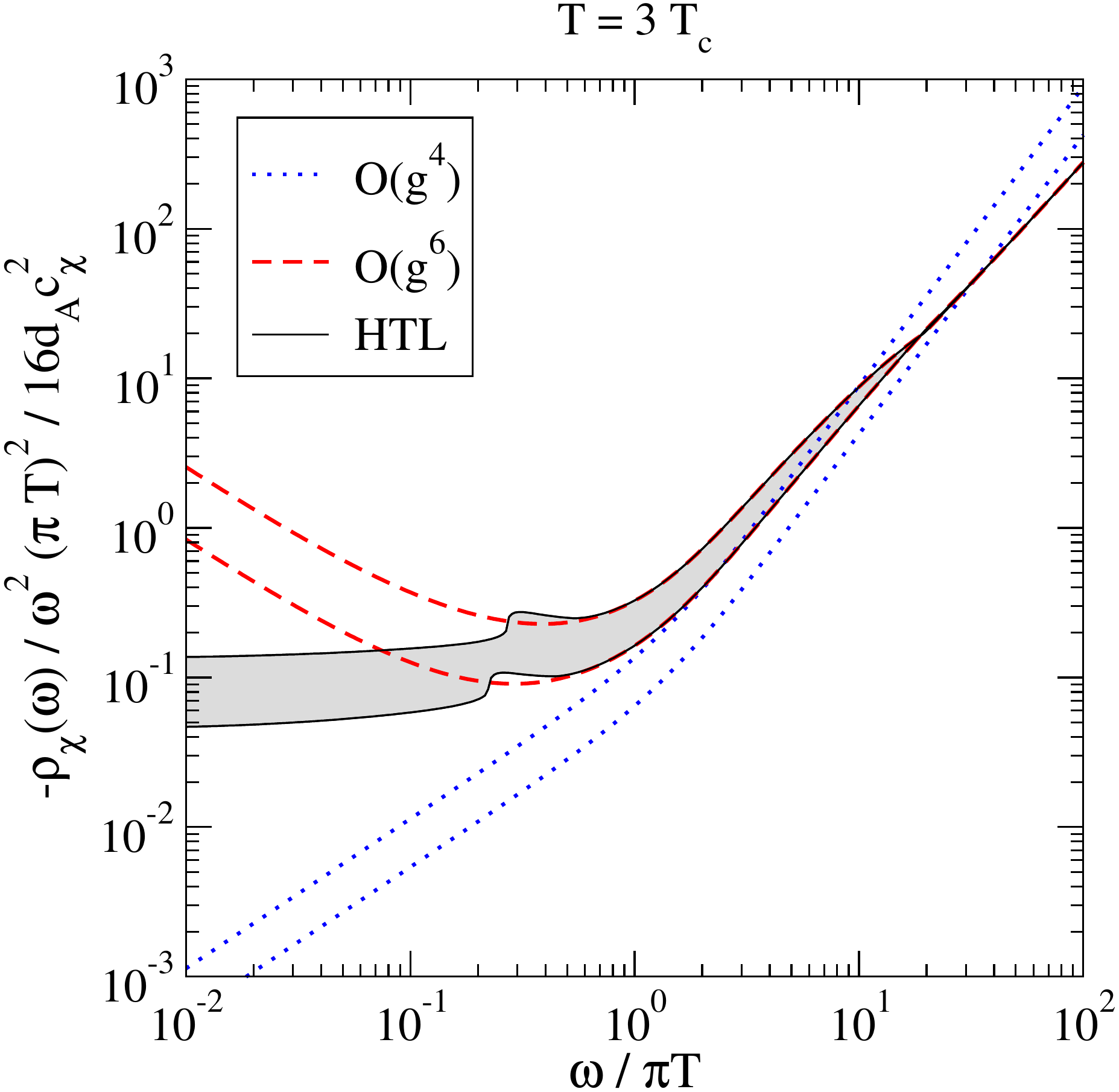}%
}
\caption[a]{\small 
A numerical evaluation of \eqs\nr{rhofinal_theta} and \nr{rhofinal_chi} (as well as the corresponding HTL expressions \nr{master_resum}) in units of $\omega^2 (\pi T)^2$, carried out for for $T = 3.75 \Lambdamsbar$ or $T = 3 \Tc$. The grey shaded areas correspond to the uncertainty associated with the value of the renormalization scale. The figure is taken from \cite{Laine:2011xm}.
 }
\la{fig:wdep}
\end{figure}

Having now fixed all appropriate constants, we display our numerical results for the bulk spectral functions in \fig\ref{fig:wdep}. We observe that the dependence of the NLO result on the scale parameter practically disappears in the ultraviolet domain, while in the IR limit a depletion due to the HTL-resummation is visible in the $\chi$-channel. A detailed comparison of our results with those of \cite{Meyer:2010ii} is problematic due to the mixing of the zero and finite temperature parts in our calculations; however, it is worth noting that the structure of the thermal part of the spectral density, $\phi^{ }_T$, clearly qualitatively agrees with \fig5 of this reference.

\subsection{Sum rules}
\la{ss:sum}

Our first concrete application of the bulk spectral densities is the verification of a nontrivial sum rule, relating an integral over these quantities to the corresponding Euclidean correlators. To this end, we first subtract the zero-temperature limits from the spectral densities and define the \textit{thermal spectral functions} 
\ba
 \frac{\Delta\rho^{ }_\theta(\omega)}{4 d_A c_\theta^2 }
  \!\! & = & \!\!  
 \frac{\pi\omega^4}{(4\pi)^2}
 \biggl\{
 2 n_{\frac{\omega}{2}} \biggl[ 
   g^4 + \frac{g^6\Nc}{(4\pi)^2}
    \biggl(
       \frac{22}{3} \ln\frac{\bmu^2}{\omega^2} + \frac{73}{3}
    \biggr) \biggr]
 \nn&+& 
 \bigl( 1 + 2 n_{\frac{\omega}{2}} \bigr)
 \frac{8 g^6\Nc \phi^{ }_T(\omega) }{(4\pi)^2}  
 \biggr\}
 \;,  \la{Delta_theta}
 \\
 \frac{-\Delta\rho^{ }_\chi(\omega)}{16 d_A c_\chi^2 }
  \!\! & = & \!\! 
 \frac{\pi\omega^4}{(4\pi)^2}
 \biggl\{
 2 n_{\frac{\omega}{2}} \biggl[ 
   g^4 + \frac{g^6\Nc}{(4\pi)^2}
    \biggl(
       \frac{22}{3} \ln\frac{\bmu^2}{\omega^2} + \frac{97}{3}
    \biggr) \biggr]
 \nn &+& 
 \bigl( 1 + 2 n_{\frac{\omega}{2}} \bigr)
 \frac{8 g^6\Nc \phi^{ }_T(\omega) }{(4\pi)^2}  
 \biggr\}
 \;.  \la{Delta_chi}
\ea
Using these expression, the sum rules, read off from \cite{Laine:2010tc}, take the form (cf.~\cite{Laine:2011xm} for details)
\ba
  \frac{\Delta \tilde G_\theta(0)}{4d_A c_\theta^2} 
   & = & 2 \int_{0}^{\infty} \! \frac{{\rm d}\omega}{\pi\omega} 
 \,  \biggl[ \frac{\Delta\rho^{ }_\theta(\omega)}{4 d_A c_\theta^2 } \biggr]
 + \lk\frac{\Delta \tilde G_\theta(P)}{4d_A c_\theta^2}\rk_\text{cont} + \rmO(g^8) 
 \;, \la{sum_theta} \\
  \frac{\Delta \tilde G_\chi(0)}{-16d_A c_\chi^2} 
  & = & 2 \int_{0}^{\infty} \! \frac{{\rm d}\omega}{\pi\omega} 
 \, \biggl[ \frac{-\Delta\rho^{ }_\chi(\omega)}{16 d_A c_\chi^2 } \biggr] 
 + \lk \frac{\Delta \tilde G_\chi(P)}{-16d_A c_\chi^2} \rk_\text{cont} + \rmO(g^8) 
 \;, \la{sum_chi}
\ea
where the quantities $\Delta\tilde{G}$ denote the vacuum-subtracted Euclidean Green's functions. The last terms on the right hand side (RHS) denote so-called contact terms, \ie~the UV limits of the Euclidean correlator, to which the spectral function is blind.

To inspect the validity of the above relations, we note that the left hand sides (LHS) of the sum rules are next to trivial to evaluate using \eqs\nr{Gtheta_bare} and \nr{Gchi_bare}, producing 
\ba
  \frac{\Delta \tilde G_\theta(0)}{4d_A c_\theta^2} 
   &=&-2g^6\Nc(D-2)^2\mathcal{I}^{0}_\rmi{a}(0)
   =   - \frac{g^6 \Nc T^4}{18}  \;, \label{LHS1} \\
  \hspace{-1cm} \frac{\Delta \tilde G_\chi(0)}{-16d_A c_\chi^2} 
  & = & 0 
 \;. \label{LHS2}
\ea
At the same time, the contact terms can be determined using the simple formulae $\int_\vec{q} q \, \nB{}(q) = \pi^2 T^4/30$, $\int_\vec{q} \nB{}(q)/q = T^2/12$ and recalling that $P=(p_n,\vec{0})$; these terms can then be read off from \eqs\nr{Gtheta_uv} and \nr{Gchi_uv} and are seen to produce
\ba
 \frac{\Delta \tilde G_\theta(P)}{4d_A c_\theta^2} 
 & \stackrel{p_n \gg \pi T }{=} 
 & - \frac{ 4 \pi^2 g^4 T^4 }{15} 
 \biggl[
       1 + \frac{g^2\Nc}{(4\pi)^2}
  \biggl(
    \frac{22}{3} \ln\frac{\bmu^2}{p_n^2} + \frac{203}{18} 
  \biggr) 
  \biggr] 
  \nn &&
 + \frac{g^6 \Nc T^4}{18} + \rmO\Bigl( \frac{g^4}{p_n^2} ,g^8 \Bigr)
 \;, \la{contact_theta} \\ 
 \frac{\Delta \tilde G_\chi(P)}{-16d_A c_\chi^2} 
 & \stackrel{p_n \gg \pi T }{=}
 & - \frac{ 4 \pi^2 g^4 T^4 }{15} 
 \biggl[
       1 + \frac{g^2\Nc}{(4\pi)^2}
  \biggl(
    \frac{22}{3} \ln\frac{\bmu^2}{p_n^2} + \frac{347}{18}
  \biggr) 
  \biggr] 
  \nn &&
 + \frac{g^6 \Nc T^4}{9} + \rmO\Bigl( \frac{g^4}{p_n^2} ,g^8 \Bigr)
 \;. \la{contact_chi} 
\ea

Collecting everything together, the RHSs of the sum rules
\nr{sum_theta} and \nr{sum_chi} obtain the forms
\ba
&&\hspace{-2cm}2 \int_{0}^{\infty} \! \frac{{\rm d}\omega}{\pi\omega} 
 \,  \biggl[ \frac{\Delta\rho^{ }_\theta(\omega)}{4 d_A c_\theta^2 } \biggr]
 + \lk\frac{\Delta \tilde G_\theta(P)}{4d_A c_\theta^2}\rk_\text{cont} 
 \nn &=&
2 \int_{0}^{\infty} \! \frac{{\rm d}\omega \, \omega^3}{(4\pi)^2} \biggl\{ 2 n_{\frac{\omega}{2}}  \biggl[  \frac{g^6\Nc}{(4\pi)^2}  \frac{22}{3} \ln\frac{\bmu^2}{\omega^2} \biggr]
 + 
 \bigl( 1 + 2 n_{\frac{\omega}{2}} \bigr)
 \frac{8 g^6\Nc \phi^{ }_T(\omega) }{(4\pi)^2}  
 \biggr\} \nn
 &-& \frac{ 4 \pi^2 g^4 T^4 }{15} 
 \biggl[
     \frac{g^2\Nc}{(4\pi)^2}
  \biggl(
    \frac{22}{3} \ln\frac{\bmu^2}{p_n^2} - \frac{235}{18} 
  \biggr) 
  \biggr] 
 + \frac{g^6 \Nc T^4}{18} 
 \;, \la{sumrhs_theta} \\
&&\hspace{-2cm} 2 \int_{0}^{\infty} \! \frac{{\rm d}\omega}{\pi\omega} 
 \, \biggl[ \frac{-\Delta\rho^{ }_\chi(\omega)}{16 d_A c_\chi^2 } \biggr] 
 + \lk \frac{\Delta \tilde G_\chi(P)}{-16d_A c_\chi^2} \rk_\text{cont}
 \nn &=&
2 \int_{0}^{\infty} \! \frac{{\rm d}\omega \, \omega^3}{(4\pi)^2} \biggl\{ 2 n_{\frac{\omega}{2}}  \biggl[  \frac{g^6\Nc}{(4\pi)^2}  \frac{22}{3} \ln\frac{\bmu^2}{\omega^2} \biggr]
 + 
 \bigl( 1 + 2 n_{\frac{\omega}{2}} \bigr)
 \frac{8 g^6\Nc \phi^{ }_T(\omega) }{(4\pi)^2}  
 \biggr\}
 \nn &-& \frac{ 4 \pi^2 g^4 T^4 }{15} 
 \biggl[
     \frac{g^2\Nc}{(4\pi)^2}
  \biggl(
    \frac{22}{3} \ln\frac{\bmu^2}{p_n^2} - \frac{235}{18}
  \biggr) 
  \biggr] 
 + \frac{g^6 \Nc T^4}{9} 
 \;. \la{sumrhs_chi}
\ea
Comparing these expressions to the LHSs, \eqs(\ref{LHS1}) and (\ref{LHS2}), we see that both sum rules reduce to the verification of the common but nontrivial relation
\ba
&&\hspace{-2cm}2 \int_{0}^{\infty} \! \frac{{\rm d}\omega \, \omega^3}{(4\pi)^2}  \biggl\{ 2 n_{\frac{\omega}{2}}  \biggl[ \fr{g^6 \Nc}{(4\pi)^2} \frac{22}{3} \ln\frac{\bmu^2}{\omega^2} \biggr]
 + 
\bigl( 1 + 2 n_{\frac{\omega}{2}} \fr{8 g^6 \Nc \phi^{ }_T (\omega)}{(4\pi)^2} \bigr)  
 \biggr\}
 \nn&=&  \frac{ 4 \pi^2 g^4 T^4 }{15} 
 \biggl[
 \fr{g^2 \Nc}{(4\pi)^2} \biggl(
    \frac{22}{3} \ln\frac{\bmu^2}{p_n^2} - \frac{335}{18}
  \biggr) \biggr]
 \;, \la{sum_com}
\ea
which however contains a divergence in the integral due to the UV limit of $\phi_T(\omega)$.

To reduce the above relation to a form that allows for numerical verification, we finally introduce a UV regulator $\Lambda_c$ in the integral and replace $\ln p_n^2\to \ln \Lambda_c^2$. Factoring out a factor $g^6 \Nc/(4\pi)^2$ from both sides of \eq\nr{sum_com}, the sum rule then transfers to
\ba
 &&\hspace{-1.2cm}\lim_{\Lambda_c\to\infty}
 \biggl\{ 
 2 \int_0^{\Lambda_c} 
 \! \frac{{\rm d}\omega \, \omega^3}{(4\pi)^2}
 \biggl[ 
   2 n_{\frac{\omega}{2}}
    \biggl(
       \frac{22}{3} \ln\frac{\bmu^2}{\omega^2} 
    \biggr)
  + 8  \bigl( 1 + 2 n_{\frac{\omega}{2}} \bigr)
  \phi^{ }_T(\omega)
 \biggr]
 - \frac{4\pi^2 T^4}{15}
 \biggl( 
       \frac{22}{3} \ln\frac{\bmu^2}{\Lambda_c^2} 
 \biggr) \biggr\} 
 \nn  
 &=& \frac{4\pi^2 T^4}{15}
 \biggl( 
        - \frac{355}{18}
 \biggr)
 \;, \hspace{1cm}
\ea
which we indeed have verified to a good numerical precision.

\begin{figure}[t]


\centerline{%
\includegraphics[width=7cm]{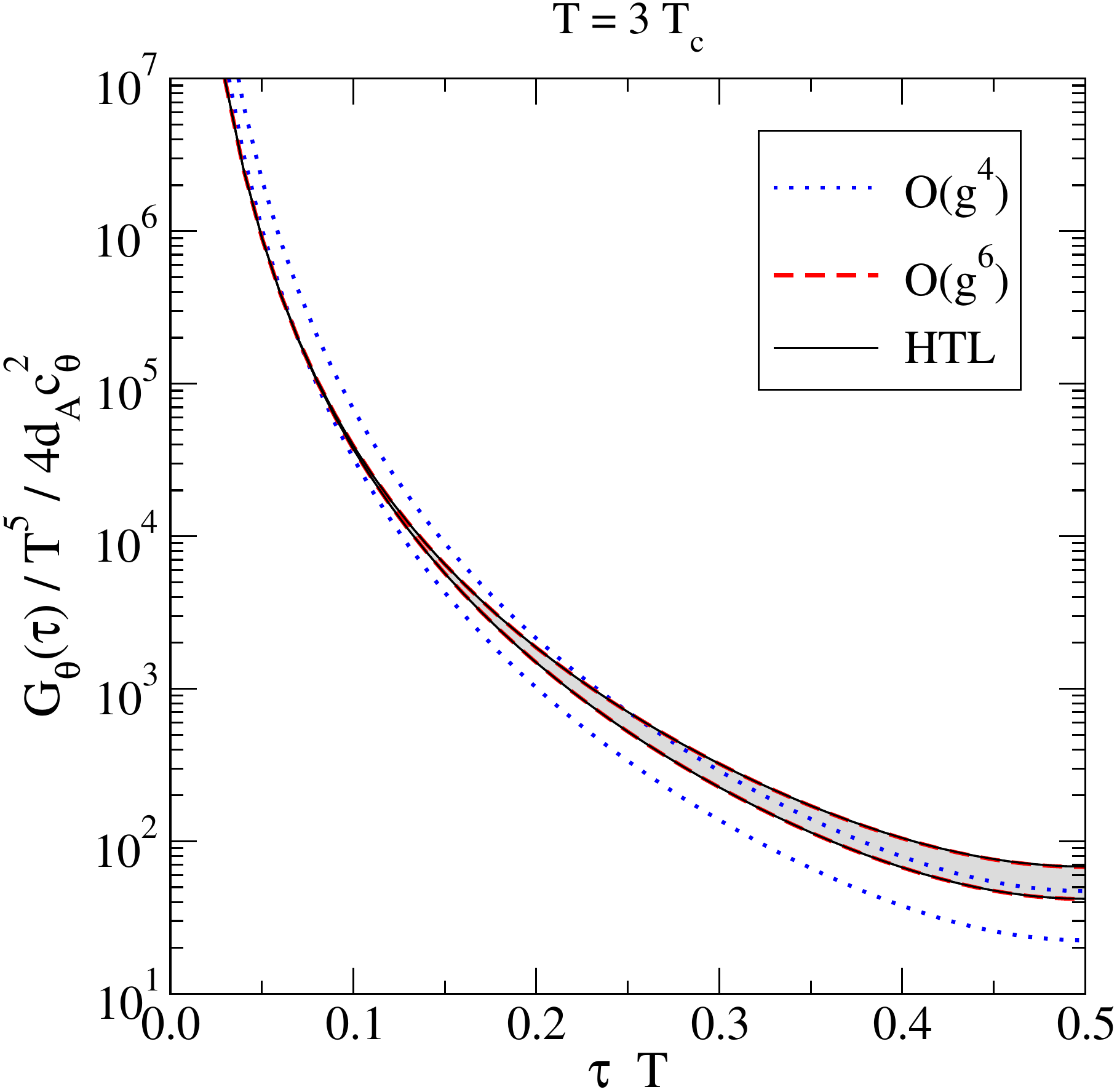}%
~~\includegraphics[width=7cm]{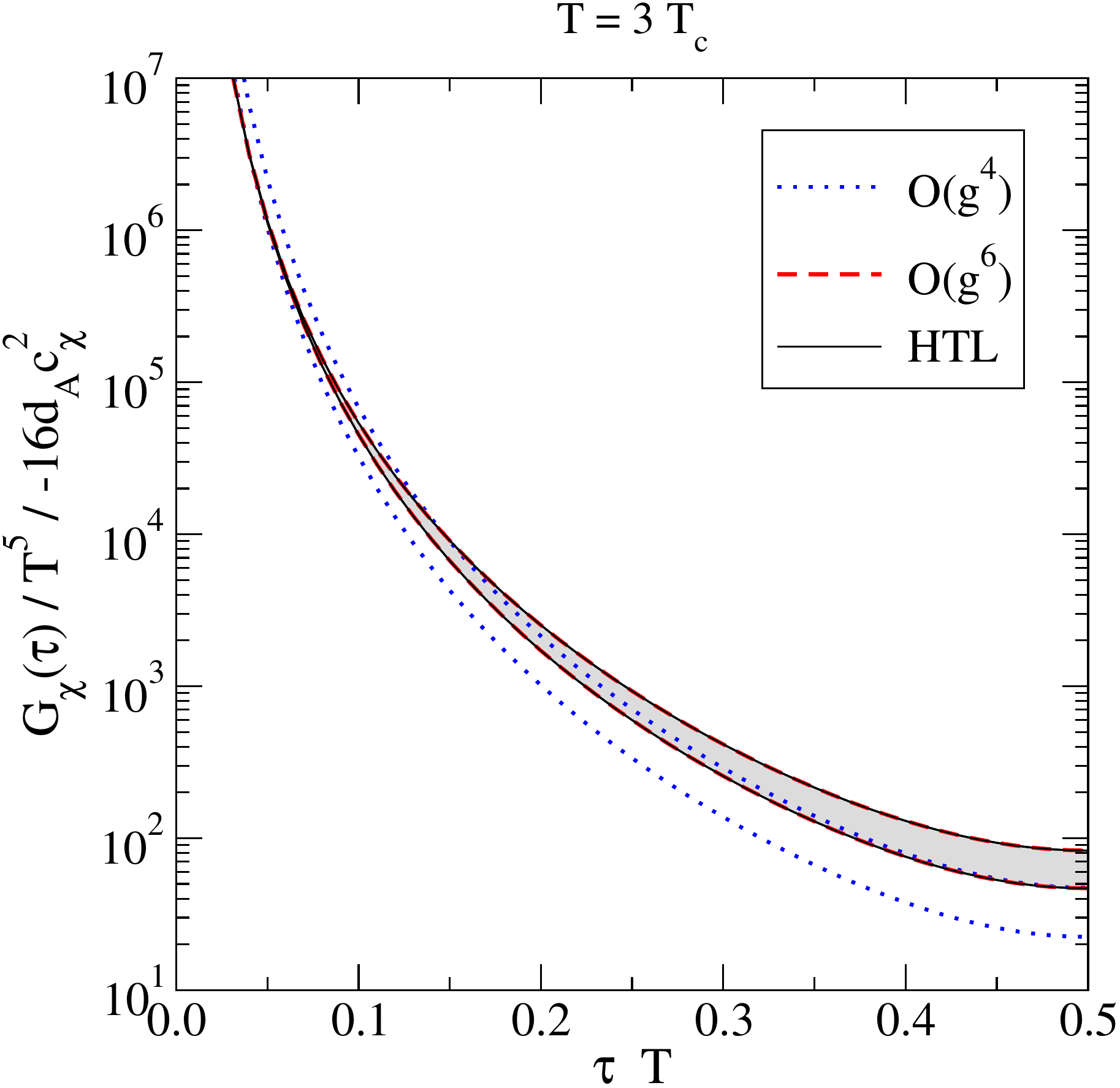}%
}

\caption[a]{\small 
  A numerical evaluation of $G_{\theta}$ (left) and $G_{\chi}$ (right)
  in units of  $T^5$, for $T = 3.75 \Lambdamsbar$, or $T = 3 \Tc$.
  The figure is taken from \cite{Laine:2011xm}.
 }
\la{fig:taudep}
\end{figure}

\begin{figure}[h!]


\centerline{%
\includegraphics[width=7cm]{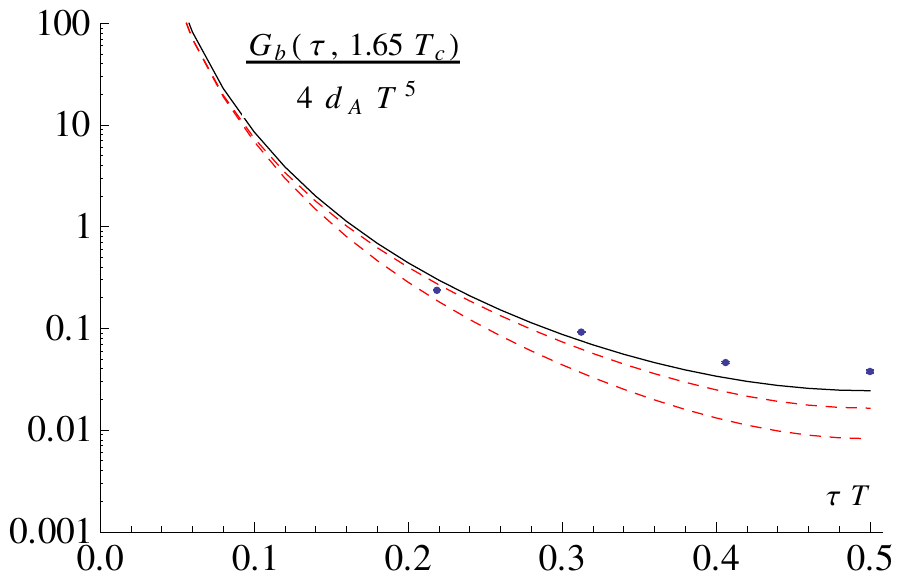}%
~~\includegraphics[width=7cm]{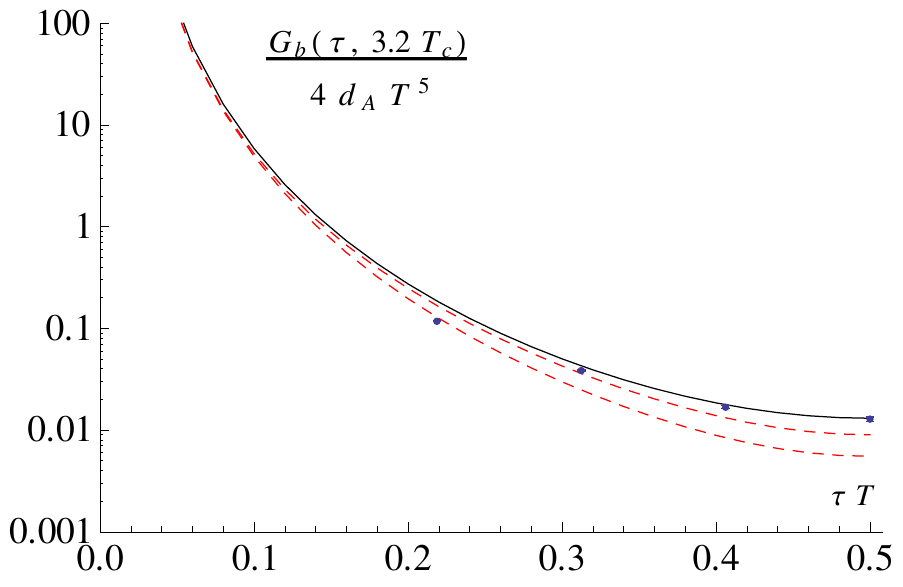}%
}
\caption[a]{\small 
The $\theta$ channel imaginary time correlator evaluated in perturbation theory (red dashed curves) and IHQCD (solid black curves) \cite{Kajantie:2013gab} for two different temperatures, and compared with the lattice data of \cite{Meyer:2010ii}. Note that the IHQCD result corresponds to the large-$N_c$ limit of the theory. The figure is taken from \cite{Kajantie:2013gab}.
 }
\la{fig:IHQCD}
\end{figure}

\subsection{Imaginary time correlators}
\la{ss:tau}

\begin{figure}[t]


\centerline{%
\includegraphics[width=7cm]{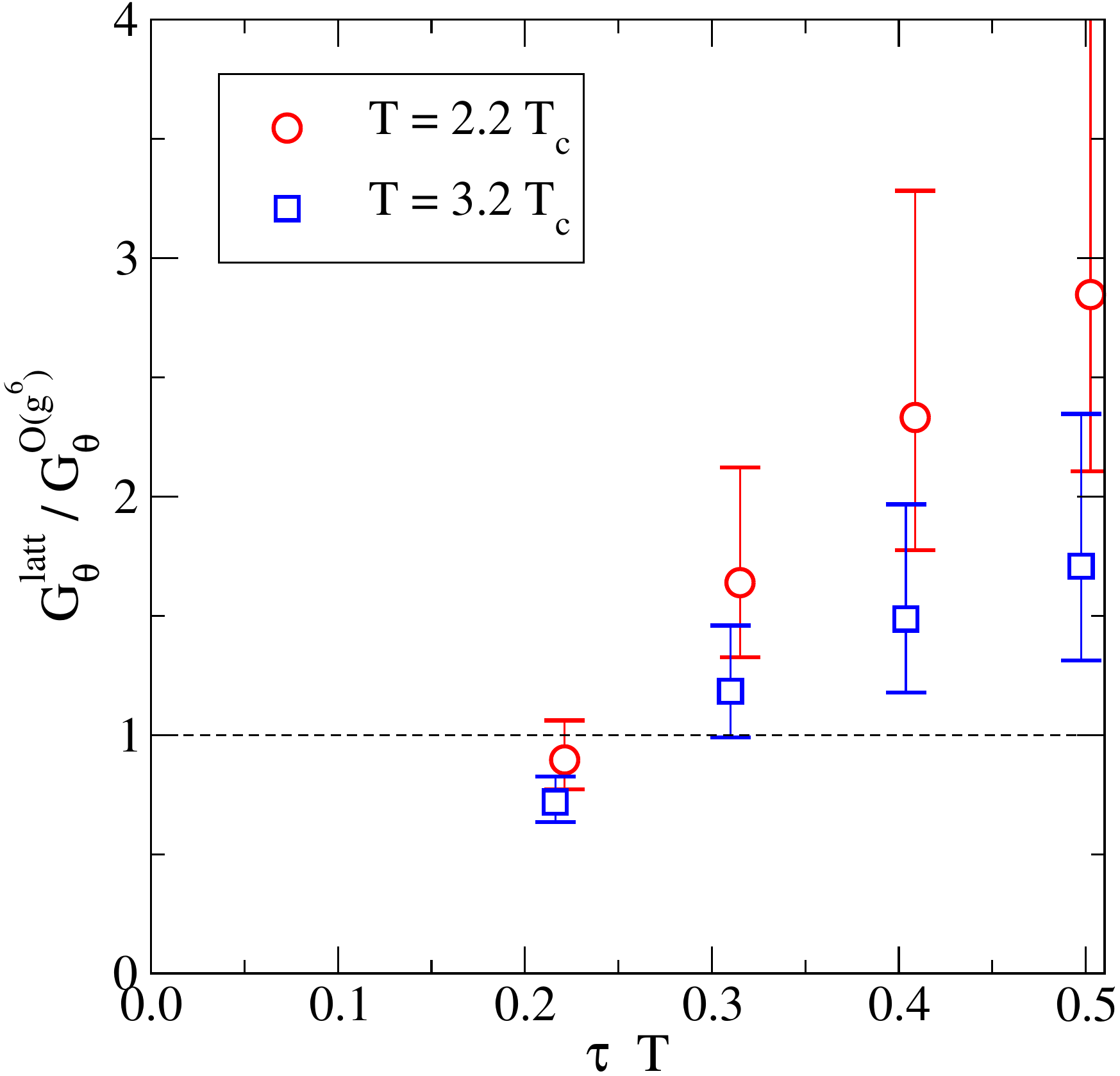}%
~~\includegraphics[width=7cm]{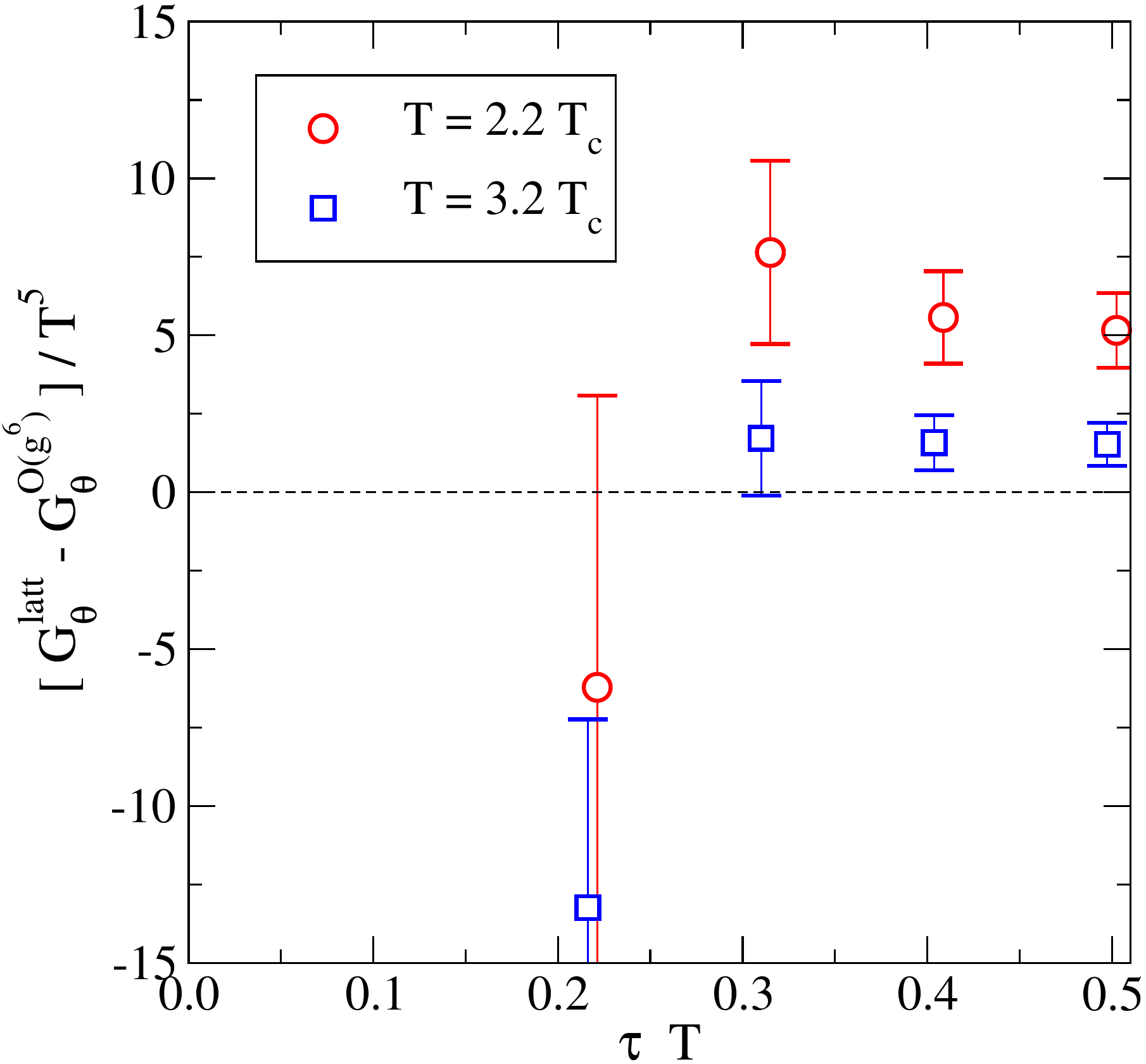}%
}
\caption[a]{\small 
  The ratio (left) and difference (right) of the lattice data of ref.~\cite{Meyer:2010ii} and our results from \fig\ref{fig:taudep} (left). The error bars here have been obtained by varying both results independently within their uncertainty bounds. The figure is taken from \cite{Laine:2011xm}.
 }
\la{fig:compare}
\end{figure}

Another obvious application of our results is the determination of Euclidean imaginary time correlators, which can be directly obtained from the spectral densities via \eq\nr{eq:intrel}. The main motivation for this exercise is to compare the results to lattice calculations such as  \cite{Meyer:2010ii}, where the Euclidean correlators are determined (see also \cite{Laine:2010fe}, where closely related Euclidean spatial correlators are evaluated perturbatively). As discussed at length in the previous sections, the ultimate purpose of these comparisons is to aid the building of nonperturbative spectral functions, from which transport coefficients such as the bulk viscosity can hopefully eventually be extracted. 

Before moving to the calculation of the imaginary time correlator, we let us stress one issue. As we have highlighted above, our calculation misses all contact terms of the form $\omega^n\delta(\omega)$, $n=1,2,...$, of which the terms with $n=1$ would in principle contribute to the imaginary time correlator. In the present NLO bulk channel calculation, these terms, however, only occur with $n>1$, so we do not need to worry about them; at one order higher, they are nevertheless expected to appear, so there one must exercise more caution \cite{Moore:2008ws,Meyer:2010ii,Romatschke:2009ng}. The same applies to the NLO shear channel calculation, which we will comment on below.

In \fig\ref{fig:taudep}, we finally show our numerical results for the imaginary time correlators. It is quite remarkable, how the significantly different IR behaviors of the spectral functions at $\rmO(g^4)$ and $\rmO(g^6)$ are seen to be almost entirely washed out by the UV contributions, leading to nearly identical results. In addition, the sizable IR suppression of $\rho^{ }_\chi$ relative to $\rho^{ }_\theta$ (cf.~\fig \ref{fig:wdep}) is no longer distinguishable from the corresponding Euclidean correlators.

Finally, we note that the bulk correlator $G^{ }_\theta / T^5$ has been studied within pure SU(3) lattice gauge theory in \cite{Meyer:2010ii}, and in large-$N_c$ Yang-Mills theory via the holographic IHQCD model in \cite{Kajantie:2013gab}. Here, we perform a simple comparison of our $N_c=3$ results with the IHQCD and lattice works, displaying in \fig\ref{fig:IHQCD} the imaginary time correlators for $T=1.65T_c$ and $T=3.2T_c$. In addition, both the ratio and difference of the perturbative and lattice correlators are shown in \fig\ref{fig:compare}. From here, we in particular see the impressive cancelation of the short distance singularity between the two results, indicating that a model independent analytic continuation of the lattice data (in the spirit of \cite{Burnier:2011jq}) could perhaps be attempted.

\mysection{Shear channel}\la{se:spf_shear}

In the shear channel, the NLO spectral function, derived in \cite{Zhu:2012be}, obtains a form very similar to that encountered in the bulk case, 
\ba
\frac{\rho_\eta(\omega)}{4d_A}&=&\frac{\omega^4}{4\pi}\bigl( 1 + 2 n_{\frac{\omega}{2}} \bigr)\Bigg\{-\frac{1}{10}+\frac{g^2N_c}{(4\pi)^2}\bigg(\frac{2}{9}+\phi_T^\eta(\omega/T)\bigg)\Bigg\}\; , \label{result1}
\ea
where $\phi_T^\eta(\omega/T)$ again denotes a rather complicated, numerically evaluatable function. The behavior of this function is displayed in fig.~\ref{res1}, while its explicit form in terms of a number of one- and  two-dimensional integrals is reproduced for convenience in the Mathematica file shearresults.nb, available at \cite{mathfile}.

\subsection{Limits and numerical evaluation}

\begin{figure}[t!]
\centering
\includegraphics[width=7.1cm]{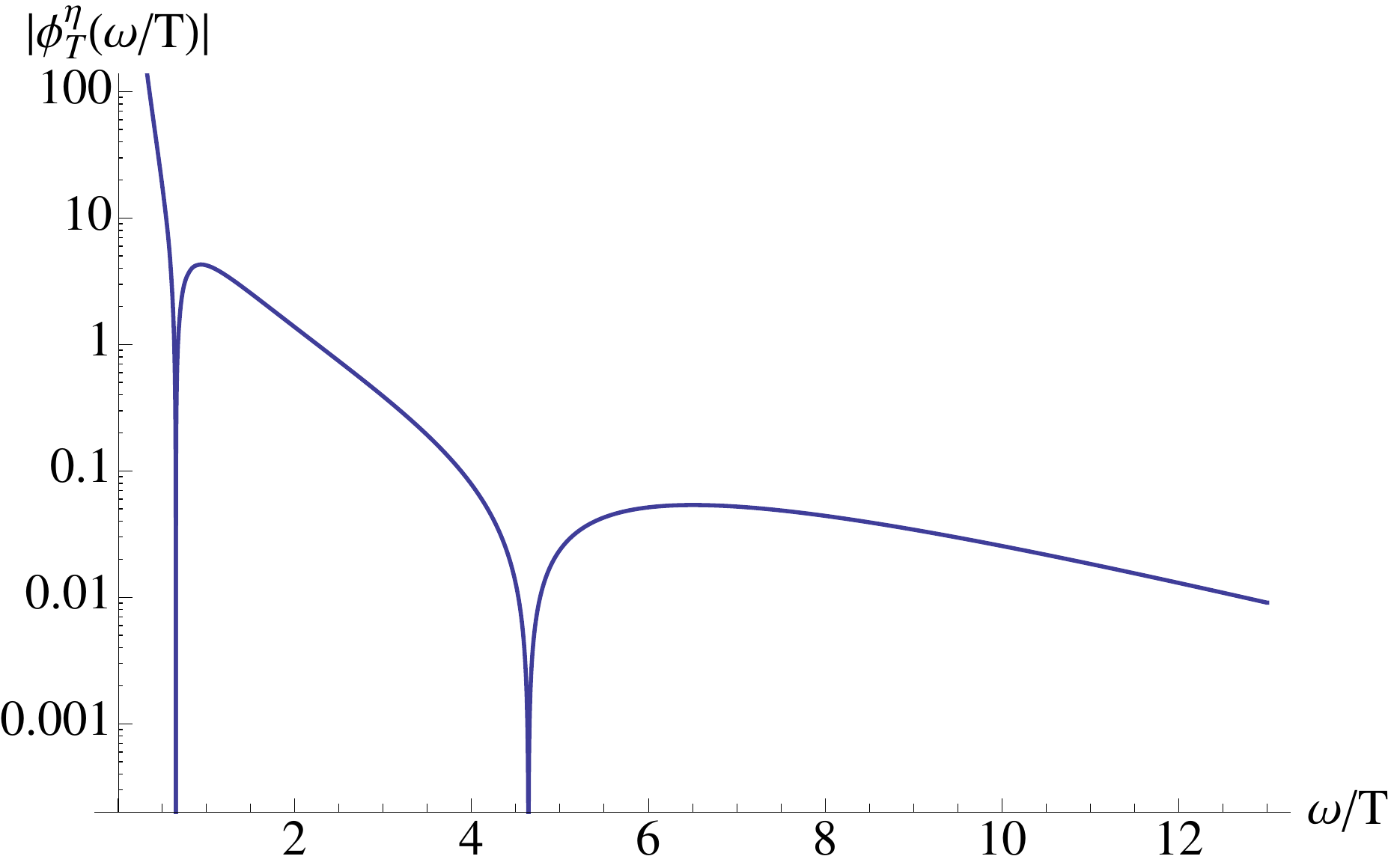}$\;\;\;$\includegraphics[width=7.1cm]{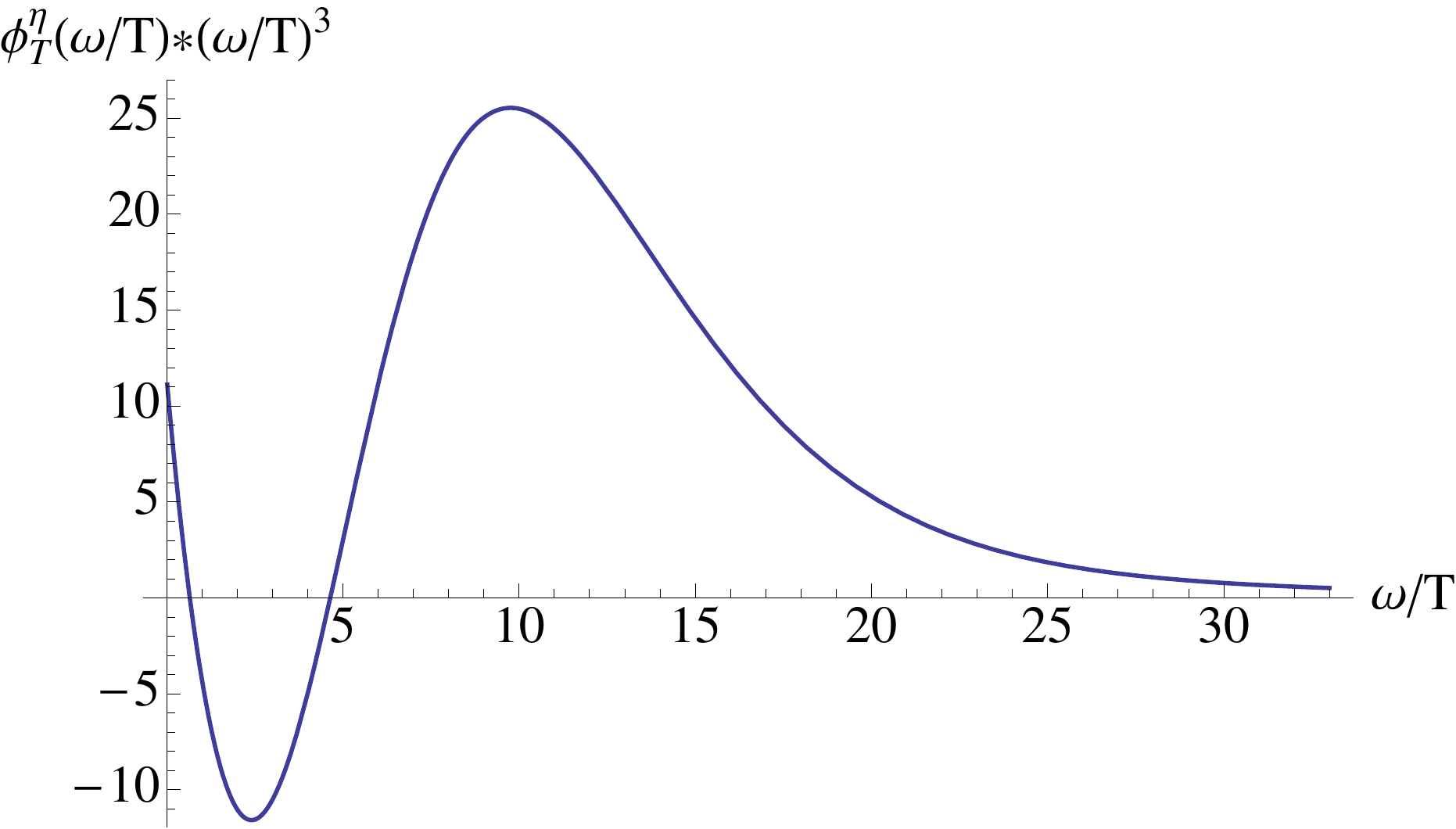}
\caption{\small The behavior of the function $\phi_T^\eta(\omega/T)$ on logarithmic and linear scales, multiplied by $(\omega/T)^3$ in the latter case. The figure is taken from \cite{Zhu:2012be}.} 
\label{res1}
\end{figure}

Just like in the bulk channel, we begin the inspection of our shear spectral function by performing a number of consistency checks on it. The first thing to note is clearly the automatic cancelation of the $1/\epsilon$ divergences in \eq(\ref{result1}), as well as the agreement of the constant terms with the $T=0$ results of \cite{Schroder:2011ht,Zoller:2012qv}. This in particular indicates that no renormalization is required at this order, as is in fact evident from the form of the shear operator. Moving then on to the UV (large $\omega$) behavior of the result, a straightforward calculation shows that the function $\phi_T^\eta(\omega)$ behaves in this limit as
\ba
\phi_T^\eta(\omega/T)&=&\frac{41\pi^6T^6}{3\omega^6} +{\mathcal O}(T^8/\omega^8)\; .
\ea
The form of this result in particular confirms the (completely independent) prediction of \cite{CaronHuot:2009ns} concerning the short distance behavior of the shear correlator, and thus serves as a highly nontrivial crosscheck of our calculation.

In the opposite limit of $\omega \ll  T$, the function $\phi_T^\eta(\omega)$ on the other hand tends to the simple limit 
\ba
\phi_T^\eta(\omega)&=&\frac{7\pi^2T^3}{6\omega^3} +{\mathcal O}(T^2/\omega^2)\; , \label{asym}
\ea
implying that the NLO shear spectral function approaches a (positive) constant of ${\mathcal O}(g^2)$ as $\omega\to 0$. This is in clear contrast with the corresponding limit of the bulk result, cf.~section \ref{se:spf_bulk}, and in fact does not represent the physical IR behavior of the quantity.\footnote{See however ref.~\cite{Kovtun:2011np}, where the authors suggest that the true $\omega\to 0$ limit of the spectral function is a constant of ${\mathcal O}(g^8)$.} As discussed already in the previous section, in order to proceed to frequencies of the order $\omega\sim gT$ (let alone to an even softer regime), one namely must implement an HTL (and ultimately even more complicated) resummation in the correlator, which we have not done. Thus, the applicability of our results is limited to frequencies of order $T$ and higher, which (as we will argue in sec.~\ref{se:imcorr_shear}) however only poses a minor limitation in the determination of the imaginary time shear correlator.

\begin{figure}
\centering
\includegraphics[width=10cm]{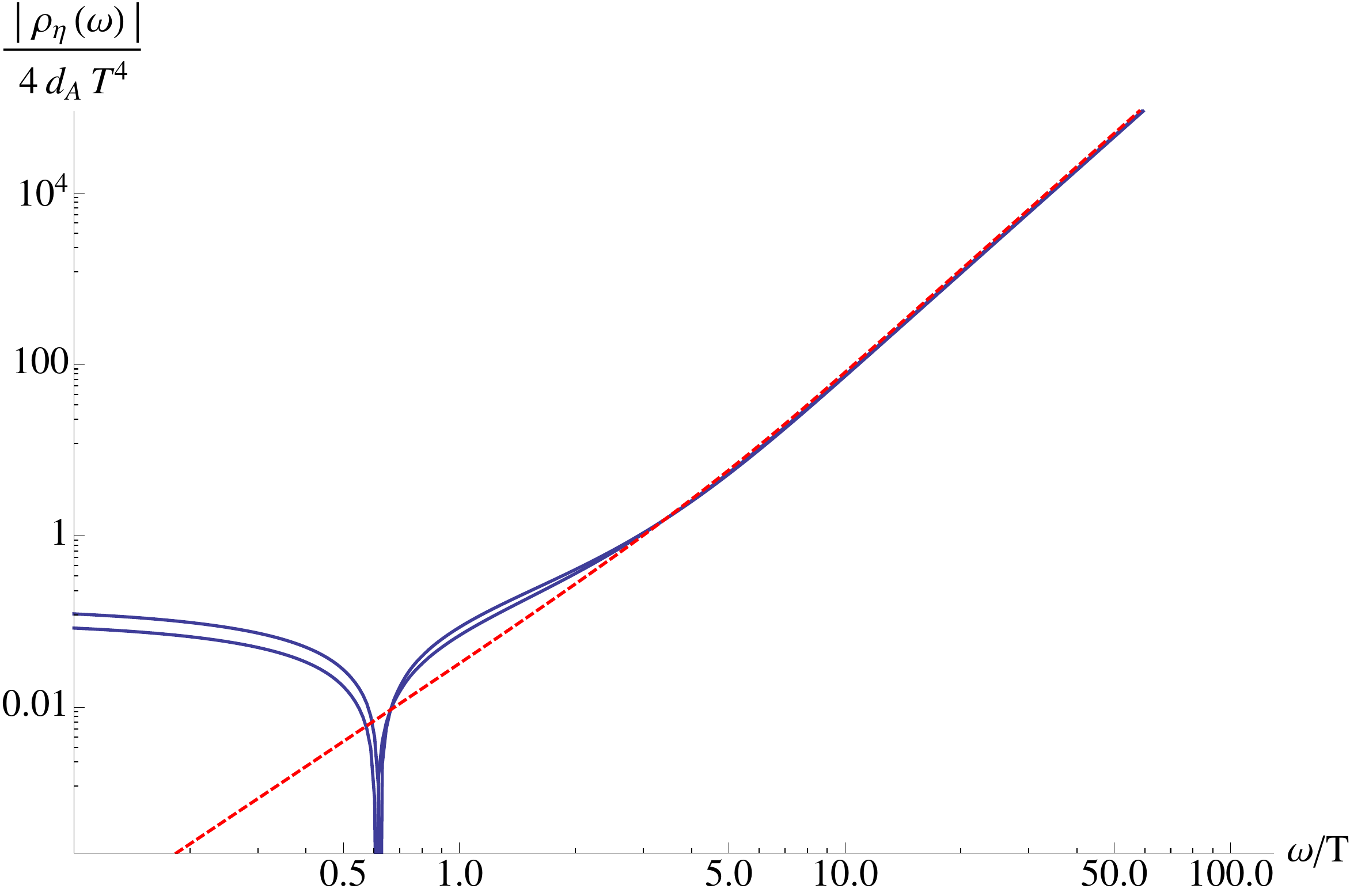}
\caption{\small The behavior of the absolute value of the shear spectral function, $|\rho_\eta(\omega)|/(4d_A T^4)$, for $T=3T_c$ (\ie~$3.75\Lambda_\tinymsbar$). The two blue curves stand for the NLO result evaluated with $\bar{\Lambda}=0.5\bar{\Lambda}^\text{opt}$ and $\bar{\Lambda}=2\bar{\Lambda}^\text{opt}$, while the dashed red curve shows the leading order (LO) result. The spikes in the NLO results correspond to changes of sign in the quantities from positive to negative (with increasing $\omega$). The figure is taken from \cite{Zhu:2012be}.}
\label{res2}
\end{figure}

Having the two limits of small and large $\omega$ under control, we next plug the numerically evaluated $\phi_T^\eta(\omega)$ into \eq\nr{result1}, and proceed to plot the spectral function. Setting $N_c=3$, using the two-loop running coupling, and choosing the renormalization scale to have the EQCD value (note that the fastest apparent convergence scheme is not available in this channel), we obtain the result shown in fig.~\ref{res2}. This time we note that the uncertainty associated with varying the renormalization scale by a factor of two is almost negligible, as is the difference between the LO and NLO results at large $\omega$. It is interesting to contrast this result to the recent IHQCD calculation of \cite{Kajantie:2013gab}, where a qualitatively highly similar behavior for the spectral function was obtained.

\subsection{Sum rule} \la{se:imcorr_shear}

In the shear channel, a sum rule similar to the one discussed above for the bulk case has been derived in \cite{Romatschke:2009ng,Meyer:2010gu}. Translated to our notation, it reads
\ba
-\fr{1}{16\pi c_\eta^2}\int_{-\infty}^{\infty}\fr{{\rm d}\omega}{\omega} \Big\{\rho_\eta(\omega)-\rho_\eta(\omega)\!\mid_{T=0}\Big\}+\lim_{\omega\rightarrow \infty}{\mathcal G}(\omega,T)=\fr{2}{3}e(T) \,, \label{rule}
\ea
where the Minkowskian correlator appearing in the contact term, ${\mathcal G}(\omega,T)$, is defined to be
\ba
{\mathcal G}(\omega,T)\equiv -\frac{\Delta\tilde G_\eta(-i\omega,\mathbf{p}=0)}{16 c_\eta^2}-\fr{2}{3}\int {\rm d}^4x\, e^{i\omega x_0}\langle T_{00}(x)T_{00}(0) \rangle_{T}\, , \label{Gdef}
\ea
with the subscript $T$ denoting the thermal part of the correlator. 

Unlike in the bulk channel, it turns out that a direct verification of the above sum rule via integration of our NLO spectral density is rather problematic. This is due to two independent reasons. First, as noted above, our result for the spectral density approaches an unphysical constant in the $\omega\to 0$ limit, rendering the integral in \eq(\ref{rule}) IR divergent (even in the principal value sense). And second, while in the bulk channel no contributions of the form $\omega\delta(\omega)$ arose, it is easy to see that in the shear channel this is no longer the case. This means that there are constant contributions to the sum rule that have been lost in our calculation. 

Fortunately, even though the verification of the above sum rule is not feasible within our current computation, it turns out to be possible to check a similar relation, cf.~\eq(4.17) of \cite{Schroder:2011ht},
\ba
\fr{1}{\pi}\int_{-\infty}^{\infty}\fr{{\rm d}\omega}{\omega} \Big\{\rho_x(\omega)-\rho_x(\omega)\!\mid_{T=0}\Big\}=
\tilde G_x(0)-\lim_{\omega\rightarrow \infty} \Delta\tilde G_x(\omega,\mathbf{p}=0),
\ea
 for a number of IR safe master spectral functions $\rho_x$. This way, we have been able to independently check the correctness of e.g.~all master integrals of j type, as well as most of the type h's --- constituting the two most complicated classes of diagrams. This serves as one of the most powerful checks of our calculation.

\subsection{Imaginary time correlator} \la{se:imcorr_shear}

\begin{figure}
\centering
\includegraphics[width=10cm]{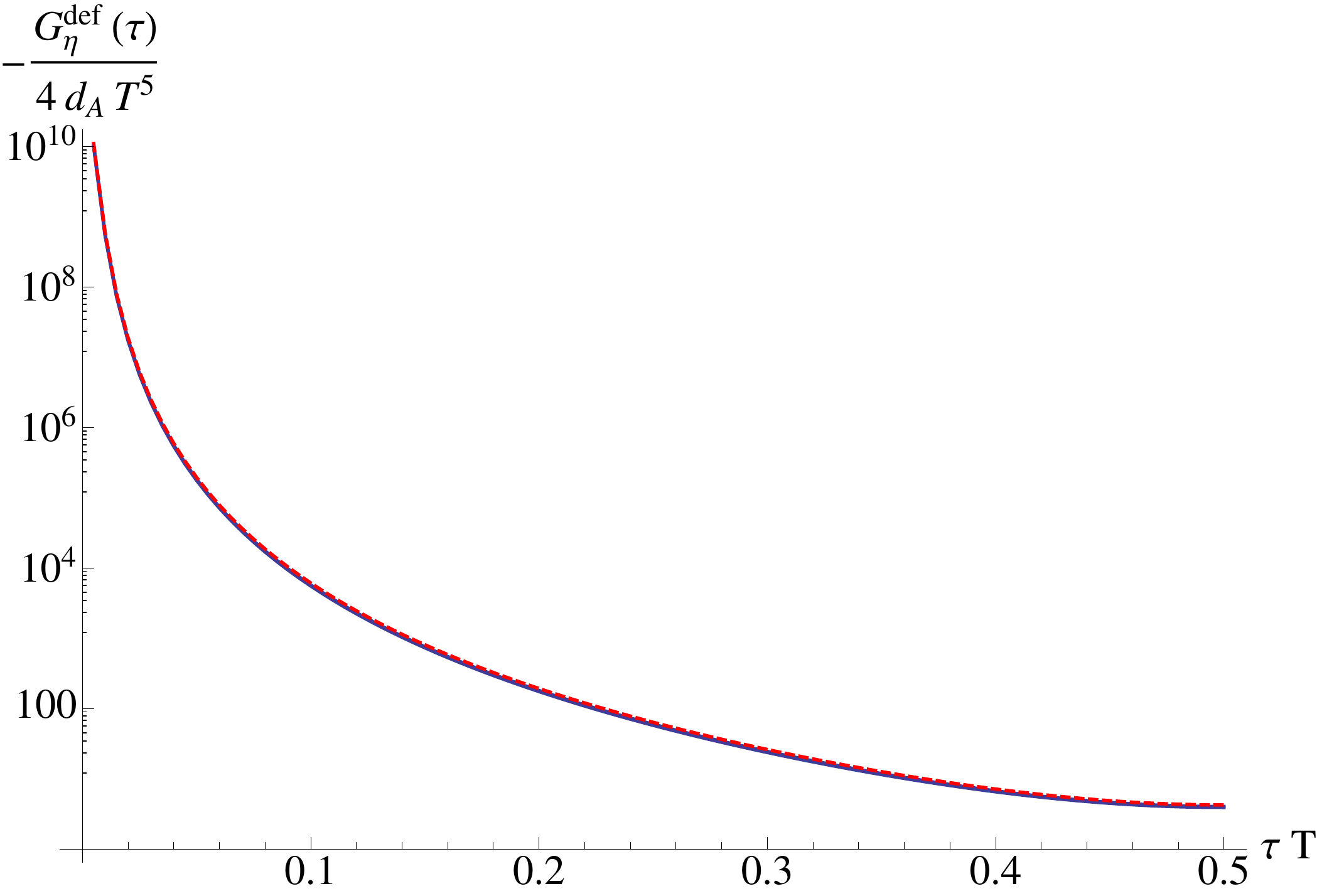}
\caption{\small The imaginary time correlator $G_\eta^\text{def}(\tau)$, defined in \eq(\ref{tau1}) and displayed on a logarithmic scale. Just as in fig.~\ref{res2}, altogether three curves are displayed here, but as is apparent form the figure, the two blue NLO curves practically overlap with the dashed red (LO) one. The figure is taken from \cite{Zhu:2012be}.}
\label{res3}
\end{figure}

Next, we apply our shear spectral function to the derivation of the corresponding Euclidean imaginary time correlator, again available by carrying out the integral in \eq(\ref{eq:intrel}). While the logarithmic IR divergence in principle again poses a problem, we may here use the insights gained in the bulk channel, suggesting that the contribution of the frequencies $\omega \lesssim T$ to the correlator is almost negligible. To this end, we define a deformed correlation function
\begin{equation}
 G_\eta^\text{def}(\tau) =
 \int_{\omega_0}^\infty
 \frac{{\rm d}\omega}{\pi} \rho_\eta(\omega)
 \frac{\cosh\Big[\! \left(\frac{\beta}{2} - \tau\right)\omega\Big]}
 {\sinh\frac{\beta \omega}{2}}\, ,\quad \quad 0<\tau <\beta \, , \label{tau1}
\end{equation}
where $\omega_0\approx 0.6T$ stands for the frequency, at which the NLO spectral function changes its sign. This is in fact a highly natural choice, as below this value the LO spectral function by definition obtains a relative correction of more than 100$\%$. It must, however, be recalled that in addition to missing the contribution of this frequency range, our Euclidean correlator will also miss the constant contribution originating from the omitted $\omega\delta(\omega)$ term.

With the above caveats in mind, we display the result for the deformed imaginary time correlator in fig.~\ref{res3}, observing a surprisingly good agreement between the LO and NLO results.
This behavior is clearly explained on one hand by the small relative size of the NLO correction to the $T=0$ spectral function, cf.~\eq(\ref{result1}), and on the other hand by the vanishing of the leading temperature dependent corrections to the quantity at large $\omega$.

%

\pagestyle{plain}
\setcounter{chapter}{5}
%
%
%
%
%
%

\mychapter{Summary and conclusions}
\label{CHAP 6}

The theoretical and experimental study of the Quark Gluon Plasma (QGP) has become a hot topic of particle and nuclear physics in the past three decades. In particular, the last ten years or so have seen a very rapid development in the field, driven by high-energy heavy ion collision experiments first at RHIC (BNL) and later at the LHC (CERN). A systematic analysis of the experimental data, such as the multiplicities and energies of the emitted hadrons, (di)leptons and photons, has been combined with both phenomenological and first principles theoretical work to build a sort of ``Standard Model'' of heavy ion physics, which is able to describe all stages of the collision process at least on a semi-quantitative level.

Despite all the progress achieved, much remains to be done on both the experimental and theory sides, before we can claim to have a quantitative understanding of the entire heavy ion collision process. In the thesis at hand, we have concentrated on one particular part of the collision, namely the evolution of a (near) thermal QGP, which can be described using numerical hydrodynamic simulations. One of the most important lessons from RHIC data has been that the success of these simulations depends crucially on the inclusion of a small, yet nonzero value for the shear viscosity, necessary to describe e.g.~the elliptic flow at large transverse momenta. Subsequently, a lot of attention has focused on the first principles determination of this parameter in thermal QCD.

Unfortunately, it has turned out that to predict the value of the shear viscosity in real life QGP is a formidable challenge. On one hand, perturbation theory is badly convergent and extremely complicated to apply for this quantity. At the same time, lattice QCD is by nature restricted to the Euclidean formulation of the theory, and is thus not available for a direct extraction of transport coefficients. As of today, the most promising approach to solve the problem appears to be a novel combination of lattice and perturbative tools \cite{Burnier:2011jq}, involving a complicated analytic continuation of Euclidean lattice data to Minkowskian signature in order to determine a nonperturbative shear spectral function. In this process, a key ingredient is obtaining as much analytic information of the UV limit of this quantity as possible, which in practice is most easily accomplished using weak coupling tools. 

The main scientific goal of the thesis at hand has been to address the very problem explained above: To develop and apply analytic machinery for the investigation of thermal spectral functions. We have done so using perturbation theory, setting as our goal the determination of Next-to-Leading Order (NLO) spectral functions in both the bulk and shear channels of SU($N$) Yang-Mills theory. This goal was indeed achieved in a series of papers \cite{Schroder:2011ht,Laine:2011xm,Zhu:2012be}, upon which the original scientific content of the thesis is based. 

In addition to introducing the calculations and results of \cite{Schroder:2011ht,Laine:2011xm,Zhu:2012be}, cf.~chapters \ref{CHAP 3} - \ref{CHAP 5}, our purpose in preparing the thesis has been to provide a concise introduction to the general topic of transport phenomena in heavy ion collisions. This has included introducing the most fundamental machinery of thermal QCD in chapter \ref{CHAP 1}, as well as explaining the basics of the hydrodynamic description of the QGP and the role that the shear and bulk viscosities play in it (cf.~chapter \ref{CHAP 2}). Through this, we have tried to in particular motivate the work performed so far, but also to offer some views for the future.

As we have repeatedly discussed in the previous chapters, the main motivation behind our efforts has been the hope that one day our results can be combined with lattice data on Euclidean correlators to yield nonperturbative predictions for the bulk and shear viscosities of Yang-Mills theory --- and later QCD. This is an ambitious task, requiring much further work in particular on lattice QCD, but we believe that the present results constitute an important step in this direction. As natural follow-up projects, it would clearly be important to supplement the present shear channel calculation with a physically correct treatment of the IR limit of the spectral function, as well as to generalize all existing results to the presence of dynamical fermions. For the latter project, a particular motivation stems from the prediction of \cite{CaronHuot:2009ns} that unlike in the case of pure Yang-Mills theory, for full QCD the leading finite-temperature correction to the shear spectral function should enter already at order $1/\omega^2$, and thus play a much more pronounced role e.g.~in the behavior of Euclidean correlators.

%

\setcounter{chapter}{0}
\pagestyle{myheadings}
\begin{appendix}
\myappendix

\mychapter{Master integrals} \label{app:master}
\pagestyle{myheadings}

The master integrals appearing in the result (\ref{Gtheta_bare}), (\ref{Gtheta_bare}) and (\ref{Gchi_bare}) read
\begin{align}
 \mathcal{J}^{0}_\rmi{a} & \equiv
 \Tint{Q} \frac{P^2}{Q^2}
 \;,  \la{m_first} \\
 \mathcal{J}_\rmi{a}^1 & \equiv
 \Tint{Q} \frac{P_T(Q)}{Q^2}
 \;, \la{Ja1}  \displaybreak[0] \\
 \mathcal{J}_\rmi{b}^0 & \equiv
 \Tint{Q} \frac{P^4}{Q^2(Q-P)^2}
 \;, \la{Jb0}  \displaybreak[0] \\
 \mathcal{J}_\rmi{b}^1 & \equiv
 \Tint{Q} \frac{P^2}{Q^2(Q-P)^2}P_T(Q)
 \;, \la{Jb1} \displaybreak[0] \\
 \mathcal{J}_\rmi{b}^2 & \equiv
 \Tint{Q} \frac{1}{Q^2(Q-P)^2}P_T(Q)^2
 \;, \la{Jb2}\displaybreak[0] \\
 \mathcal{I}^{0}_\rmi{a} & \equiv
 \Tint{Q,R} \frac{1}{Q^2R^2}
 \;,\\
 \mathcal{I}_\rmi{b}^0 & \equiv
 \Tint{Q,R} \frac{P^2}{Q^2R^2(R-P)^2}
 \;,\displaybreak[0] \\
 \mathcal{I}_\rmi{b}^1 & \equiv
 \Tint{Q,R} \frac{1}{Q^2R^2(R-P)^2}P_T(Q)
 \;,\displaybreak[0] \\
 \mathcal{I}_\rmi{b}^2 & \equiv
 \Tint{Q,R} \frac{1}{Q^2R^2(R-P)^2}P_T(R)
 \;,\displaybreak[0] \\
 \mathcal{I}^{0}_\rmi{c} & \equiv 
 \Tint{Q,R} \frac{P^2}{Q^2R^4}
 \;, \\
 \mathcal{I}_\rmi{d}^0 & \equiv
 \Tint{Q,R} \frac{P^4}{Q^2R^4(R-P)^2}
 \;, \displaybreak[0] \\
 \mathcal{I}_\rmi{d}^1 & \equiv
 \Tint{Q,R} \frac{P^2}{Q^2R^4(R-P)^2}P_T(Q)
 \;,\displaybreak[0] \\
 \mathcal{I}_\rmi{d}^2 & \equiv
 \Tint{Q,R} \frac{P^2}{Q^2R^4(R-P)^2}P_T(R)
 \;,\displaybreak[0] \\
 \mathcal{I}_\rmi{d}^3 & \equiv
 \Tint{Q,R} \frac{1}{Q^2R^4(R-P)^2}P_T(R)^2
 \;,\displaybreak[0] \\
 \mathcal{I}^{0}_\rmi{e} & \equiv
 \Tint{Q,R} \frac{P^2}{Q^2R^2(Q-R)^2}
 \;, \\
\mathcal{I}_\rmi{f}^0 & \equiv
 \Tint{Q,R} \frac{P^2}{Q^2(Q-R)^2(R-P)^2}
 \;, \displaybreak[0] \\
 \mathcal{I}_\rmi{f}^1 & \equiv
 \Tint{Q,R} \frac{1}{Q^2(Q-R)^2(R-P)^2}P_T(Q)
 \;,\displaybreak[0] \\
 \mathcal{I}^{0}_\rmi{g} & \equiv
 \Tint{Q,R} \frac{P^4}{Q^2(Q-P)^2R^2(R-P)^2}
 \;, \\
\mathcal{I}_\rmi{h}^0 & \equiv
 \Tint{Q,R} \frac{P^4}{Q^2R^2(Q-R)^2(R-P)^2}
 \;, \la{def_Ih0} \displaybreak[0] \\
\mathcal{I}_\rmi{h}^1 & \equiv
 \Tint{Q,R} \frac{P^2}{Q^2R^2(Q-R)^2(R-P)^2}P_T(Q)
 \;, \la{def_Ih1} \displaybreak[0] \\
\mathcal{I}_\rmi{h}^2 & \equiv
 \Tint{Q,R} \frac{P^2}{Q^2R^2(Q-R)^2(R-P)^2}P_T(R)
 \;, \la{def_Ih2} \displaybreak[0] \\
\mathcal{I}_\rmi{h}^3 & \equiv
 \Tint{Q,R} \frac{P^4}{Q^2R^4(Q-R)^2(R-P)^2}P_T(R)
 \;, \la{def_Ih3} \displaybreak[0] \\
\mathcal{I}_\rmi{h}^4 & \equiv
 \Tint{Q,R} \frac{P^2}{Q^2R^4(Q-R)^2(R-P)^2}P_T(Q)^2
 \;, \la{def_Ih4} \displaybreak[0] \\
\mathcal{I}_\rmi{h}^5 & \equiv
 \Tint{Q,R} \frac{P^2}{Q^2R^4(Q-R)^2(R-P)^2}P_T(R)^2
 \;, \la{def_Ih5} \displaybreak[0] \\
\mathcal{I}_\rmi{h}^6 & \equiv
 \Tint{Q,R} \frac{P^2}{Q^2R^4(Q-R)^2(R-P)^2}P_T(Q)P_T(R)
 \;, \la{def_Ih6} \displaybreak[0] \\
\mathcal{I}_\rmi{h}^7 & \equiv
 \Tint{Q,R} \frac{P^2}{Q^2R^4(Q-R)^2(R-P)^2}P_T(Q)P_T(Q-R)
 \;, \la{def_Ih7} \displaybreak[0] \\
 \mathcal{I}_\rmi{i}^0 & \equiv
 \Tint{Q,R} \frac{(Q-P)^4}{Q^2R^2(Q-R)^2(R-P)^2}
 \;,\displaybreak[0] \\
 \mathcal{I}_\rmi{i}^1 & \equiv
 \Tint{Q,R} \frac{(Q-P)^2}{Q^2R^2(Q-R)^2(R-P)^2}P_T(Q)
 \;,\displaybreak[0] \\
 \mathcal{I}_\rmi{i}^2 & \equiv
 \Tint{Q,R} \frac{P^2(Q-P)^2}{Q^2R^4(Q-R)^2(R-P)^2}P_T(Q)
 \;,\displaybreak[0] \\
 \mathcal{I}_\rmi{i}^3 & \equiv
 \Tint{Q,R} \frac{(Q-P)^4}{Q^2R^4(Q-R)^2(R-P)^2}P_T(R)
 \;,\displaybreak[0] \\
 \mathcal{I}_\rmi{i'} & \equiv
 \Tint{Q,R} \frac{4(Q\cdot P)^2}{Q^2R^2(Q-R)^2(R-P)^2}
 \;, \displaybreak[0]  \label{ii'}\\
 \mathcal{I}_\rmi{j}^0 & \equiv
 \Tint{Q,R} \frac{P^6}{Q^2R^2(Q-R)^2(Q-P)^2(R-P)^2}
 \;, \la{m_last}\displaybreak[0] \\
 \mathcal{I}_\rmi{j}^1 & \equiv
 \Tint{Q,R} \frac{P^4}{Q^2R^2(Q-R)^2(Q-P)^2(R-P)^2}P_T(Q)
 \;,\displaybreak[0] \\
 \mathcal{I}_\rmi{j}^2 & \equiv
 \Tint{Q,R} \frac{P^4}{Q^2R^2(Q-R)^2(Q-P)^2(R-P)^2}P_T(Q-R)
 \;,\displaybreak[0] \\
 \mathcal{I}_\rmi{j}^3 & \equiv
 \Tint{Q,R} \frac{P^2}{Q^2R^2(Q-R)^2(Q-P)^2(R-P)^2}P_T(Q)^2
 \;,\displaybreak[0] \\
 \mathcal{I}_\rmi{j}^4 & \equiv
 \Tint{Q,R} \frac{P^2}{Q^2R^2(Q-R)^2(Q-P)^2(R-P)^2}P_T(Q-R)^2
 \;,\displaybreak[0] \\
 \mathcal{I}_\rmi{j}^5 & \equiv
 \Tint{Q,R} \frac{P^2}{Q^2R^2(Q-R)^2(Q-P)^2(R-P)^2}P_T(Q)P_T(R)
 \;,\displaybreak[0] \\
\mathcal{I}_\rmi{j}^6 & \equiv
 \Tint{Q,R} \frac{P^2}{Q^2R^2(Q-R)^2(Q-P)^2(R-P)^2}P_T(Q)P_T(Q-R)
 \;.
\end{align}
In these expressions, we have defined $P_T(Q) \equiv Q_\mu Q_\nu P^T_{\mu\nu}(P) = \mathbf{q}^2-(\mathbf{q\cdot \hat p})^2$, where $P^T_{\mu\nu}(P)$ is the transverse projection operator discussed in section \ref{subs:corre_shear}.

%
%
\mychapter{Integrands in the shear channel}
\pagestyle{myheadings}
\la{fs}
In section \ref{se:spf_shear}, we reported a number of relations between the integrands $f_{\mathcal{I}^{n}_\rmii{x}}$ of different master spectral functions, relating them to cases encountered in \cite{Laine:2011xm} (albeit with a different IR regulator). Using a notation, where $E_r\equiv \sqrt{r^2+m^2}$, $E_{qr}\equiv |\mathbf{q}-\mathbf{r}|$, and $n_{qr}\equiv n_{E_{qr}}$, the latter read:
\small{
\ba
 f^{ }_{\mathcal{J}^{0}_\rmii{b}} &=&  
  \frac{\omega^4\pi}{4 q^2}
 \Bigl[ \delta(\omega - 2 q) - \delta(\omega + 2 q) \Bigr]
 (1+2 n_q)
 \;, \la{fJb0} \\
 f^{ }_{\mathcal{I}^{0}_\rmii{b}} &=&  
  -\frac{\omega^2\pi}{8 q r^2}
 \Bigl[ \delta(\omega - 2 r) - \delta(\omega + 2 r) \Bigr]
 (1+2 n_q) (1+2 n_r)
 \;, \la{fIb0} \\
 f^{ }_{\mathcal{I}^{0}_\rmii{d}} &=&
 -\fr12\lim_{m\to 0} 
 \Bigg\{\frac{{\rm d}}{{\rm d}m^2} \frac{\omega^4\pi}{8 q E_r^2}
 \Bigl[ \delta(\omega - 2 E_r) - \delta(\omega + 2 E_r) \Bigr]
 (1+2 n_q) \bigl( 1+2 n_{E_r}\, \bigr)\Biggr\}
 \;, \la{fId0} \\
f^{ }_{\mathcal{I}^{0}_\rmii{f}} &=& 
 \frac{\omega^2 \pi }{8 q r E_{qr}} \Bigg\{ 
 \la{If} \\
 \nn & - & \!\!
 \Bigl[\delta(\omega - q - r -E_{qr}) - \delta(\omega+q+r+E_{qr}) \Bigr]
 \Bigl[ {(1+n_{qr})(1+n_q+n_r)+n_q n_r} \Bigr]
 \nn & - &  \!\!
 \Bigl[\delta(\omega-q-r+E_{qr}) - \delta(\omega + q + r -E_{qr}) \Bigr]
 \Bigl[ {n_{qr}(1+n_q + n_r )-n_qn_r} \Bigr]
 \nn & - &  \!\!
 \Bigl[\delta(\omega - q + r -E_{qr}) - \delta(\omega+q-r+E_{qr}) \Bigr]
 \Bigl[ {n_r(1+n_q+n_{qr})-n_q n_{qr}} \Bigr]
 \nn & - &  \!\!
 \Bigl[\delta(\omega + q - r -E_{qr}) - \delta(\omega-q+r+E_{qr}) \Bigr]
 \Bigl[ {n_q(1+n_r+n_{qr})-n_r n_{qr}} \Bigr]
 \Biggr\} \;, \la{fIf0} \nn
 f_{\mathcal{I}^{0}_\rmii{h}} &=& 
 \frac{\omega^4 \pi }{8 q r E_{qr}} \Bigg\{  \la{fIh0} \\ 
 & & \!\!
 \frac{1}{2E_r} 
 \Bigl[\delta(\omega - E_{r} - r) - \delta(\omega + E_{r}+r) \Bigr]
 \times 
 \nn & & \times \biggl[
 \biggl( 
 \frac{1}{q+E_{r}-E_{qr}} +  
 \frac{1}{q-E_{r}-E_{qr}} 
 \biggr)
 (1+n_{E_r}+n_r)(n_{qr}-n_q)
 \nn & & \;\; 
 +
 \biggl(
 \frac{1}{q+E_{r}+E_{qr}} +  
 \frac{1}{q-E_{r}+E_{qr}} 
 \biggr)
 (1+n_{E_r}+n_r)(1 + n_{qr}+n_q)
 \biggr]
 \nn &+& \!\!
 \frac{1}{2E_r} 
 \Bigl[\delta(\omega + E_{r} - r) - \delta(\omega - E_{r}+r) \Bigr]
 \times 
 \nn & & \times \biggl[
 \biggl( 
 \frac{1}{q+E_{r}-E_{qr}} +  
 \frac{1}{q-E_{r}-E_{qr}} 
 \biggr)
 (n_{E_r}-n_r)(n_{qr}-n_q)
 \nn & & \;\; 
 +
 \biggl(
 \frac{1}{q+E_{r}+E_{qr}} +  
 \frac{1}{q-E_{r}+E_{qr}} 
 \biggr)
 (n_{E_r}-n_r)(1 + n_{qr}+n_q)
 \biggr]
 \nn & - & \!\!
 \Bigl[\delta(\omega - q - r -E_{qr}) - \delta(\omega+q+r+E_{qr}) \Bigr]
 \frac{(1+n_{qr})(1+n_q+n_r)+n_q n_r}
      {(q+E_{r}+E_{qr})(q-E_{r}+E_{qr})}
 \nn & - &  \!\!
 \Bigl[ \delta(\omega-q-r+E_{qr}) - \delta(\omega + q + r -E_{qr}) \Bigr]
 \frac{n_{qr}(1+n_q + n_r )-n_qn_r}
      {(q+E_{r}-E_{qr})(q-E_{r}-E_{qr})}
 \nn & - &  \!\!
 \Bigl[\delta(\omega - q + r -E_{qr}) - \delta(\omega+q-r+E_{qr}) \Bigr]
 \frac{n_r(1+n_q+n_{qr})-n_q n_{qr}}
      {(q-E_{r}+E_{qr})(q+E_{r}+E_{qr})}
 \nn & - &  \!\!
 \Bigl[\delta(\omega + q - r -E_{qr}) - \delta(\omega-q+r+E_{qr}) \Bigr]
 \frac{n_q(1+n_r+n_{qr})-n_r n_{qr}}
      {(q-E_{r}-E_{qr})(q+E_{r}-E_{qr})}
 \Biggr\}
 \;,  \nonumber
 \\
 f_{\mathcal{I}^{}_\rmii{i'}}
 &=& 
 \frac{\omega^2 \pi q }{2r E_{qr}} \Bigg\{  \la{fIip}
 \\ 
 & & \!\!
 \frac{1}{2E_r} 
 \Bigl[\delta(\omega - E_{r} - r) - \delta(\omega + E_{r}+r) \Bigr]
 \times 
 \nn & & \times 
 \biggl[-\frac{2}{q}(1+n_{E_r}+n_r)(1+2n_{qr})
 \nn & & \;\; 
 +
 \biggl( 
 \frac{1}{q+E_{r}-E_{qr}} +  
 \frac{1}{q-E_{r}-E_{qr}} 
 \biggr)
 (1+n_{E_r}+n_r)(n_{qr}-n_q)
 \nn & & \;\; 
 +
 \biggl(
 \frac{1}{q+E_{r}+E_{qr}} +  
 \frac{1}{q-E_{r}+E_{qr}} 
 \biggr)
 (1+n_{E_r}+n_r)(1 + n_{qr}+n_q)
 \biggr]
 \nn &+& \!\!
 \frac{1}{2E_r} 
 \Bigl[\delta(\omega + E_{r} - r) - \delta(\omega - E_{r}+r) \Bigr]
 \times 
 \nn & & \times 
 \biggl[-\frac{2}{q}(n_{E_r}-n_r)(1+2n_{qr})
 \nn & & \;\; 
 +
 \biggl( 
 \frac{1}{q+E_{r}-E_{qr}} +  
 \frac{1}{q-E_{r}-E_{qr}} 
 \biggr)
 (n_{E_r}-n_r)(n_{qr}-n_q)
 \nn & & \;\; 
 +
 \biggl(
 \frac{1}{q+E_{r}+E_{qr}} +  
 \frac{1}{q-E_{r}+E_{qr}} 
 \biggr)
 (n_{E_r}-n_r)(1 + n_{qr}+n_q)
 \biggr]
 \nn & - & \!\!
 \Bigl[\delta(\omega - q - r -E_{qr}) - \delta(\omega+q+r+E_{qr}) \Bigr]
 \frac{(1+n_{qr})(1+n_q+n_r)+n_q n_r}
      {(q+E_{r}+E_{qr})(q-E_{r}+E_{qr})}
 \nn & - &  \!\!
 \Bigl[ \delta(\omega-q-r+E_{qr}) - \delta(\omega + q + r -E_{qr}) \Bigr]
 \frac{n_{qr}(1+n_q + n_r )-n_qn_r}
      {(q+E_{r}-E_{qr})(q-E_{r}-E_{qr})}
 \nn & - &  \!\!
 \Bigl[\delta(\omega - q + r -E_{qr}) - \delta(\omega+q-r+E_{qr}) \Bigr]
 \frac{n_r(1+n_q+n_{qr})-n_q n_{qr}}
      {(q-E_{r}+E_{qr})(q+E_{r}+E_{qr})}
 \nn & - &  \!\!
 \Bigl[\delta(\omega + q - r -E_{qr}) - \delta(\omega-q+r+E_{qr}) \Bigr]
 \frac{n_q(1+n_r+n_{qr})-n_r n_{qr}}
      {(q-E_{r}-E_{qr})(q+E_{r}-E_{qr})}
 \Biggr\} \;, \hspace*{1cm} \nonumber
\\
 f_{\mathcal{I}^{1}_\rmii{i}} 
 &=& \frac{D-2}{D-1}
 \frac{\omega \pi q }{8 r E_{qr}} \Bigg\{  \la{fIi1} \\
 & & \!\!
 \frac{1}{2E_r} 
 \Bigl[\delta(\omega - E_{r} - r)\Bigr]
 \biggl[\frac{2q}{\omega}(1+n_{E_r}+n_r)(1+2n_{qr})
 \nn & & \;\; 
 -
 \biggl( 
 \frac{\omega+2q}{q+E_{r}-E_{qr}} +  
 \frac{\omega-2q}{q-E_{r}-E_{qr}} 
 \biggr)
 (1+n_{E_r}+n_r)(n_{qr}-n_q)
 \nn & & \;\; 
 -
 \biggl(
 \frac{\omega+2q}{q+E_{r}+E_{qr}} +  
 \frac{\omega-2q}{q-E_{r}+E_{qr}} 
 \biggr)
 (1+n_{E_r}+n_r)(1 + n_{qr}+n_q)
 \biggr]
 \nn &-& \!\!
 \frac{1}{2E_r} 
 \Bigl[\delta(\omega + E_{r}+r) \Bigr] 
 \biggl[\frac{2q}{\omega}(1+n_{E_r}+n_r)(1+2n_{qr})
 \nn & & \;\; 
 -
 \biggl( 
 \frac{\omega-2q}{q+E_{r}-E_{qr}} +  
 \frac{\omega+2q}{q-E_{r}-E_{qr}} 
 \biggr)
 (1+n_{E_r}+n_r)(n_{qr}-n_q)
 \nn & & \;\; 
 -
 \biggl(
 \frac{\omega-2q}{q+E_{r}+E_{qr}} +  
 \frac{\omega+2q}{q-E_{r}+E_{qr}} 
 \biggr)
 (1+n_{E_r}+n_r)(1 + n_{qr}+n_q)
 \biggr]
 \nn &+& \!\! 
 \frac{1}{2E_r} 
 \Bigl[\delta(\omega + E_{r} - r)\Bigr] 
 \biggl[\frac{2q}{\omega}(n_{E_r}-n_r)(1+2n_{qr})
 \nn & & \;\; 
 -
 \biggl( 
 \frac{\omega-2q}{q+E_{r}-E_{qr}} +  
 \frac{\omega+2q}{q-E_{r}-E_{qr}} 
 \biggr)
 (n_{E_r}-n_r)(n_{qr}-n_q)
 \nn & & \;\; 
 -
 \biggl(
 \frac{\omega-2q}{q+E_{r}+E_{qr}} +  
 \frac{\omega+2q}{q-E_{r}+E_{qr}} 
 \biggr)
 (n_{E_r}-n_r)(1 + n_{qr}+n_q)
 \biggr]
  \nn &-& \!\!
 \frac{1}{2E_r} 
 \Bigl[\delta(\omega - E_{r}+r) \Bigr]
 \biggl[\frac{2q}{\omega}(n_{E_r}-n_r)(1+2n_{qr})
 \nn & & \;\; 
 -
 \biggl( 
 \frac{\omega+2q}{q+E_{r}-E_{qr}} +  
 \frac{\omega-2q}{q-E_{r}-E_{qr}} 
 \biggr)
 (n_{E_r}-n_r)(n_{qr}-n_q)
 \nn & & \;\; 
 -
 \biggl(
 \frac{\omega+2q}{q+E_{r}+E_{qr}} +  
 \frac{\omega-2q}{q-E_{r}+E_{qr}} 
 \biggr)
 (n_{E_r}-n_r)(1 + n_{qr}+n_q)
 \biggr]
 \nn & - & \!\!
 \Bigl[\delta(\omega - q - r -E_{qr})(2q-\omega) + \delta(\omega+q+r+E_{qr}) (2q+\omega)\Bigr]
 \nn &&\times
 \frac{(1+n_{qr})(1+n_q+n_r)+n_q n_r}
      {(q+E_{r}+E_{qr})(q-E_{r}+E_{qr})}
 \nn & - &  \!\!
 \Bigl[ \delta(\omega-q-r+E_{qr})(2q-\omega) + \delta(\omega + q + r -E_{qr})(2q+\omega) \Bigr]
 \nn &&\times
 \frac{n_{qr}(1+n_q + n_r )-n_qn_r}
      {(q+E_{r}-E_{qr})(q-E_{r}-E_{qr})}
 \nn & - &  \!\!
 \Bigl[\delta(\omega - q + r -E_{qr})(2q-\omega) + \delta(\omega+q-r+E_{qr}) (2q+\omega)\Bigr]
 \nn &&\times
 \frac{n_r(1+n_q+n_{qr})-n_q n_{qr}}
      {(q-E_{r}+E_{qr})(q+E_{r}+E_{qr})}
 \nn & + &  \!\!
 \Bigl[\delta(\omega + q - r -E_{qr})(2q+\omega) + \delta(\omega-q+r+E_{qr})(2q-\omega) \Bigr]
 \nn &&\times
 \frac{n_q(1+n_r+n_{qr})-n_r n_{qr}}
      {(q-E_{r}-E_{qr})(q+E_{r}-E_{qr})}
 \Biggr\} \;, \nonumber
\\
 f_{\mathcal{I}^{0}_\rmii{j}}& =& 
 \frac{\omega^6 \pi }{4 q r E_{qr}} \Bigg\{ 
 \la{fIj0} \\
 & & \!\!
 \frac{1}{8q^2} 
 \Bigl[\delta(\omega - 2 q) - \delta(\omega+2 q) \Bigr]
 \times 
 \nn & & \times \biggl[
 \biggl( 
 \frac{1}{(q+r-E_{qr})(q+r)} -  
 \frac{1}{(q-r+E_{qr})(q-r)} 
 \biggr)
 (1 + 2 n_q) (n_{qr}-n_r)
 \nn & & \;\; 
 +
 \biggl(
 \frac{1}{(q+r+E_{qr})(q+r)} -  
 \frac{1}{(q-r-E_{qr})(q-r)} 
 \biggr)
 (1 + 2 n_q) (1 + n_{qr}+n_r)
 \biggr]
 \nn & + & \!\!
 \frac{1}{8r^2} 
 \Bigl[\delta(\omega - 2 r) - \delta(\omega+2 r) \Bigr]
 \times 
 \nn & & \times \biggl[
 \biggl( 
 \frac{1}{(q+r-E_{qr})(q+r)} -  
 \frac{1}{(q-r-E_{qr})(q-r)} 
 \biggr)
 (1+2n_r)(n_{qr}-n_q)
 \nn & & \;\; 
 +
 \biggl(
 \frac{1}{(q+r+E_{qr})(q+r)} -  
 \frac{1}{(q-r+E_{qr})(q-r)} 
 \biggr)
 (1+2n_r)(1 + n_{qr}+n_q)
 \biggr]
 \nn & + & \!\!
 \Bigl[\delta(\omega - q - r -E_{qr}) - \delta(\omega+q+r+E_{qr}) \Bigr]
 \frac{(1+n_{qr})(1+n_q+n_r)+n_q n_r}
      {(q+r+E_{qr})^2(q-r+E_{qr})(q-r-E_{qr})}
 \nn & + &  \!\!
 \Bigl[\delta(\omega-q-r+E_{qr}) - \delta(\omega + q + r -E_{qr}) \Bigr]
 \frac{n_{qr}(1+n_q + n_r ) - n_qn_r}
      {(q+r-E_{qr})^2(q-r+E_{qr})(q-r-E_{qr})}
 \nn & + &  \!\!
 \Bigl[\delta(\omega - q + r -E_{qr}) - \delta(\omega+q-r+E_{qr}) \Bigr]
 \frac{n_r(1+n_q+n_{qr})-n_q n_{qr}}
      {(q-r+E_{qr})^2(q+r+E_{qr})(q+r-E_{qr})}
 \nn & + &  \!\!
 \Bigl[\delta(\omega + q - r -E_{qr}) - \delta(\omega-q+r+E_{qr}) \Bigr]
 \frac{n_q(1+n_r+n_{qr})-n_r n_{qr}}
      {(q-r-E_{qr})^2(q+r+E_{qr})(q+r-E_{qr})}
 \Biggr\} \;. \nonumber
\ea
}

\end{appendix}

\pagestyle{empty}


\end{document}